\documentclass[longauth]{aa}
\usepackage{graphicx}
\usepackage{txfonts}
\usepackage[colorlinks=true,citecolor=blue,backref=page]{hyperref}
\usepackage{xspace}
\usepackage{supertabular}
\usepackage{longtable}
\usepackage{rotating}
\newcommand{\idm}{\mathrm}

\newcommand{\host}{K2-223}
\newcommand{\hostb}{{K2-223\,b}}
\newcommand{\hostc}{{K2-223\,c}}
\newcommand{\hostd}{{K2-223\,d}}
\newcommand{\planetb}{{K2-223\,b}}
\newcommand{\planetc}{{K2-223\,c}}

\newcommand{\steff}{$T_\mathrm{eff}$}
\newcommand{\logsg}{$\log\,g_{\star}$}
\newcommand{\sfeh}{$\mathrm{[Fe/H]}$}

\newcommand{\sm}{$M_{\star}$}
\newcommand{\sr}{$R_{\star}$}
\newcommand{\sden}{$\rho_{\star}$}
\newcommand{\slum}{$L_{\star}$}

\newcommand{\svrotsini}{$\mathit{v}_\mathrm{rot} \sin\,i_{\star}$}

\newcommand{\sprot}{$P_\mathrm{rot}$}

\newcommand{\logrhk}{$\log\,R^{'}_\mathrm{HK}$}
\newcommand{\soamp}{$O_\mathrm{amp}$}
\newcommand{\spdec}{$P_\mathrm{dec}$}

\newcommand{\Msun}{$\mathrm{M_{\odot}}$}
\newcommand{\Rsun}{$\mathrm{R_{\odot}}$}

\newcommand{\Lsun}{$\mathrm{L_{\odot}}$}

\newcommand{\mps}{$\mathrm{m\,s^{-1}}$}

\newcommand{\kmps}{$\mathrm{km\,s^{-1}}$}

\newcommand{\cmpss}{$\mathrm{cm\,s^{-2}}$}

\newcommand{\dex}{$\mathrm{dex}$}
\newcommand{\pxtzero}{$T_{0}$}
\newcommand{\pxporb}{$P$}
\newcommand{\pxporbR}{$P_{\rm orb}$}
\newcommand{\pxe}{$e$}
\newcommand{\pxw}{$\omega$}
\newcommand{\pxpsrr}{$R/R_\star$}
\newcommand{\pxb}{$b$}
\newcommand{\pxar}{$a/R_\star$}
\newcommand{\pxa}{$a$}
\newcommand{\pxi}{$i$}

\newcommand{\pxk}{$K$}
\newcommand{\pxr}{$R$}
\newcommand{\pxm}{$M$}
\newcommand{\pxmsini}{$M \sin\,i$}
\newcommand{\pxden}{$\rho$}

\newcommand{\pxteq}{$T_\mathrm{eq}$}

\newcommand{\ppa}{$a_p$}

\newcommand{\ppr}{$R_p$}

\newcommand{\pbporb}{$P_\mathrm{b}$}

\newcommand{\pbar}{$a_\mathrm{b}/R_\star$}

\newcommand{\pbi}{$i_{b}$}

\newcommand{\pbk}{$K_\mathrm{b}$}
\newcommand{\pbr}{$R_\mathrm{b}$}
\newcommand{\pbm}{$M_\mathrm{b}$}

\newcommand{\pcporb}{$P_\mathrm{c}$}
\newcommand{\pce}{$e_\mathrm{c}$}

\newcommand{\pcttof}{$T_{1,4}$}
\newcommand{\pctttt}{$T_{2,3}$}
\newcommand{\pck}{$K_\mathrm{c}$}
\newcommand{\pcr}{$R_\mathrm{c}$}
\newcommand{\pcm}{$M_\mathrm{c}$}

\newcommand{\pcden}{$\rho_\mathrm{c}$}

\newcommand{\pdporb}{$P_\mathrm{d}$}

\newcommand{\pdi}{$i_{d}$}

\newcommand{\pdmsini}{$M_\mathrm{d} \sin\,i_\mathrm{d}$}

\newcommand{\pbav}[1][$\mathrm{AU}$]{$0.01271_{-0.00028}^{+0.00027}$~#1}

\newcommand{\pbrv}[1][$\mathrm{R_{\oplus}}$]{$0.79$\,$\pm$\,$0.10$~#1}

\newcommand{\pbmtsulv}[1][$\mathrm{M_{\oplus}}$]{${2.8}$~#1}

\newcommand{\pbteqv}[1][$\mathrm{K}$]{$2480^{+103}_{-72}$~#1}
\newcommand{\pbtsmv}[1][dimensionless]{$2.020^{+9.871}_{-1.357}$~#1}
\newcommand{\pbesmv}[1][dimensionless]{$0.760^{+0.451}_{-0.287}$~#1}
\newcommand{\pcporbv}[1][$\mathrm{d}$]{$4.56395_{-0.00033}^{+0.00034}$~#1}
\newcommand{\pcporbprochev}{$29.0$\,$\pm$\,$4.9$}

\newcommand{\pcav}[1][$\mathrm{AU}$]{$0.0551$\,$\pm$\,$0.0012$~#1}

\newcommand{\pcrv}[1][$\mathrm{R_{\oplus}}$]{$1.41$\,$\pm$\,$0.15$~#1}
\newcommand{\pcmv}[1][$\mathrm{M_{\oplus}}$]{$4.2$\,$\pm$\,$1.3$~#1}
\newcommand{\pcdenv}[1][$\mathrm{g\,cm^{-3}}$]{$8.3$\,$\pm$\,$3.7$~#1}
\newcommand{\pcdenEv}[1][$\mathrm{\rho_{\oplus}}$]{$1.50$\,$\pm$\,$0.67$~#1}
\newcommand{\pcteqv}[1][$\mathrm{K}$]{$1191^{+50}_{-34}$~#1}
\newcommand{\pctsmv}[1][dimensionless]{$0.844^{+1.062}_{-0.432}$~#1}
\newcommand{\pcesmv}[1][dimensionless]{$0.724^{+0.365}_{-0.240}$~#1}
\newcommand{\pdporbv}[1][$\mathrm{years}$]{$4.53^{+0.32}_{-0.34}$~#1}

\newcommand{\pdav}[1][$\mathrm{AU}$]{$2.80$\,$\pm$\,$0.15$~#1}

\newcommand{\pdmsiniv}[1][$\mathrm{M_{\rm Jup}}$]{$1.29^{+0.23}_{-0.17}$~#1}

\newcommand{\Mea}{$\mathrm{M_{\oplus}}$}
\newcommand{\Rea}{$\mathrm{R_{\oplus}}$}
\newcommand{\Dea}{$\mathrm{\rho_{\oplus}}$}
\newcommand{\Mjup}{$\mathrm{M_\mathrm{Jup}}$}

\newcommand{\ppdenu}{$\mathrm{g\,cm^{-3}}$}

\newcommand{\corot}{CoRoT}
\newcommand{\kepler}{\emph{{\it Kepler}}}
\newcommand{\ktwo}{K2}
\newcommand{\tess}{TESS}

\begin{document} 

\title{Mass constraints for the K2-223 system planets}
\subtitle{An ultra-short-period sub-Earth, a short-period super-Earth, and a tentative long-period giant planet}

\author{Dawid~Jankowski\inst{\ref{ia_ncu}}
\and Grzegorz~Nowak \inst{\ref{ia_ncu}}
\and Gaia~Lacedelli\inst{\ref{iac}, \ref{ull}}
\and Enric~Pall{\'e}\inst{\ref{iac}, \ref{ull}}
\and Krzysztof~Go\'zdziewski\inst{\ref{ia_ncu}}
\and Thomas~Masseron\inst{\ref{iac}, \ref{ull}}
\and Ilaria~Carleo\inst{\ref{inaf-oato}}
\and Rafael~Luque\inst{\ref{iaa-scic}}
\and Felipe~Murgas\inst{\ref{iac},\ref{ull}}
\and Davide~Gandolfi\inst{\ref{unito}}
\and Artie~P.~Hatzes \inst{\ref{tautenburg}}
\and William~Cochran \inst{\ref{uot1},\ref{uot2}}
\and Pedro~Figueira\inst{\ref{iaa-scic}}
\and Rafael~A.~Garc\'ia\inst{\ref{cea-saclay}}
\and Samuel~Gerald{\'i}a-Gonz{\'a}lez \inst{\ref{iac},\ref{ull}}
\and Judith~Korth \inst{\ref{unige}}
\and Pierrot~Lamontagne \inst{\ref{iaa-scic}}
\and John~H.~Livingston \inst{\ref{tokyo},\ref{nrao},\ref{sokendai}}
\and Savita~Mathur\inst{\ref{iac},\ref{ull}}
\and Giuseppe~Morello\inst{\ref{iaa-scic},\ref{inaf-oapa}}
\and Jaume~Orell-Miquel\inst{\ref{utexas}}
\and Dinil~B.~Palakkatharappil\inst{\ref{cea-saclay}}
\and Carina~M.~Persson\inst{\ref{oso}}
\and Seth~Redfield\inst{\ref{wesleyan}}
\and Atanas~K.~Stefanov\inst{\ref{iac}, \ref{ull}}
\and Vincent~Van~Eylen\inst{\ref{mullard}}
}

\institute{
Institute of Astronomy, Faculty of Physics, Astronomy and Informatics, Nicolaus Copernicus University, Grudzi\c{a}dzka 5, 87-100 Toru\'n, Poland\label{ia_ncu},
\email{jankowski@doktorant.umk.pl} 
\and Instituto de Astrof\'{i}sica de Canarias (IAC), 38205 La Laguna, Tenerife, Spain\label{iac}
\and Departamento de Astrof\'isica, Universidad de La Laguna (ULL), E-38206 La Laguna, Tenerife, Spain\label{ull}
\and INAF -- Osservatorio Astrofisico di Torino, Via Osservatorio 20, I-10025, Pino Torinese, Italy\label{inaf-oato}
\and Instituto de Astrof\'isica de Andaluc\'ia (IAA-CSIC), Glorieta de la Astronom\'ia s/n, 18008 Granada, Spain\label{iaa-scic}
\and Dipartimento di Fisica, Universit{\'a} degli Studi di Torino, via Pietro Giuria 1, I-10125, Torino, Italy\label{unito}
\and Th\"uringer Landessternwarte Sternwarte 5 D-07778, Tautenburg, Germany\label{tautenburg}
\and McDonald Observatory, The University of Texas, Austin Texas USA\label{uot1}
\and Center for Planetary Systems Habitability, The University of Texas, Austin Texas USA\label{uot2}
\and Universit\'e Paris-Saclay, Universit\'e Paris Cit\'e, CEA, CNRS, AIM, 91191, Gif-sur-Yvette, France\label{cea-saclay}
\and Astronomical Observatory, University of Geneva, Chemin Pegasi 51b, CH-1290 Versoix, Switzerland\label{unige}
\and Astrobiology Center, 2-21-1 Osawa, Mitaka, Tokyo 181-8588, Japan\label{tokyo}
\and National Astronomical Observatory of Japan, 2-21-1 Osawa, Mitaka, Tokyo 181-8588, Japan\label{nrao}
\and Department of Astronomy, The Graduate University for Advanced Studies (SOKENDAI), 2-21-1 Osawa, Mitaka, Tokyo, Japan\label{sokendai}
\and INAF - Osservatorio Astronomico di Palermo, Piazza del Parlamento, 1, 90134 Palermo, Italy\label{inaf-oapa}
\and The University of Texas at Austin, 2515 Speedway, Stop C1402, Austin, Texas 78712-1206\label{utexas}
\and Department of Physics and Astronomy, Chalmers University of Technology, Onsala Space Observatory, SE-439 92 Onsala, Sweden\label{oso}
\and Astronomy Department and Van Vleck Observatory, Wesleyan University, Middletown, CT 06459, USA\label{wesleyan}
\and Mullard Space Science Laboratory, University College London, Dorking, Surrey, RH5 6NT\label{mullard}
}

\authorrunning{D. Jankowski et al.}
\titlerunning{Mass constraints for the \host{} system planets}
\date{Received 31 October, 2025 / Accepted 31 July, 2026}
\abstract
{We present mass constraints of two short-period Earth-sized planets transiting K2-223 based on high-precision radial velocity measurements from HARPS-N and ESPRESSO, as well as a tentative indication of an outer Jupiter-like planet orbiting the G1V dwarf K2-223. With a radius of $R_\mathrm{b}$\,=\,$0.79$\,$\pm$\,$0.10$ $\mathrm{R_{\oplus}}$ and a 3$\sigma$ upper mass limit $M_\mathrm{b}$\,${<}$\,${2.8}$~$\mathrm{M_{\oplus}}$, K2-223\,b belongs to the small group of known sub-Earth planets and is currently the smallest known ultra-short-period planet ($P_\mathrm{b}$\,$\approx$\,0.5\,day) transiting a solar-type star. With a radius of $R_\mathrm{c}$\,=\,$1.41$\,$\pm$\,$0.15$ $\mathrm{R_{\oplus}}$, a mass of~$M_\mathrm{c}$\,=\,$4.2$\,$\pm$\,$1.3$~$\mathrm{M_{\oplus}}$, and a density of $\rho_\mathrm{c}$\,=\,$8.3$\,$\pm$\,$3.7$ $\mathrm{g\,cm^{-3}}$, K2-223\,c is a short-period ($P_\mathrm{c}$\,$\approx$\,4.5\,days) super-Earth. Based on almost six and a~half years of radial velocity monitoring of K2-223 with the HARPS-N spectrograph, we identified a tentative giant planet with an orbital period of~$P_\mathrm{d}$\,=\,$4.53^{+0.32}_{-0.34}$ $\mathrm{years}$ and a minimum mass of $1.29^{+0.23}_{-0.17}$ $\mathrm{M_{\rm Jup}}$. Two close-in small planets accompanied by a distant candidate Jupiter-like companion would make K2-223 a system with a rare architecture. This is valuable for testing scenarios of planetary formation and evolution. The extreme proximity of K2-223\,b to the parent star means that relativistic and tidal perturbations to Newtonian gravity need to be considered. We discuss the timescales and amplitudes of these effects in the context of the radial velocity model and dynamical simulations of the K2-223 multiple-planet system. We also present a parametrisation of the planets with directly measured masses, radii, and bulk densities in terms of the orbital period normalised by the Roche period. This provides an alternative representation in the context of the Neptune desert.
}

\maketitle

\section{Introduction}
\label{sec-introduction}
Space-based missions such as Convection, Rotation and planetary Transits \citep[\corot{}:][]{corot-2009A&A...506..411A}, \kepler{} \citep{kepler-2010Sci...327..977B}, and the Transiting Exoplanet Survey Satellite \citep[\tess{}:][]{tess-2015JATIS...1a4003R} revolutionised the field of extrasolar planetary systems by revealing the great diversity of planets. One of the most intriguing groups of planets are ultra-short-period (USP) and short-period planets, which are not present in our Solar System. We define USP planets as having orbital periods shorter than 1 day and short-period planets as having periods shorter than 10 days, but other definitions are also recognised \citep[e.g.][]{Goyal-Songhu-2025AJ....169..191G}. The USP planets are especially interesting as they allow us not only to verify planetary formation scenarios \citep[e.g.][]{Winn-2017AJ....154...60W, Zhu-2025ApJ...991..206Z}, but through their proximity to the parent stars also to study interactions with the hosts \citep{Lanza-2021A&A...653A.112L} and to monitor the final stages of planetary evolution in real time, including decaying orbits \citep{Dai-2024AJ....168..101D} and the process of planetary disintegration \citep{Rappaport-2012ApJ...752....1R,Sanchis-Ojeda-2015ApJ...812..112S}. The occurrence rate of USP planets strongly decreases with increasing mass of the host star from 1.1\,$\pm$\,0.4\,\% for M dwarfs through 0.51\,$\pm$\,0.07\,\% for G dwarfs and 0.15\,$\pm$\,0.05\,\% for F dwarfs \citep{Winn-2018NewAR..83...37W}, and it increases with the age of the parent stars \citep{Schmidt-2024AJ....168..109S,Tu-2025NatAs...9..995T}. It is a very diverse group of planets composed of ultra-hot Jupiters, such as WASP-18\,b \citep{Hellier-2009Natur.460.1098H}, ultra-hot Neptunes and sub-Neptunes, such as TOI-3261\,b \citep{Nabbie-2024AJ....168..132N}, and ultra-hot Earth-sized (\ppr{}\,$\leq$\,2\,\Rea{}) planets, such as HD 20329\,b \citep{Murgas-2022A&A...668A.158M}. Earth-sized USPs are thought form in the process of photoevaporation of sub-Neptunes \citep{Lopez-2017MNRAS.472..245L,Dai-2021AJ....162...62D}, and this includes scenarios involving tidal disruption \citep{Matsakos2016} or high-eccentricity tidal migration of rocky planets driven by secular chaos from periods of $\sim$5-10 days \citep{Petrovich-2019AJ....157..180P} rather than by stripping envelopes of giant planets \citep{Konigl-2017ApJ...846L..13K,Uzsoy-2021ApJ...919...26U}. However, other migration pathways have also been proposed in the literature, including low-eccentricity migration {\citep{Pu2019}} and obliquity-driven migration \citep{Millholland2020}. The Earth-sized USP planets are also very often accompanied by planetary companions \citep{Sanchis-Ojeda-2014ApJ...787...47S,Adams2021}. Furthermore, recent discoveries have highlighted the diversity of these extreme worlds, including ultra-dense planets such as those found in GJ 367 \citep{Lam2021, Goffo2023} and K2-360 \citep{Livingston2024}, and systems with multiple non-transiting outer companions such as TOI-500 \citep{Serrano2022} and K2-157 \citep{Castro-Gonzalez2025}. USPs associated with longer-period planetary companions are usually dynamically detached from the other planets. This is evident at least in terms of the orbital period ratio with the nearby outer planet, which is almost always greater than 3 \citep{Winn-2018NewAR..83...37W}, as well as in the higher mutual inclinations \citep{Dai-2018ApJ...864L..38D,Hua-2025ApJ...980L..46H}. Systems with USPs are therefore classified as a different subpopulation of planetary system architectures. To date, only four systems with USPs are known to host confirmed Jupiter-like companions on orbits with \ppa{}\,>\,1\,AU. 

We present the characterisation of the multi-planet system around a solar-type star \host{} known to host two transiting short-period Earth-sized planets, namely \planetb{} (\pbporb{}\,$\approx$\,0.5\,d, \pbr{}\,$\approx$\,0.8\,\Rea{}) and \planetc{} (\pcporb{}\,$\approx$\,4.5\,d, \pcr{}\,$\approx$\,1.4\,\Rea{}). The discovery of these planets in the \ktwo{} \citep{Howell2014} photometry was independently reported and statistically validated by \citet{Mayo2018} and \citet{Livingston2018} and was subsequently reanalysed by \citet{Adams2021}. The radial velocity (RV) monitoring with the High Accuracy Radial velocity Planet Searcher for the Northern hemisphere \citep[HARPS-N:][]{2012SPIE.8446E..1VC} and Echelle SPectrograph for Rocky Exoplanets and Stable Spectroscopic Observations \citep[ESPRESSO:][]{Pepe2021} spectrographs allowed us to place an upper limit on the mass of planet b (\pbm{}\,${<}$\,\pbmtsulv{}), measure the mass of planet c (\pcm{}\,=\,\pcmv{}), and to tentatively identify an~outer Jupiter-like planet, \hostd{}, with an orbital period of~\pdporb{}\,=\,\pdporbv{} and a minimum mass of~\pdmsiniv{}. We also used high-resolution spectral observations to determine precise stellar parameters of \host.

The paper is organised as follows. Sect.~\ref{sec-observations} presents the photometric and spectroscopic data we used for the analysis. Sect.~\ref{sec-stellar-modelling} describes our determination of the fundamental parameters of the \host{} star. In Sect.~\ref{sec-modelling_and_results} we present the data modelling and report the derived planetary masses and orbital parameters. Sect.~\ref{sec-non-newton} describes tidal and relativistic perturbations to Newtonian gravity in the context of estimating the validity of the Keplerian RV model. Sect.~\ref{sec-dynamic_analysis} presents the dynamical analysis of the system. In Sect.~\ref{sec-demography} we discuss \host{} in the context of planetary populations, focusing on planets with a measured mass, radius, and density. Finally, we consider the internal composition of the transiting planets \hostb{} and c, and we present our conclusions in~Sect.~\ref{sec-discussion}.

\section{Observations}
\label{sec-observations}

\subsection{K2 photometry}
\label{sec-observations-pht-k2}
\host{} was observed by the \ktwo{} mission in Campaign~10, spanning $\sim$69 days between 13 July 2016 and 20 September 2016 (UT). To perform the photometric curve analysis, we used data processed by the EVEREST pipeline \citep{Luger2016, Luger2018}. The set consisted of $\sim$2,700 measurements taken from the Mikulski Archive for Space Telescopes\footnote{\url{https://mast.stsci.edu/portal/Mashup/Clients/Mast/Portal.html}}, each with an exposure time of 1800~s in long-cadence mode. The photometric dataset is not continuous. It contains a $\sim$14-day gap between 20~July 2016 and 03 August 2016 (UT)\footnote{Caused by a pointing error in the initial 6 days and the subsequent failure of module 4 of the instrument.}.

\subsection{\tess{} photometry}
\label{sec-observations-pht-tess}
\host{} was observed by \tess{} in Sectors 36 (camera 1) between 7 March 2021 and 1 April 2021 (UT), 46 (camera 4) between 3~December 2021 and 30 December 2021 (UT), and 91 (camera~1) between 9 April 2025 and 7 May 2025 (UT). All measurements were taken with an exposure time of 120 seconds and were downloaded from the Mikulski Archive for Space Telescopes. We included two versions of the light curve in our analysis. The raw simple aperture photometry (SAP) flux was used to search for stellar activity, while the processed pre-search data conditioned SAP (PDCSAP) flux, corrected for instrumental and systematic effects, was used to model the transit signals \citep{Smith2012, Stumpe2012, Stumpe2014}. Each sector was normalised using a weighted average.

\subsection{TNG/HARPS-N spectroscopy}
\label{sec-observations-rv-harpsn}
Between 10 January 2019 (UT) and 8 June 2025 (UT), we collected 75 spectra with HARPS-N mounted at the 3.58-m Telescopio Nazionale Galileo (TNG) of Roque de los Muchachos Observatory in La Palma, Spain\footnote{Observing programs: OPT18B\_52, A38TAC\_26, CAT19A\_162, CAT21A\_119, CAT23A\_52, CAT23B\_74, CAT24B\_20, CAT25A\_76.}. Four of the spectra were excluded from the analysis because of contamination by too intense scattered Moon light ($\mathrm{BJD_{TBD}}$\,=\,2460010.64064 and 2460367.54661) or by poor and variable seeing conditions ($\mathrm{BJD_{TBD}}$\,=\,2459664.605088 and 2460072.45030) that in all four cases resulted in imprecise radial velocity measurements. The exposure time was set to 1800--3000 seconds based on weather conditions and scheduling constraints, leading to a signal-to-noise ratio (S/N) per pixel of 22.8--61.4 at 550\,nm. The spectra were extracted using the off-line version 3.7 of the HARPS-N DRS pipeline \citep{2014SPIE.9147E..8CC}. Doppler measurements and spectral activity indicators of cross-correlation functions (CCFs: CCF\_FWHM, the full width at half maximum of CCF, CCF\_CTR - CCF contrast, CCF\_BIS - CCF bisector inverse slope) were measured using an on-line version of the DRS, the YABI tool \citep{yabi}\footnote{Available at \url{http://ia2-harps.oats.inaf.it:8000}.}, by cross-correlating the extracted spectra with a G2 mask \citep{1996A&AS..119..373B}. The CCFs were constructed in a range of 30\,\kmps{} and a step of~0.25\,\kmps{}. Using YABI, we also measured the Mount Wilson S-index and R$^{'}_{HK}$. We corrected the RVs for Moon contamination using the CCFs of the sky spectra acquired with fiber B and following the recipe described in \citet{Malavolta-2017AJ....153..224M}. The corrected RV dataset has a median uncertainty of~2.3\,\mps{} and a root-mean-square (RMS) of 13.4\,\mps{} (see Table~\ref{table-C10_1452-tng_harpn-0071-drs-complete_output}). We also used the template-matching {\tt serval} code \citep{2018A&A...609A..12Z} to measure the relative RVs, chromatic index (CRX), differential line width (dLW), and the H$\alpha$, sodium Na~D1, and Na~D2 indexes. This RV dataset has a median uncertainty of~1.9\,\mps{} and an RMS of 13.9\,\mps{} (see Table~\ref{table-C10_1452-tng_harpn-0071-srv-complete_output}). We used absolute RVs in the joint RV and transit analysis presented in Sect.~\ref{sec-modelling_and_results}.

\subsection{VLT/ESPRESSO spectroscopy}
\label{sec-observations-rv-espresso}
Between 17 December 2023 (UT) and 12 March 2024 (UT), 25~measurements were collected using ESPRESSO (program ID: 112.25F2). These observations were obtained within the framework of the THIRSTEE radial velocity program \citep{Lacedelli2024}, aimed at measuring the mass of the radius-gap planet c. The exposure time was set to 900 seconds, leading to an S/N per pixel of 41.3--72.2 at 573\,nm. As with HARPS-N, we measured absolute RVs, CCF\_FWHM, CCF\_CTR, CCF\_BIS, CCF\_ASYM (CCF skewness), and the Mount Wilson S-index (see Table~\ref{table-C10_1452-vlt_espresso-0025-drs-complete_output}) using DRS version 3.1.0 and the relative RVs, CRX, dLW, H$\alpha$, and sodium Na~D1 and Na~D2 indexes using {\tt serval} (see Table~\ref{table-C10_1452-vlt_espresso-0025-srv-complete_output}). As in the case of HARPS-N, we corrected the ESPRESSO absolute RVs for Moon contamination following the recipe described in \citet{Malavolta-2017AJ....153..224M}. The ESPRESSO relative RVs have a median uncertainty of~0.6\,\mps{}, and the corrected absolute RVs have a median uncertainty of 0.9\,\mps{}. The RMS values of the relative and absolute ESPRESSO RVs are $\sim$3.8\,\mps{}. The RMS of ESPRESSO RVs is significantly lower than the RMS of HARPS-N mainly because the ESPRESSO RVs cover only a small part of the long-term variability in \host{} RVs (see Sect.~\ref{sec-modelling_and_results}). In the modelling, we used all absolute ESPRESSO RVs.

\section{Stellar characterisation}
\label{sec-stellar-modelling}
K2-223 is a solar-type star located $207.85 \pm 0.94$ parsecs away, based on the Gaia Data Release 3 parallax corrected for the zero-point bias following \cite{Lindegren2021}. The basic astrometric and photometric properties of the target, along with the photospheric and physical stellar parameters computed from available spectroscopic data (Sect.~\ref{sec-stellar-modelling-stellar_params}) are listed in Table~\ref{tab:stellar_params}. We paid particular attention to identifying the stellar rotation period, which is discussed in Sect.~\ref{sec-stellar-modelling-stellar_rotation}.

\subsection{Stellar parameters}
\label{sec-stellar-modelling-stellar_params}
We analysed the co-added HARPS-N spectrum using the code \texttt{BACCHUS}  \citep{2016ascl.soft05004M,Hayes2022} and relied on the MARCS model atmospheres \citep{Gustafsson2008} and the line lists from \citet{Heiter2021}. The effective temperature was derived  by requiring no trend of the \ion{Fe}{I} line abundances against their respective excitation potential. The surface gravity was determined by requiring ionisation balance of \ion{Fe}{I} lines and \ion{Fe}{II} line. The microturbulence velocity value was also derived by requiring no trend of Fe line abundances against their equivalent widths. The output metallicity is represented by the average abundance of the \ion{Fe}{I} lines. 

We then used the Bayesian tool PARAM \citep{2012MNRAS.427..127B,2017MNRAS.467.1433R} to derive the stellar masses and radii using our spectroscopic temperature and the stellar luminosity obtained with the Gaia parallax, Gaia G magnitude, and a bolometric correction consistent with our spectroscopic parameters. However, these Bayesian tools underestimate the error budget as they do not take the systematic errors between different sets of isochrones into account due to the various underlying assumptions in the respective stellar evolutionary codes. In order to take these systematic errors into account, we combined the results of the two sets of isochrones provided by PARAM (i.e. MESA and Parsec) and added the difference between the two sets of results to the error budget provided by PARAM. We emphasise that although using two sets of isochrones might mitigate underlying systematic errors, our formal error budget for the radius and luminosity might still be underestimated, as demonstrated by \citet{Tayar2022}. 

We used the HARPS-N spectra to attempt measuring the stellar rotation (\svrotsini{}) using the average of the Fe line broadening after subtracting the instrument resolution and a macroturbulence value of 3.1\,\kmps{} according to \citet{Doyle2014}. We estimated the system age by averaging the values obtained from the gyrochronological relations of \citet{2008ApJ...687.1264M} and \citet{2015MNRAS.450.1787A}. The value of \svrotsini{} presented in Table~\ref{tab:stellar_params} ($2.0_{-1.2}^{+0.8}$\,\kmps{}) agrees with a value determined by \cite{Rainer+2023A&A...676A..90R} based on empirical relations to directly convert the HARPS-N CCF-FWHM obtained with the G2 and K5 mask into a \svrotsini{} value ($2.24 \pm 0.65$\,\kmps{}).

The broad wavelength coverage of the HARPS spectrum also allowed us to derive an R$^{'}_{HK}$ index and led to an alternative estimate of the rotation period and age using the \citet{2015MNRAS.452.2745S, Mascareno2016} parameters for the R$^{'}_{HK}$-rotation relation and \citet{2008ApJ...687.1264M} for the rotation-age relation. However, we stress that no emission line is observed at~the core of the Ca II H and K lines, indicating that the star has a very low chromospheric activity (if any). This implies that these rotation periods and the age are probably lower limits. We note that all rotation indicators from spectroscopy as described in this section have large uncertainties, and we refer to the next section for a more reliable estimate of the stellar rotation period.

Finally, we were able to measure a lithium abundance from the HARPS-N spectrum for which we obtained A(Li)\,=\,1.72${\,\pm\,0.05}$. The rotational period and Li value are consistent with literature values of main-sequence stars with similar temperatures and ages \citep{2018A&A...614A..55A}, supporting the robustness of our analysis.

All the obtained parameters are presented in Table~\ref{tab:stellar_params}. These parameters clearly indicate that \host{} is a metal-rich solar-like star. Its effective temperature, mass, and radius indicate that \host{} is only slightly hotter and larger than the Sun, representing a star of spectral type G1V\footnote{\url{https://www.pas.rochester.edu/~emamajek/EEM_dwarf_UBVIJHK_colors_Teff.txt}} \citep{Pecaut-Mamajek-2013ApJS..208....9P}.

\begin{table}
    \centering
    \caption{K2-223 stellar parameters.}
    \begin{tabular}{lclc}
    \hline
    \hline
    \noalign{\smallskip}
    Parameter & Unit & Value & Reference \\
    \noalign{\smallskip}
    \hline
    \noalign{\smallskip}
    EPIC ID & & 228721452 & 1 \\
    TIC ID & & 98594138 & 2 \\
    \noalign{\smallskip}
    \hline
    \noalign{\smallskip}
        \multicolumn{4}{c}{Astrometric and photometric properties} \\
    \noalign{\smallskip}
    \hline
    \noalign{\smallskip}
    $\alpha_{\rm J2016}$ & & $\rm 12^{h}21^{m}13.48^{s}$ & 3 \\
    $\delta_{\rm J2016}$ & & $-10^{\circ}16'55.54''$ & 3 \\
    $\mu_\alpha \cos \delta$ & [mas yr$^{-1}$] & $7.058\pm0.026$ & 3 \\
    $\mu_\delta$ & [mas yr$^{-1}$] & $-14.838\pm0.018$ & 3 \\
    $\pi^{(a)}$ & [mas] & $4.811\pm0.022$ & 7 \\
    $d^{(a)}$ & [pc] & $207.85 \pm 0.94$ & 7 \\
    $m_{TESS}$ & [mag] & $10.8242 \pm 0.0061$ & 4 \\
    $m_{\rm Gaia}$ & [mag] & $11.2712 \pm 0.0003$ & 3 \\
    $m_{\rm Gaia,\ BP}$ & [mag] & $11.6109 \pm 0.0006$ & 3 \\
    $m_{\rm Gaia,\ RP}$ & [mag] & $10.7650 \pm 0.0005$ & 3 \\
    $m_{\rm Gaia,\ BP-RP}$ & [mag] & $0.8458 \pm 0.0008$ & 3 \\
    $m_{\rm B}$ & [mag] & $12.13\pm0.03$ & 5 \\
    $m_{\rm V}$ & [mag] & $11.43\pm0.02$ & 5 \\
    $m_{\rm R}$ & [mag] & $11.35\pm0.05$ & 5 \\
    $m_{\rm J}$ & [mag] & $10.177\pm0.024$ & 6 \\
    $m_{\rm H}$ & [mag] & $9.946\pm0.024$ & 6 \\
    $m_{\rm K}$ & [mag] & $9.835\pm0.024$ & 6 \\
    $m_{\rm g}$ & [mag] & $11.74\pm0.04$ & 5 \\
    $m_{\rm r}$ & [mag] & $11.26\pm0.01$ & 5 \\
    $m_{\rm i}$ & [mag] & $11.15\pm0.10$ & 5 \\
    $\gamma$ & [\mps] & $9354.7 \pm 0.7$ & 7 \\ 
    \noalign{\smallskip}
    \hline
    \noalign{\smallskip}
        \multicolumn{4}{c}{Photospheric and physical parameters} \\
    \noalign{\smallskip}
    \hline
    \noalign{\smallskip}
    \steff{} & [K] & $5858 \pm 26$ & 7 \\
    \logsg{} & [$\log$ \cmpss{}] & $4.56 \pm 0.16$ & 7 \\
    \sfeh{} & [\dex{}] & $0.18 \pm 0.06$ & 7 \\
    $v_t$ & [\kmps{}] & $1.05 \pm 0.02$ & 7 \\
    \slum{}$^{(b)}$ & [\Lsun{}] & $0.99\pm 0.16$ & 7 \\
    \sm{}$^{(c)}$ & [\Msun{}] & $1.07 \pm 0.07$ & 7 \\
    \sr{}$^{(c)}$ & [\Rsun{}] & $1.05 \pm 0.10$ & 7 \\
    \svrotsini{} & [\kmps{}] & $2.0_{-1.2}^{+0.8}$ & 7 \\
    \logrhk{} & & $-4.847 \pm 0.049$ & 7 \\
    $A(\rm Li)$ & & $1.72 \pm 0.05$ & 7 \\
    \sprot{}$^{(d)}$ & [d] & $25_{-7}^{+35}$ & 7 \\
    \sprot{}$^{(e)}$ & [d] & $26 \pm 10$ & 7 \\
    Age$^{(c)}$ & [Gyr] & $2.8 \pm 2.8$ & 7 \\
    Age$^{(d)}$ & [Gyr] & $3.5_{-1.3}^{+10}$ & 7 \\
    Age$^{(e)}$ & [Gyr] & $4.8 \pm 2.9$ & 7 \\
    \noalign{\smallskip}
    \hline
    \hline
    \end{tabular}
    \label{tab:stellar_params}
    \tablefoot{
    1)~\cite{Huber2016}. 2)~\cite{Stassun2018}. 3)~\cite{Gaia2023}. 4)~\cite{Paegert2022}. 5)~\cite{Zacharias2012}. 6)~\cite{Cutri2003}. 7)~This work.
    \tablefoottext{a}{From Gaia with zero-point correction.}
    \tablefoottext{b}{From Gaia with updated \steff.}
    \tablefoottext{c}{From isochrone.}
    \tablefoottext{d}{From \svrotsini.}
    \tablefoottext{e}{From \logrhk.}
    }
\end{table}

\subsection{Stellar rotation}
\label{sec-stellar-modelling-stellar_rotation}
Rotation periods were derived from photometric time series, where the measurements rely on dark spots or bright faculae on the stellar surface. This technique has produced extensive catalogues of stellar rotation periods \citep[e.g.][]{McQuillan2014, Santos2019, Santos2021, Reinhold2020, Gordon2021}.

To determine the rotation period of \host{}, we used the EVEREST light curve, discarded the first few days, and interpolated the remaining small gaps through a multiscale discrete sine transform, following the inpainting technique described by~\cite{Garcia2014a} and \cite{Pires2015}. The periodic modulation in the light curve was investigated using three complementary methods. (1) A time–period analysis based on Morlet wavelets \citep[e.g.][]{Torrence1998, Mathur2010}; (2)~the auto-correlation function \citep[ACF; e.g.][]{McQuillan2013, Garcia2014b}; and (3) the composite spectrum, defined as the product of the wavelet power spectrum and the ACF \citep{Ceillier2017}. The combined use of these techniques provides robust assessments of~the reliability of the detected period and has been, for example, successfully applied to Kepler data \cite{Santos2019, Santos2021}.

Our analysis yields a rotation period of ${19.5 \pm 2.3}$ days, consistent with the predictions in Sect.~\ref{sec-stellar-modelling-stellar_params}. However, this value lies near the detection limit because the total time span of the light curve is 47 days and the inferred value exceeds one-third of this duration. Independent verification using TESS data is not possible, because they are only available for one sector at a time, separated by at least 9 months, which prevents the inference of~a~reliable rotation period longer than $\sim$10 days \citep[for more details, see][]{Palakkatharappil2024}.

In order to estimate a more accurate rotation period, we performed a periodogram analysis of the remaining datasets listed in Sect.~\ref{sec-observations}. We used the generalised Lomb-Scargle (GLS) periodogram as implemented in {\tt Astropy} \citep{Astropy2022}, and the false-alarm probability (FAP) levels were estimated via a bootstrap resampling with 10$^4$ iterations. Thresholds of 10\%, 5\%, and 1\% were used to indicate the significance of the detected peaks. We also examined the window function to assess sampling effects.

We checked long-term ground-based photometry. We used the All Sky Automated Survey for SuperNovae \citep[ASAS-SN:][]{Shappee2014, Kochanek2017}\footnote{\url{https://asas-sn.osu.edu/photometry/}} database for this purpose. The available photometry in the V band from three cameras (bd, bf, and bh) does not show any significant signals. 
These data were not included in the global modelling.

\begin{figure}[!h]
\centering
\includegraphics[width=1.000\linewidth]{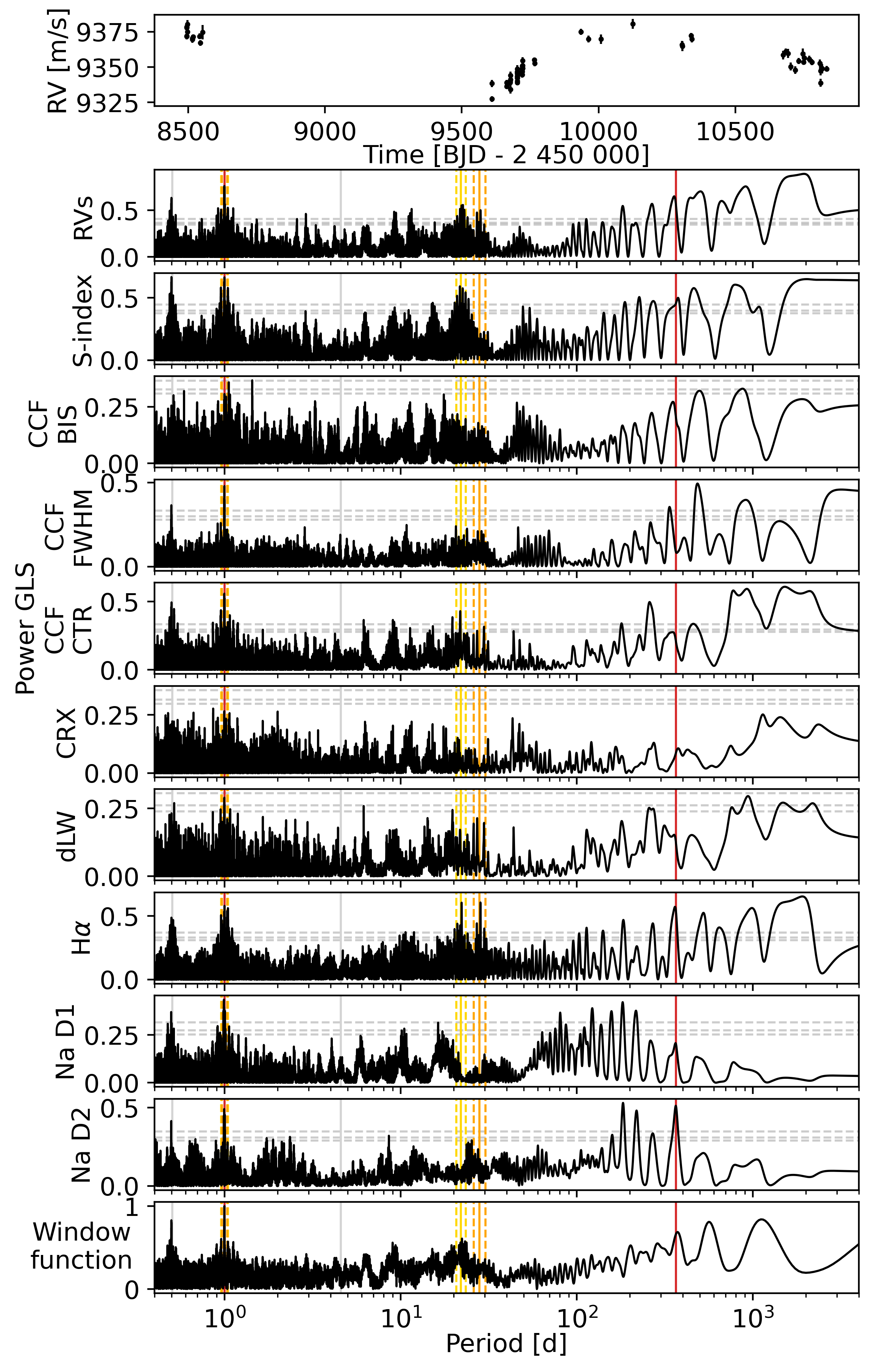}
    \caption{\textit{Top}: HARPS-N RVs and all available activity indicators for \host{}, with their corresponding GLS periodograms. \textit{Bottom}: Window function for this dataset. The dotted horizontal grey lines indicate FAP levels of 10\%, 5\%, and 1\%. The vertical grey lines represent the orbital periods of planets b and c, and the vertical red lines mark 1-day and 1-year periods. The vertical yellow and orange lines indicate the two possible stellar rotation periods, and the dashed lines of the same colours show their corresponding daily and yearly aliases.
    \label{figure-periodogram-harpsn}}
\end{figure}

Figure~\ref{figure-periodogram-harpsn} presents the GLS periodograms of RVs and spectral activity indicators measured from the HARPS-N spectra, computed over the period range 0.4–4000 days, together with the window function for this dataset. The selected range covers all signals of interest, that is, the orbital periods of planets b and c ($\sim$0.506 and $\sim$4.564 days), the frequencies of 1 day and 1 year, and the preliminarily estimated stellar activity periods of $\sim$22 and $\sim$28 days, including their daily and yearly aliases. The strongest signal visible in the RVs periodogram (but also in CCF\_CTR and H$\alpha$) has a period of $\sim$1950 days. It might be related to the stellar activity and to another planet, K2-223 d, as further discussed in Sect.~\ref{sec-modelling_and_results-planets_in_periodograms}. In addition to the annual and daily periods and their harmonics visible in almost all cases, other significant peaks can also be observed, which might indicate the rotation period of the star under study. One of the strongest periods visible in the S-index periodogram is $\sim$21.7 days (see panel three of Fig.~\ref{figure-periodogram-harpsn}), as also seen in the window function, indicating a possible sampling effect. Similar less distinct signals are also visible for CCF\_CTR, H$_{\alpha}$, and Na D2 at $\sim$21.8, $\sim$22.2 (but also $\sim$28.4), and $\sim$25.3 days, respectively. Their harmonics are also present. Comparing the obtained values with the parameters presented in Table~\ref{tab:stellar_params}, we conclude that this periodicity is related to the rotation period of the star, which is most likely in the range~16--36 days, with two tentative values visible in the RVs: $\sim$22 and $\sim$28~days.

\begin{figure}[!h]
\centering
\includegraphics[width=1.000\linewidth]{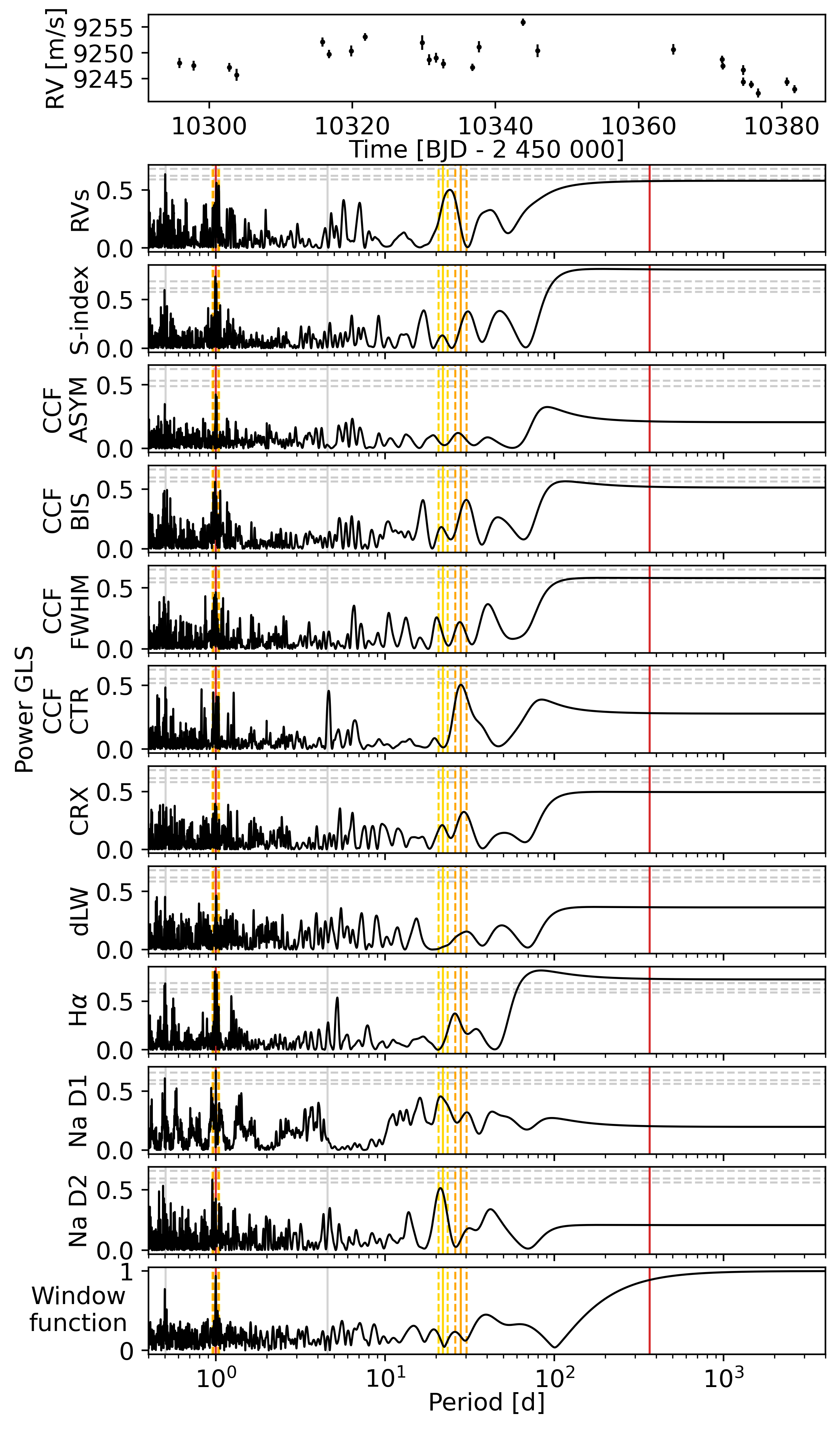}
    \caption{Same as Fig.~\ref{figure-periodogram-harpsn}, but for the ESPRESSO dataset.
    \label{figure-periodogram-espresso}}
\end{figure}

Similar signals are also tentatively seen in the GLS periodograms of the ESPRESSO dataset, presented in Fig.~\ref{figure-periodogram-espresso}, which show visible peaks that might indicate the rotation period of the star, that is, $\sim$24.1 days for RVs, $\sim$28.1 days for CCF\_CTR, and $\sim$21.3 days for Na D2. However, because all these peaks remain below the 1\% FAP threshold and the time span between the first and last measurements is shorter ($\sim$86 days), we cannot independently analyse the long-term signals in the HARPS-N periodograms (see Fig.~\ref{figure-periodogram-harpsn}). 

The periods we examined are close to the rotation period of \host{} estimated from \citet{1984ApJ...279..763N} and \citet{2008ApJ...687.1264M} activity-rotation relations. Using (B-V) of 0.70 and the \logrhk{} measured with YABI (-4.862\,$\pm$\,0.055), we estimated a rotation period of \host{} for 26.9\,$\pm$\,6.1\,days and 27.0\,$\pm$\,3.6\,days, respectively. Using the activity-age relation of \citet{2008ApJ...687.1264M}, we also found the age of \host{} to be in a range of 2.64--7.43\,Gyr.

\section{Orbital model, numerical setup, and results}
\label{sec-modelling_and_results}
As an initial stage in modelling the \host{} system, we analysed periodograms of multiple photometric and RV datasets to constrain the space of known and candidate periodic signals. The details of this step are described in 
Sect.~\ref{sec-modelling_and_results-planets_in_periodograms} below. Then, we used the \texttt{PyORBIT} package\footnote{\url{https://github.com/LucaMalavolta/PyORBIT}} \citep{Malavolta2018,Malavolta2016} to first model only the photometric data and to subsequently perform a~joint model to all available measurements and constraints.

The \texttt{PyORBIT} package enables the simultaneous modelling of transit curves, radial velocities, and stellar activity indicators. The code is built around Bayesian statistics and a Markov chain Monte Carlo (MCMC) sampling of the posterior parameter distribution.

The usual sampling run starts from predetermining best-fit model parameters with differential evolution \cite[DE,][]{Price2005} implemented in \texttt{PyDE}\footnote{\url{https://github.com/hpparvi/PyDE}} package. The second step is the MCMC sampling, initialised with a number of initial parameters (so-called walkers) selected in a small hyper-cube around the best-fitting model found through the DE step. To sample the posterior distribution, we chose the \texttt{emcee} package\footnote{\url{https://github.com/dfm/emcee}} \citep{Foreman-Mackey2013} implementing ensemble samplers with affine invariance \citep{GoodmanWeare2010}. The two steps of~the modelling pipeline are based on the same priors and parameter bounds. This hybrid optimisation is an efficient approach to determine the best-fitting final parameters and their uncertainties.

The data model defined in \texttt{PyORBIT} combines the Keplerian parametrisation of elliptic orbits with stellar and transit parameters and with other parameters required to explain the observations. These parameters are listed in Tabs.~\ref{tab:planet_bc_phot}--\ref{tab:planet_bcd} and are defined as the orbital period of the planet \pxporb{}, the RV semi-amplitude \pxk{}, the eccentricity \pxe{}, the argument (or longitude) of periastron \pxw{} ($\varpi$), the central time of the first transit \pxtzero{} expressing the mean anomaly ${\cal M}$ at the given epoch, the planet radius in stellar radii \pxpsrr{}, and the impact parameter \pxb{}. In cases of nearly circular orbits, the eccentricity and the argument (or longitude) of periastron were replaced by non-singular Poincar\'e  variables $x=\sqrt{e} \cos \omega$ and $y=\sqrt{e} \sin \omega$, respectively, in order to avoid undetermined \pxw{} (or $\varpi$) and to remove \pxe{} $=0$ wall.  The density of the star (\sden{}), the quadratic limb-darkening coefficients ($q_1$, $q_2$), the RV offsets ($\gamma$) and jitters ($\sigma$) for each dataset were also included.

We monitored basic diagnostics of the MCMC sampling, such as the integrated autocorrelation time for each parameter $\tau_{\rm MCMC}$ and the acceptance fraction $\alpha_{\rm MCMC}$ reported by the \texttt{emcee} code. Specifically, $\alpha_{\rm MCMC}$ should be approximately within the range of (0.2--0.5) for an optimal sampling. The auto-correlation time is crucial for determining the chain length (number of iterations) that ensures proper convergence and coverage of the posterior distribution. Optimally, the chain length should not be shorter than approximately $50\, \tau_{\rm MCMC}$ for each parameter \citep{Foreman-Mackey2013}.

In our MCMC experiments, we maintained these requirements. The number of walkers was set to five times the number of free parameters. For the final model with 34 parameters (see Table~\ref{tab:planet_bcd}), we conducted the MCMC sampling for $2 \times 10^6$ iterations per walker. This ensured that we exceeded the required $50\, \tau_{\rm MCMC}$ iterations for each parameter; for some of them, the factor was as large as 300 because $\tau_{\rm MCMC}$ is not uniform. We observed an acceptance fraction $\alpha_{\rm MCMC} \simeq 0.1$. This low value (below the expected lower limit of $\simeq 0.2$) is likely caused by the orbital period $P_{\rm b}$ of \hostb{}, which proved most difficult to sample due to the multi-modal character of its posterior distribution (see Fig.~\ref{figure-planet_bcd-mcmc}). Although the standard \texttt{emcee} sampler might not be the optimal choice in these cases, which would benefit from parallel tempering, we retained it because a constrained $P_{\rm b}$ appears to imply the only posterior with a complex shape in the parameter set. With these long chains, we were also able to safely discard 20\% of the initial iterations, which represent the burn-in phase of the MCMC sampling.

\subsection{Detection of planet signals in the periodograms}
\label{sec-modelling_and_results-planets_in_periodograms}
No signals from planets b and c are detected as transits in the \ktwo{} mission in the GLS periodogram of TNG/HARPS-N and VLT/ESPRESSO RVs (Figs.~\ref{figure-periodogram-harpsn} and \ref{figure-periodogram-espresso}). The apparent peak near the period of \planetb{} is a 0.5-day alias and not the planetary signal. No clear signals are detected in the \tess{} photometry either; this is verified more thoroughly during the modelling of the transit curves in Sect.~\ref{sec-modelling_and_results-transits}.

As mentioned in Sect.~\ref{sec-stellar-modelling-stellar_rotation}, the strongest signal visible in~the TNG/HARPS-N RV data have a period of $\sim$1950 days (see Fig.~\ref{figure-periodogram-harpsn}). The half-amplitude of this peak is $\sim$27\,\mps{} (half of~the peak-to-peak variation of the RVs). Similar frequencies are visible in periodograms of activity indicators such as CCF\_CTR and H$\alpha$. The correlation analysis (Fig.~\ref{figure-ind-coeff}) shows weak but non-negligible Pearson and Spearman correlations between the RVs and both CCF\_CTR and H$\alpha$, consistent with the periodogram comparison. The dLW indicator is also notable in the correlations, although its periodogram peaks lie at or slightly above the lowest adopted FAP levels, making their significance uncertain. The limited time span between the first and last RV measurements, covering $\sim$1.2 orbital periods of the potential planet \host{} d, further limits a reliable interpretation of all indicators. To~use the TNG/HARPS-N RVs data for modelling planets b and c, we took this signal into account in the final modelling (Sect.~\ref{sec-modelling_and_results-planet_bcd}), giving it a planetary character with the eccentricity set as a free parameter.

\subsection{Transit modelling}
\label{sec-modelling_and_results-transits}
The transit curves were fitted using the {\tt batman}\footnote{\url{https://github.com/lkreidberg/batman}} \citep{Kreidberg2015} package included with \texttt{PyORBIT}. For the \ktwo{} photometry, we included the supersample factor parameter with a value of 10 to more accurately capture the transit shapes. In order to establish better convergence of the chains, we set Gaussian priors on~the limb-darkening coefficients. We chose a quadratic model using predicted values calculated using the {\tt LDTk}\footnote{\url{https://github.com/hpparvi/ldtk}} \citep{Parviainen2015} package. The obtained 1$\sigma$ uncertainties were multiplied by a factor of 100 and included in the model together with the estimated values of the coefficients. For \ktwo, $q_1=0.51 \pm 0.10$ and $q_2=0.15 \pm 0.17$, and for \tess{}, $q_1=0.41 \pm 0.07$ and $q_2=0.16 \pm 0.13$.

In order to correctly identify transits in the \ktwo{} and \tess{} photometric data, it is necessary to compensate for changes in~the normalised flux caused by instrumental effects and stellar variability. We did this with {\tt wotan}\footnote{\url{https://github.com/hippke/wotan}} \citep{Hippke2019} using the bi-weight method, with a window length of 0.5 days, a~break tolerance of 0.1 days, and an edge cutoff of 0.1 days. The \ktwo{} data that deviated by more than 7.5$\sigma$ from the normalised brightness curve were removed during the modelling, as they indicated outliers. In the case of TESS data, using the PDCSAP flux values facilitated the subsequent removal of the presented effects.

The results of the transit modelling of the \ktwo{} data of \host{} b and c are presented in Table~\ref{tab:planet_bc_phot}. We found no other regular decreases in brightness, which indicates additional transiting candidates. The outcomes are consistent with the latest literature value \citep{Adams2021, Mayo2018, Livingston2018}. The strongest difference lies in the inclination of planet b, but this value is estimated with a high uncertainty ($76.3_{-3.6}^{+3.9}$\,deg). \cite{Adams2021} determined the parameter \pbi{}\,=\,89.7\,$\pm$\,10\,deg, and it falls within the $1\sigma$ range. The lowest possible inclination of \host{} b allowing transits is $68.9^{+1.4}_{-1.9}$\,deg (using \pbar{} from Table~\ref{tab:planet_bc_phot}).

The search for transit signals in the \tess{} photometric data was unsuccessful. We performed three independent modelling runs, one per sector (36, 46, and 91). In each case, the transit signals of planets b and c did not rise above the measurement scatter and were thus undetectable. A simultaneous modelling of all photometric data (\ktwo{} + \tess{}) did not improve the results presented in Table~\ref{tab:planet_bc_phot}. Therefore, \tess{} measurements were not included in the following analysis.

\subsection{Determining the mass of planet b}
\label{sec-modelling_and_results-planet_b}
To verify the possibility of detecting planet \hostb{} in the available radial velocity measurements, we used the floating chunk offset method \citep[FCO:][]{Hatzes2014}. It involves extracting sets of~radial velocities consisting of at least two measurements taken on the same night and adding independent offsets to them. This approach eliminates long-term components caused by stellar variability and allows USP planets to be characterised.

In this section, we focus on determining the mass of \hostb{}. The data selection allowed us to collect 12 datasets (12 nights when more than one measurement was taken) of~TNG/HARPS-N RVs and 2 datasets of VLT/ESPRESSO RVs. In total, this amounts to 40 and 4 points, respectively. We also included photometric data from the \ktwo{} mission in this modelling to give appropriate priors imposed by the visible transits. 

The results shown in Table~\ref{tab:planet_bcd} indicate that the mass of the inner planet \hostb{} is poorly constrained, resulting in a 99th percentile upper limit of $M_{\rm b}$\,<\,$4.7$ \Mea{}. This is illustrated by the MCMC result for planet b presented in Fig.~\ref{figure-planet_b-mcmc} and the phased curves shown in Fig.~\ref{figure-planet_b-phased}. The posterior distribution peaks essentially at $K$\,=\,0 and remains heavily skewed toward zero over most of its support. The values of the other parameters are consistent with the modelling outcomes obtained using photometric data alone (see Table~\ref{tab:planet_bc_phot}).

\subsection{Final joint fit}
\label{sec-modelling_and_results-planet_bcd}
The identification of the rotation period of \host{} (see Sect.~\ref{sec-stellar-modelling-stellar_rotation}) enabled the use of Gaussian processes (GP). For this purpose, the {\tt tinygp}\footnote{\url{https://github.com/dfm/tinygp}} \citep{Foreman-Mackey2023} package was used, more specifically, its quasi-periodic kernel \citep{Rajpaul2015}. Stellar activity is then described using four parameters: the stellar rotation period ($P_{\rm rot}$), the coherence scale ($O_{\rm amp}$), the decay scale of active regions ($P_{\rm dec}$), and the amplitude of the kernel ($H_{\rm amp}$).

Based on the periodograms in Sect.~\ref{sec-stellar-modelling-stellar_rotation}, we imposed a uniform prior on $P_{\rm rot}$ equal to [16, 36] days in our modelling. The choice was dictated by clear signals in radial velocities and the TNG/HARPS-N S-index. Additionally, we followed the recommendations for $O_{\rm amp}$ of \cite{Perger2021} (centred at 0.311, 1$\sigma$ = 0.016, boundaries [0.2, 0.5]) and adopted a Gaussian prior with a relaxed 5$\sigma$ uncertainty and expanded boundaries of [0.01, 0.6]. This approach allowed us to describe the effect of spots on~the RVs data more realistically. All other stellar activity contributions were modelled by the jitter.

\begin{figure*}[!h]
\centering
    \includegraphics[width=0.35\linewidth]{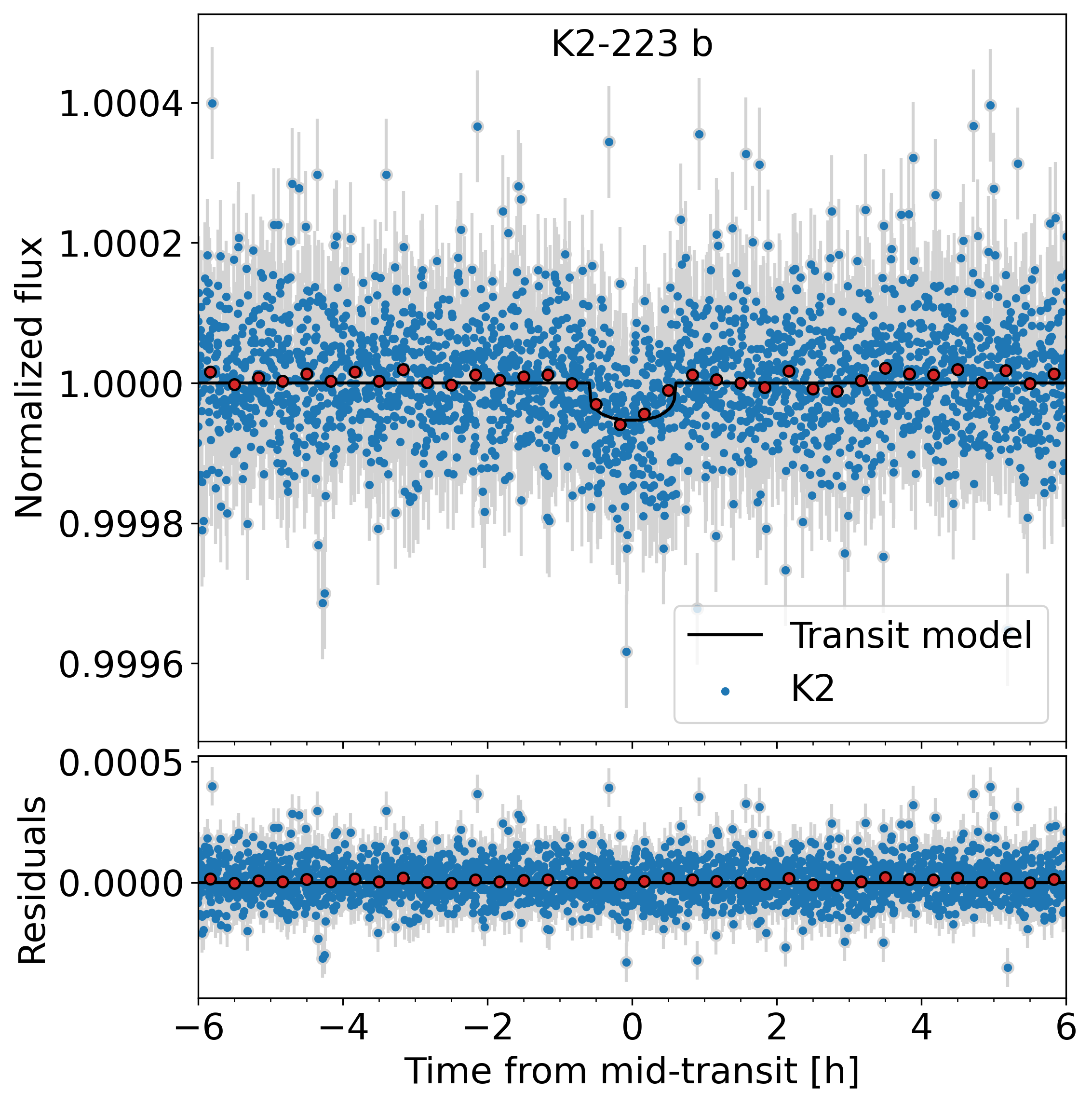}
    \includegraphics[width=0.35\linewidth]{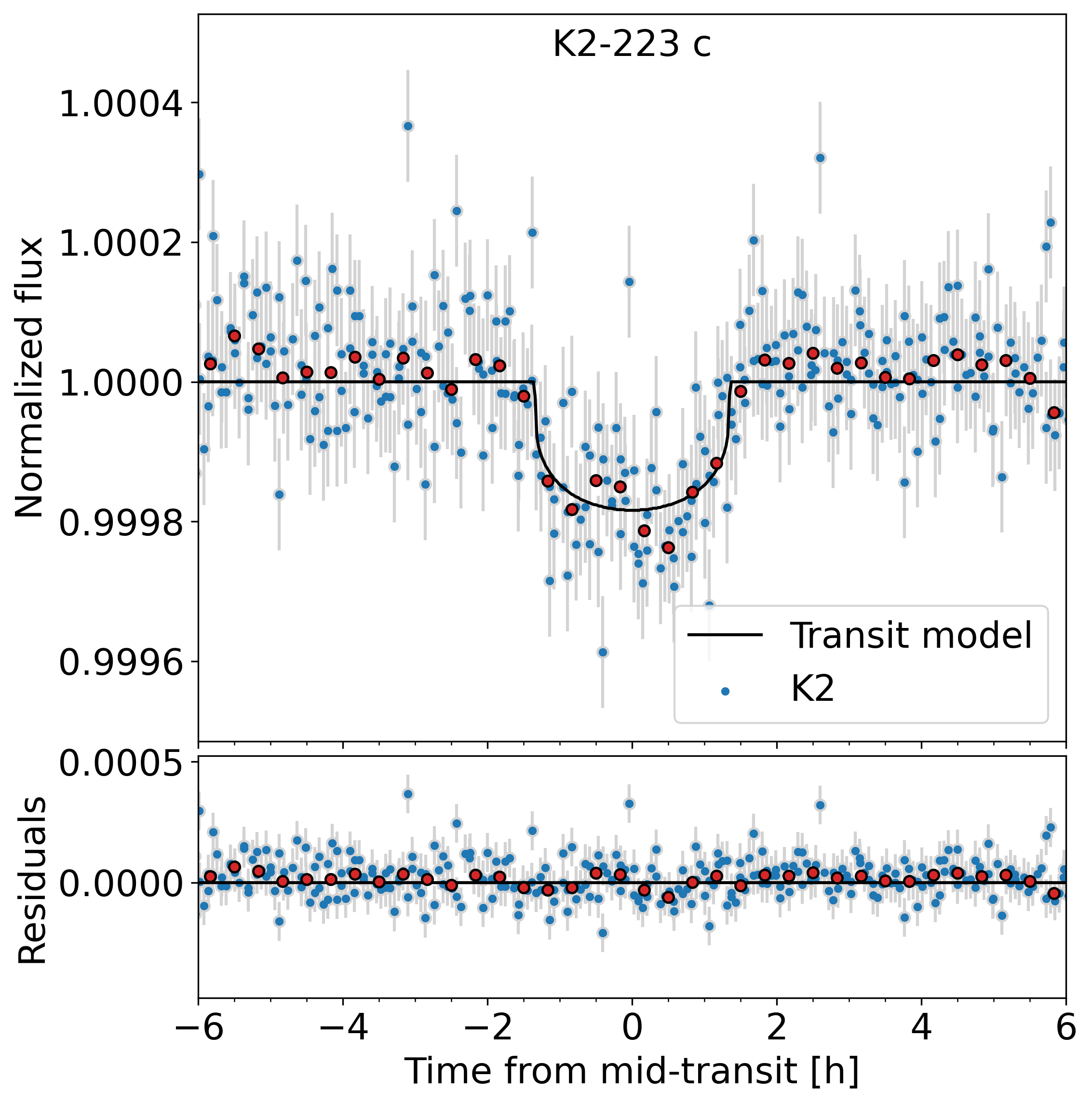}
    \includegraphics[width=0.33\linewidth]{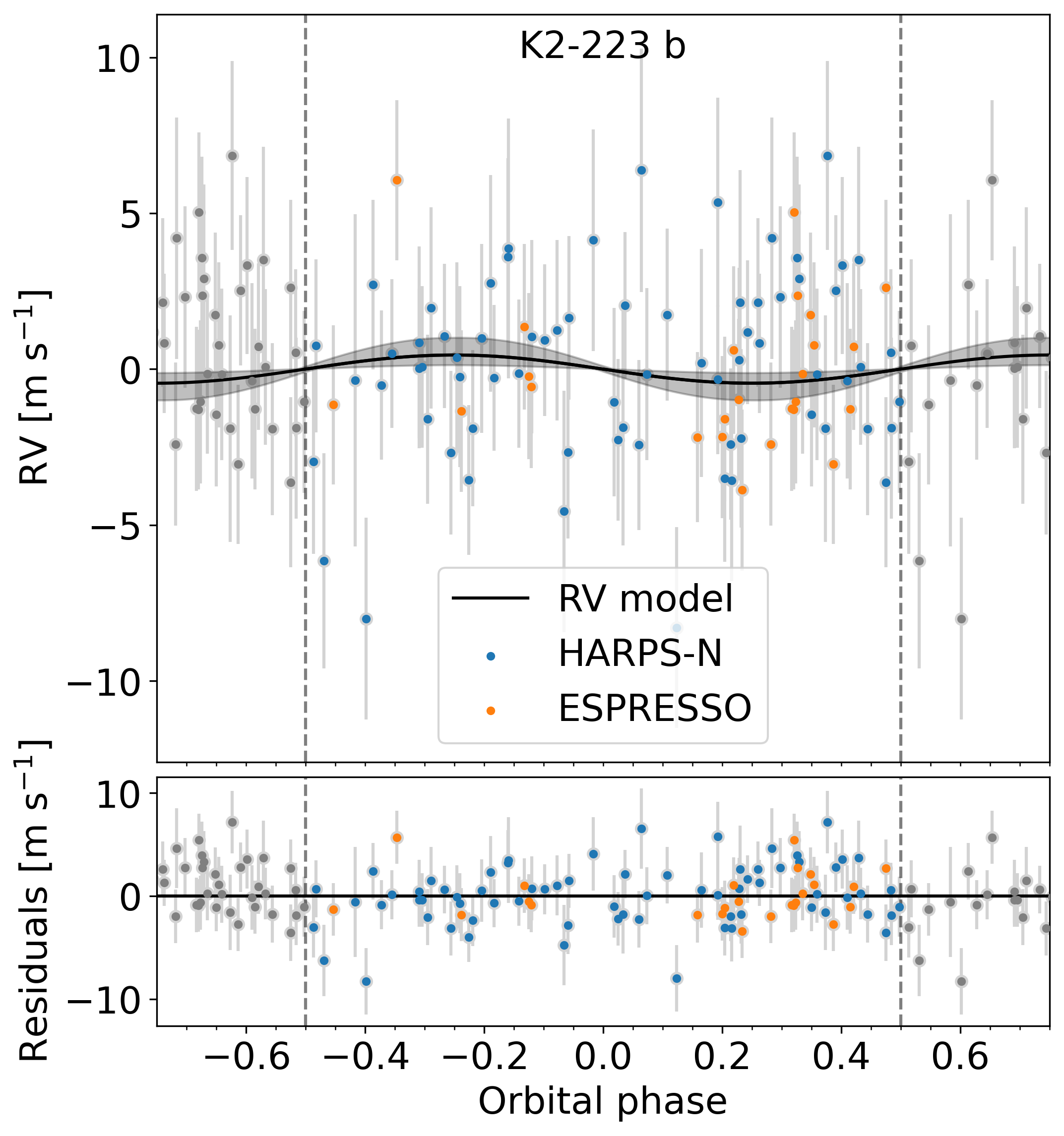}
    \includegraphics[width=0.33\linewidth]{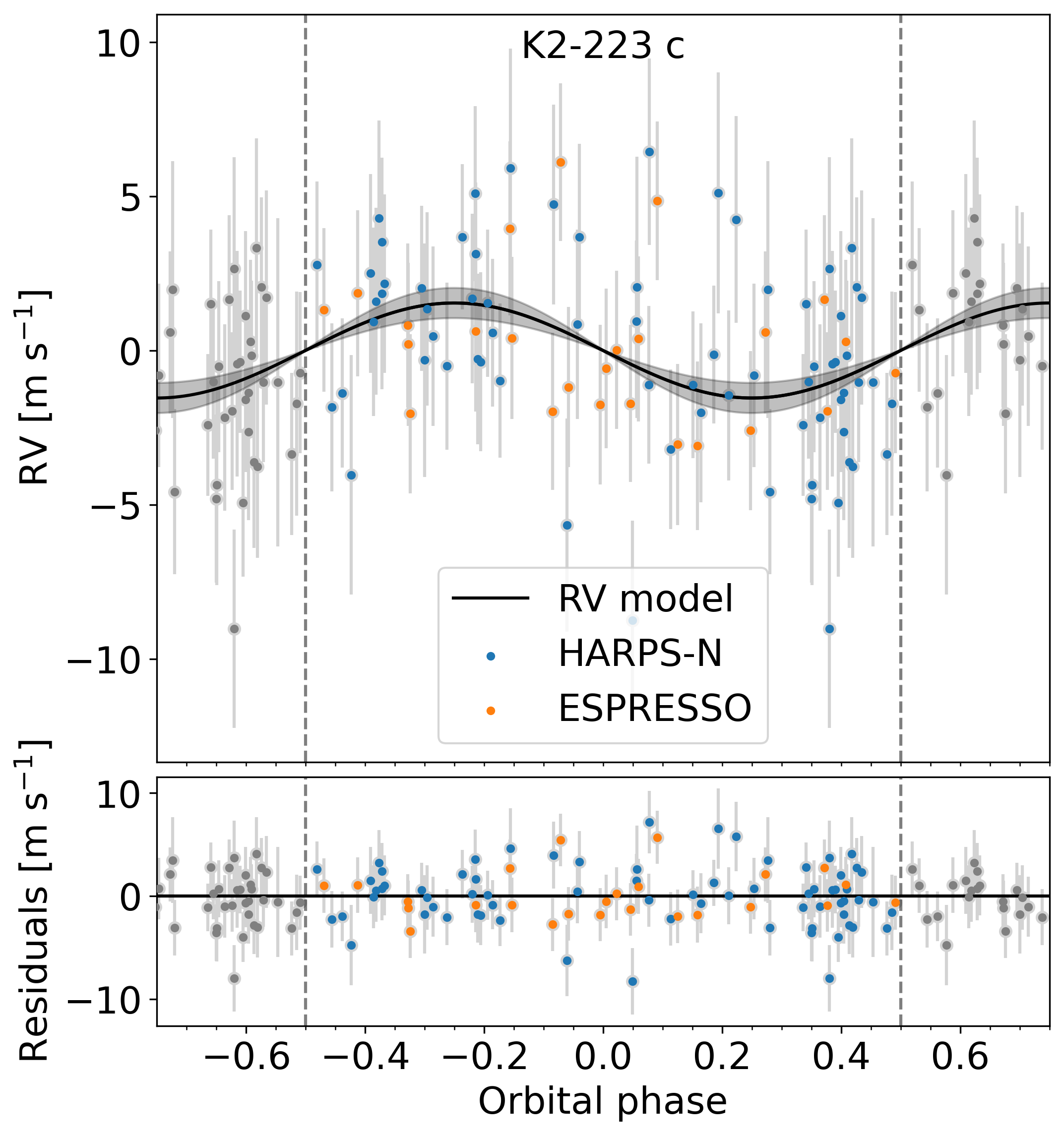}
    \includegraphics[width=0.33\linewidth]{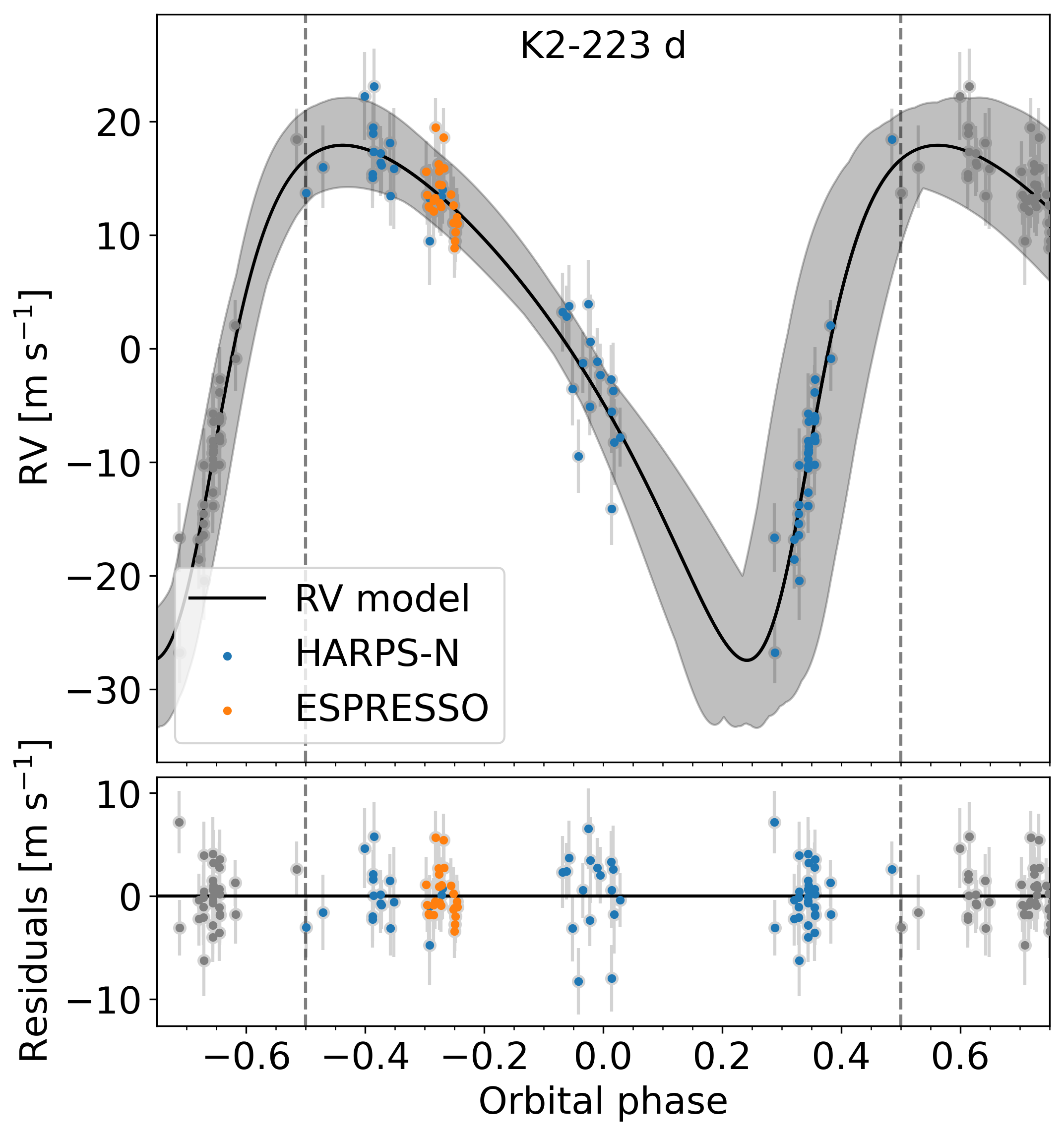}
    \caption{Phased model marked with a black line for the photometric curves of planets b and c (\textit{top panels}) and radial velocities of planets b, c, and d using GPs for stellar rotation period (\textit{bottom panels}). The red dots represent the median value calculated every 20 minutes from the centre of the transit. The grey area for radial velocities represents the 1$\sigma$ uncertainties of all fitted orbital parameters.\label{figure-planet_bcd-phased}}
\end{figure*}

\begin{figure*}[!h]
\centering
    \includegraphics[width=0.33\linewidth]{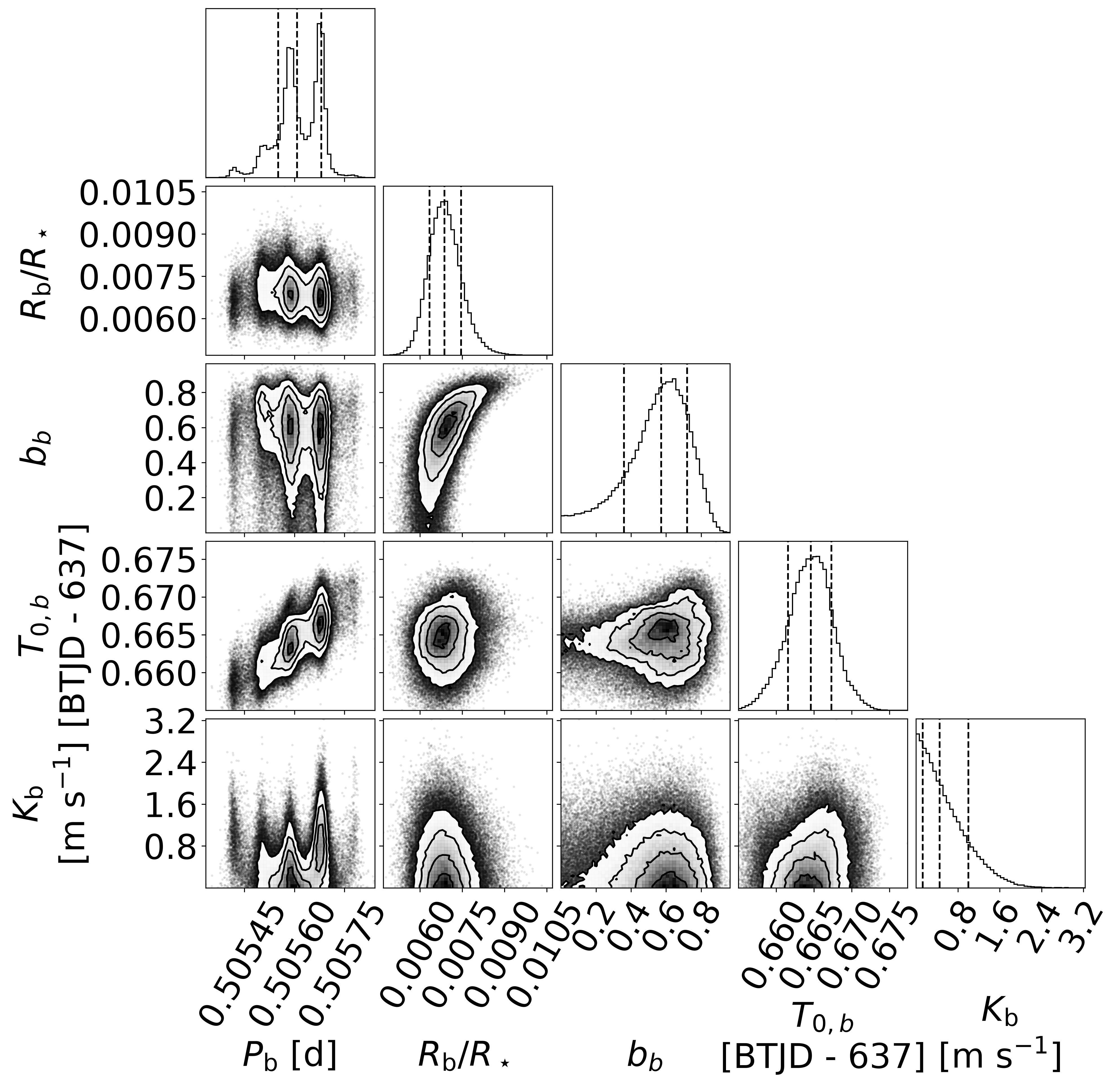}
    \includegraphics[width=0.33\linewidth]{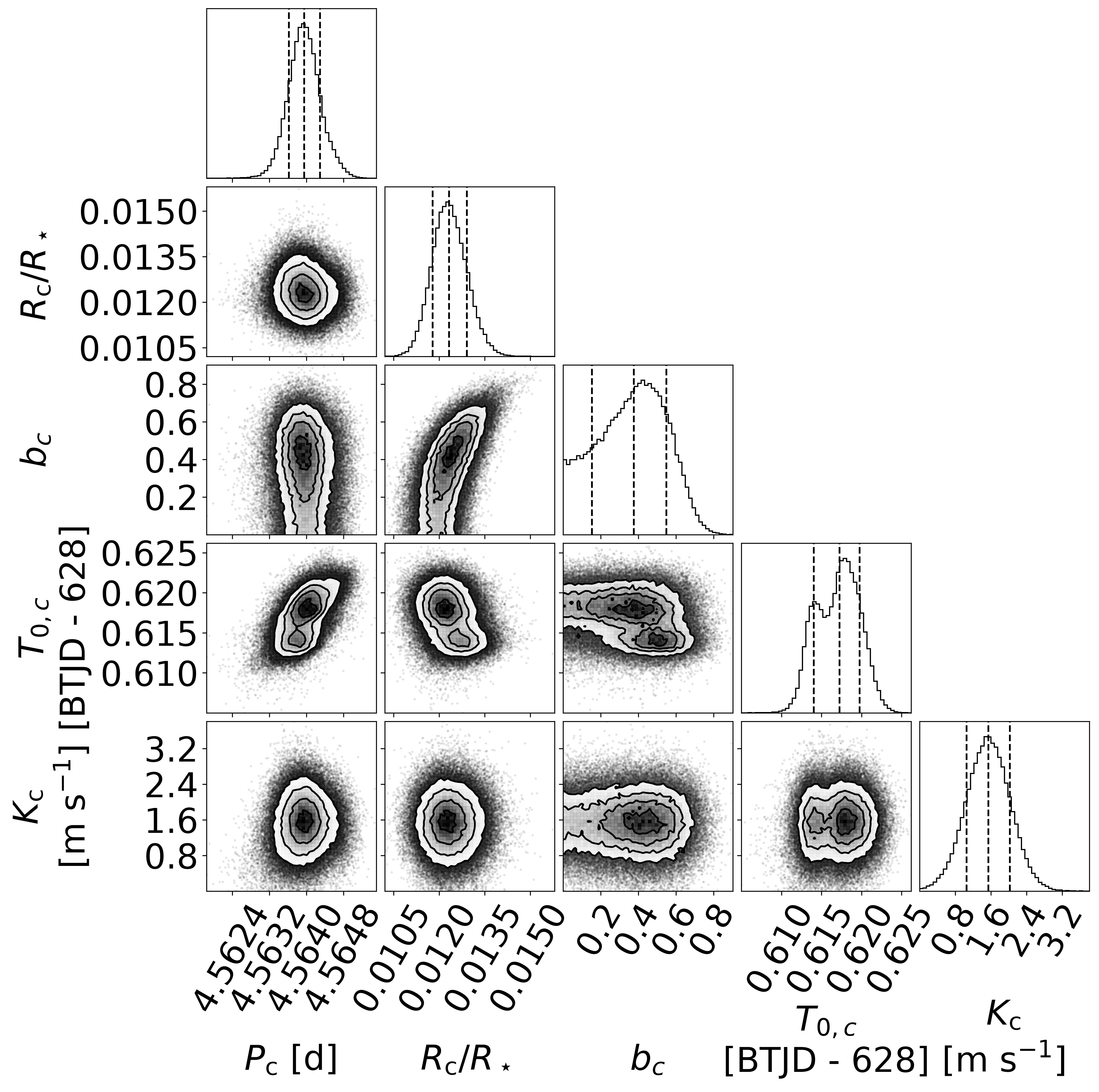}
    \includegraphics[width=0.33\linewidth]{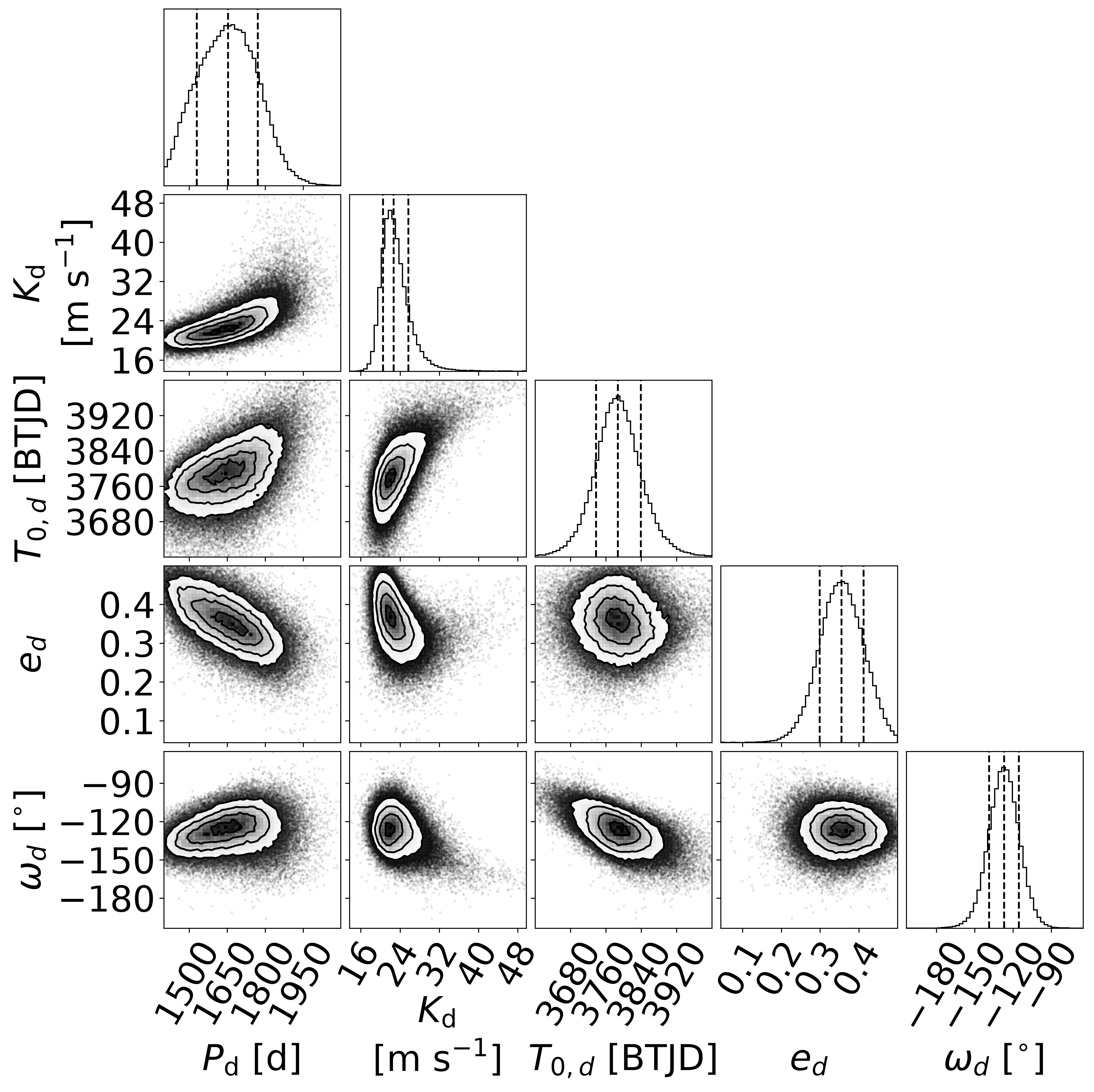}
    \caption{MCMC histograms for planets \host{} b and c and tentative d using GPs for a stellar rotation period. The dashed lines mark the 16th, 50th, and 84th percentiles of each marginalised posterior distribution. \label{figure-planet_bcd-mcmc}}
\end{figure*}

The final model included signals of \host{}\,b, c, the additional \hostd{}, and the stellar activity modelled using GP regression. Photometry, VLT/ESPRESSO RVs, TNG/HARPS-N RVs, and the TNG/HARPS-N S-index were included in a~standard GP framework implemented in \texttt{PyORBIT}, in which datasets share the same covariance kernel and hyperparameters while allowing dataset-dependent amplitudes. The S-index was thus used to jointly constrain the properties of correlated noise and the stellar rotation period. The results are presented in Table~\ref{tab:planet_bcd}, the phased curves in Figure~\ref{figure-planet_bcd-phased}, and the MCMC histograms in~Fig.~\ref{figure-planet_bcd-mcmc}. The obtained parameter values describing planets b and c coincide with those calculated on the basis of~photometry (see Table~\ref{tab:planet_bc_phot}) and using the FCO method (see Table~\ref{tab:planet_bcd}). The derived mass constraints, in combination with the estimated radii, provide guidance on the possible composition of~\planetc{} alone (see Sect.~\ref{sec-discussion-composition}).

All obtained values of parameters \pbk{}, \pck{}, \pbm{} and \pcm{} in~Table~\ref{tab:planet_bcd} coincide with predictions from previous works. According to Table 5 in \cite{Livingston2018}, they should be \pbk{}\,=\,$0.7_{-0.4}^{+0.6}$\,\mps{} (\pbm{}\,=\,$0.9_{-0.6}^{+0.8}$\,\Mea{}) and \pck{}\,=\,$1.8_{-0.8}^{+0.9}$\,\mps{} (\pcm{}\,=\,$5.0_{-2.2}^{+2.4}$\,\Mea{}).

The modelling we carried out allowed us to determine the mass of planet \host{}\,c to a precision of $\sim$3$\sigma$. However, we were only able to obtain an upper limit to the semi-amplitude and mass of planet \host{}\,b, without a well-defined average value, and the signal is only marginally significant at $\sim$1.3$\sigma$ (see Fig.~\ref{figure-planet_bcd-mcmc}). The large uncertainties make this result consistent with the previous one obtained using the FCO method (see Sect.~\ref{sec-modelling_and_results-planet_b}).

\section{The K2-223 system in the demography of planets}
\label{sec-demography}

\subsection{\planetb{} and the population of USP planets}
\label{sec-demography-K2-223_b}
Since \host{} b is a USP planet, we initially focused on planets with an orbital period shorter than one day. According to the NASA Exoplanet Archive as of 9 June 2026 \citep{Christiansen2025}, there are 158 (non-controversial) USP planets. Of these, 49 ($\sim$31\%) have observationally constrained masses with at least 3$\sigma$ uncertainties. \planetb{} is not included in this subsample because our analysis only yielded an upper limit on its mass, and therefore, further observations and characterisation are required.

\subsection{\planetc{} among the measured planets}
\label{sec-demography-K2-223_c}
With its radius measured and its mass constrained at the $\sim$3$\sigma$ level, \planetc{} joins a growing sample of planets for which mean densities can be derived. The availability of such estimates enables the use of density-based diagnostics and population studies \citep[e.g.][]{Luque2022}. Comparisons with the broader exoplanet population in parameter spaces involving orbital period, mass, radius, and density have become a common tool in the characterisation of individual planets \citep[e.g.][]{Carleo2026b, Carleo2026a, Lacedelli2026, Persson2025}.

One density-dependent quantity particularly relevant for~close-in planets is the Roche period, the minimum orbital period below which a planet would be tidally disrupted by its host star. As shown by \cite{Rappaport2013}, the Roche period can be expressed in terms of the planet's mean density ($\rho_{\rm p}$) and depends on its compressibility. For a body comprised of an incompressible fluid, the corresponding period is given by \citep[Eq.~2 in][]{Rappaport2013}
\begin{equation}
P_{\rm Roche} \simeq 12.6\ {\rm h} \left( \frac{\rho_{\rm p}}{1\ {\rm g\ cm^{-3}}}\right)^{-1/2}.
\label{eq:Roche1}
\end{equation}
At the opposite extreme, for a planet composed of a highly compressible fluid, the expression becomes \citep[Eq. 4 in][]{Rappaport2013}\begin{equation}
P_{\rm Roche} \simeq 9.6\ {\rm h} \left( \frac{\rho_{\rm p}}{1\ {\rm g\ cm^{-3}}}\right)^{-1/2}.
\label{eq:Roche2}
\end{equation}
For intermediate regimes, where planets are neither incompressible nor highly compressible, the Roche period can be approximated as \citep[Eq. 5 in][]{Rappaport2013}\begin{equation}
P_{\rm Roche} \simeq 12.6\ {\rm h} \left( \frac{\rho_{\rm p}}{1\ {\rm g\ cm^{-3}}}\right)^{-1/2} \left( \frac{\rho_{\rm 0p}}{\rho_{\rm p}}  \right)^{-0.16},
\label{eq:Roche3}
\end{equation}
where $\rho_{\rm 0p}$ is the central density of the planet, and the ratio $\rho_{\rm 0p}/\rho_{\rm p}$ reflects its composition and internal structure. This formula interpolates between the two limiting cases above, reducing to the incompressible fluid expression for $\rho_{\rm 0p}/\rho_{\rm p}$\,=\,1 (Eq.~\ref{eq:Roche1}) and to the highly compressible one for $\rho_{\rm 0p}/\rho_{\rm p}$\,$\approx$\,6 (Eq.~\ref{eq:Roche2}), beyond which the approximation is no longer valid.

To place \planetc{} in the context of the Roche period introduced above, we compared it with well-characterised planets orbiting main-sequence stars from the NASA Exoplanet Archive, selecting all objects with a mass, radius, and density determined to a precision of 2$\sigma$. For the entire sample, we computed $P_{\rm Roche}$ using Eq.~\ref{eq:Roche3}. Since the ratio $\rho_{\rm 0p}/\rho_{\rm p}$ cannot be measured directly and depends on the internal planet structure, we divided the sample into three categories using limiting radii at 2 \Rea{} and 6.5 \Rea{}. These boundaries approximately correspond to the upper edge of~the radius valley \citep{Fulton2017} and the transition to~giant planets \citep{Kopparapu2018}, defining the Earth- and super-Earth-sized planets (E\&sE), sub-Neptune- and Neptune-sized planets (sN\&N), and Jupiter-sized planets (J). For each class, we assigned fiducial values of $\rho_{\rm 0p}/\rho_{\rm p}$ of 2.5, 4, and 6, respectively. For E\&sE planets, 2.5 is the upper value of the range given by \citet{Rappaport2013} for bodies composed of iron and silicates. For the J class, 6 is the upper limit of the compressible regime introduced above, consistent with the upper end of the polytropic range for gas-giant interiors ($n$\,=\,1--1.5, $\rho_{\rm 0p}/\rho_{\rm p}$\,$\approx$\,3.3--6; \citealt{Wei2024}). The sN\&N class, lying between the two in composition and structure, is assigned an intermediate value of~4.

\begin{figure*}[!h]
\centering
    \includegraphics[width=0.49\linewidth]{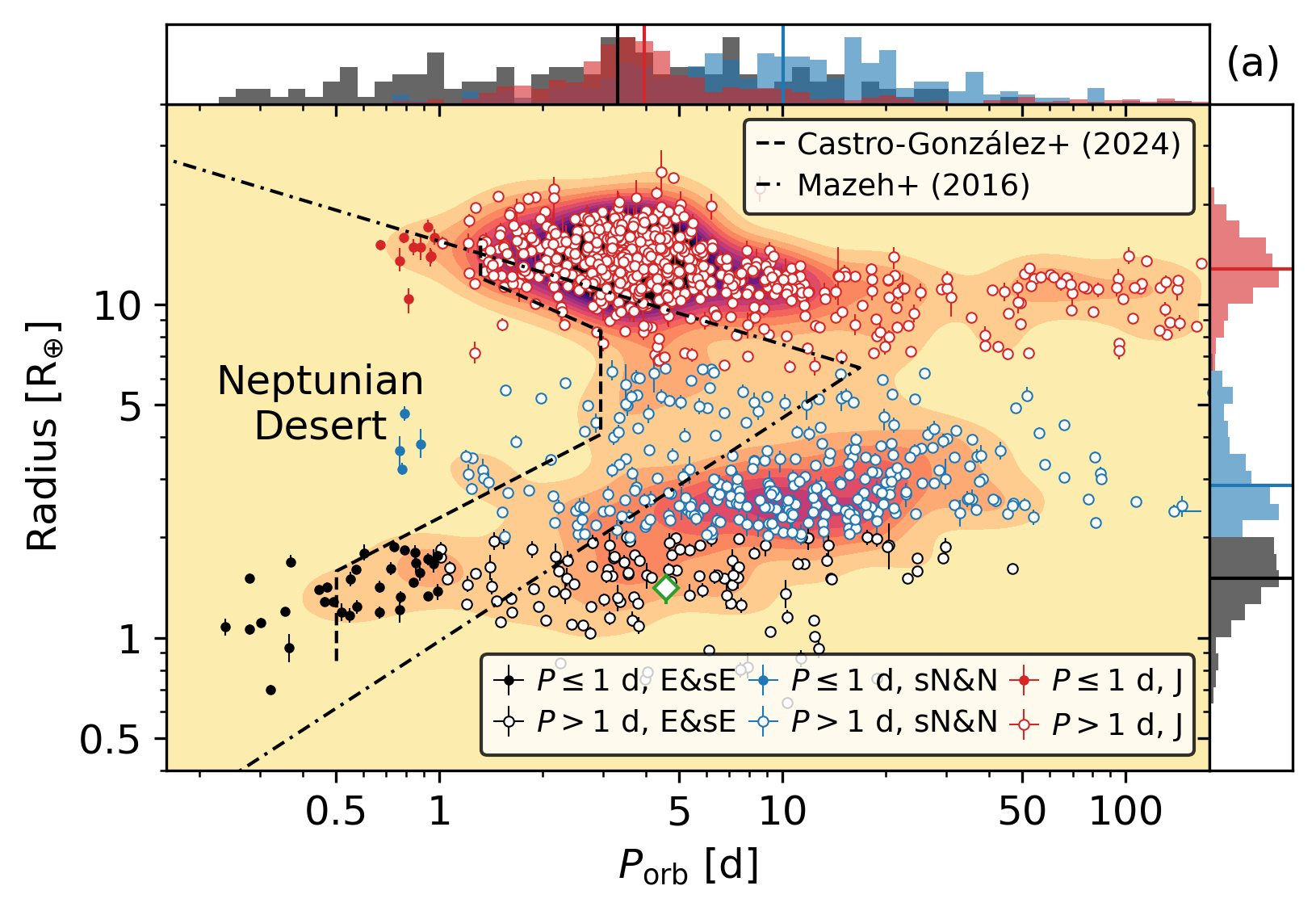}
    \includegraphics[width=0.49\linewidth]{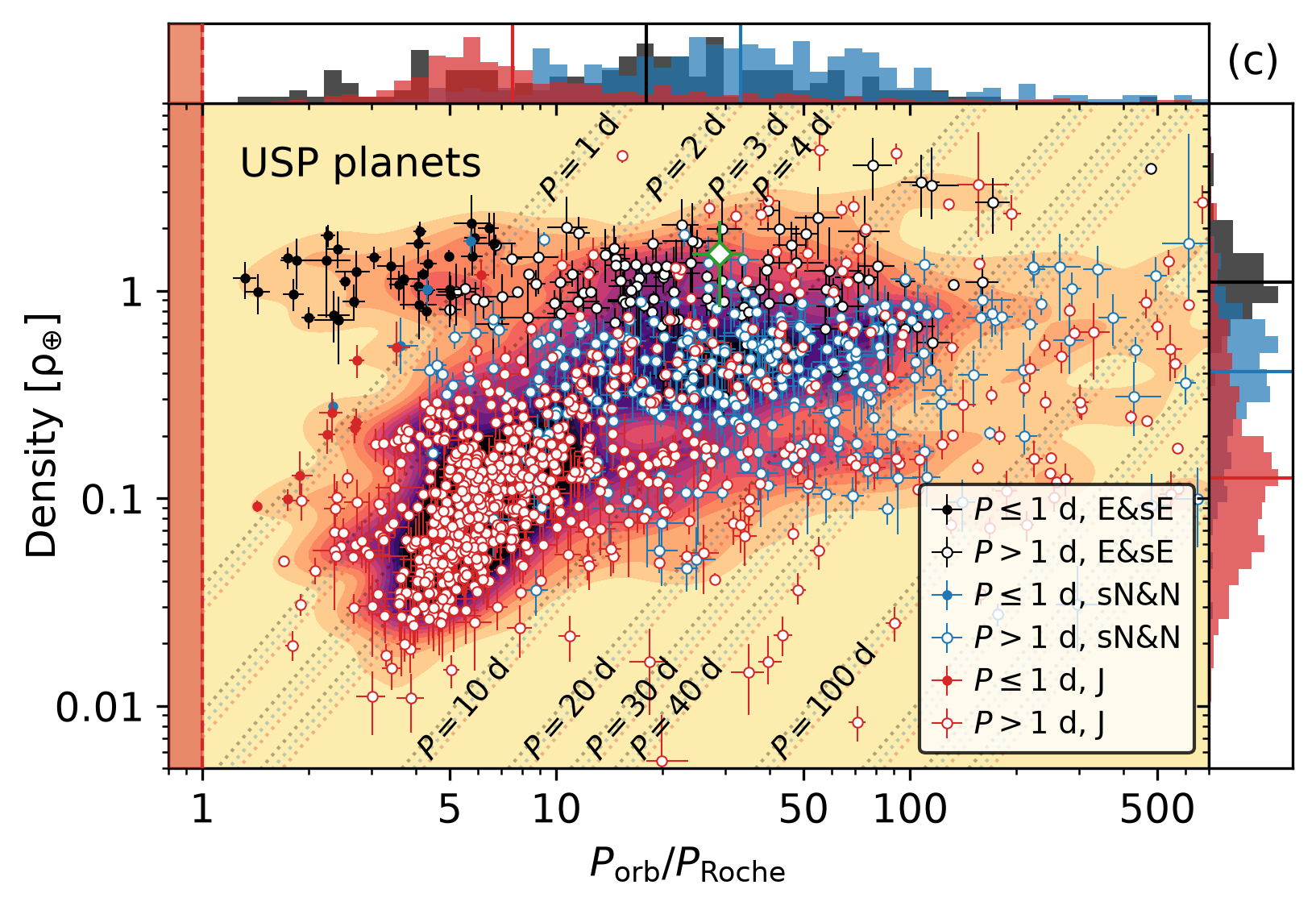}
    \includegraphics[width=0.49\linewidth]{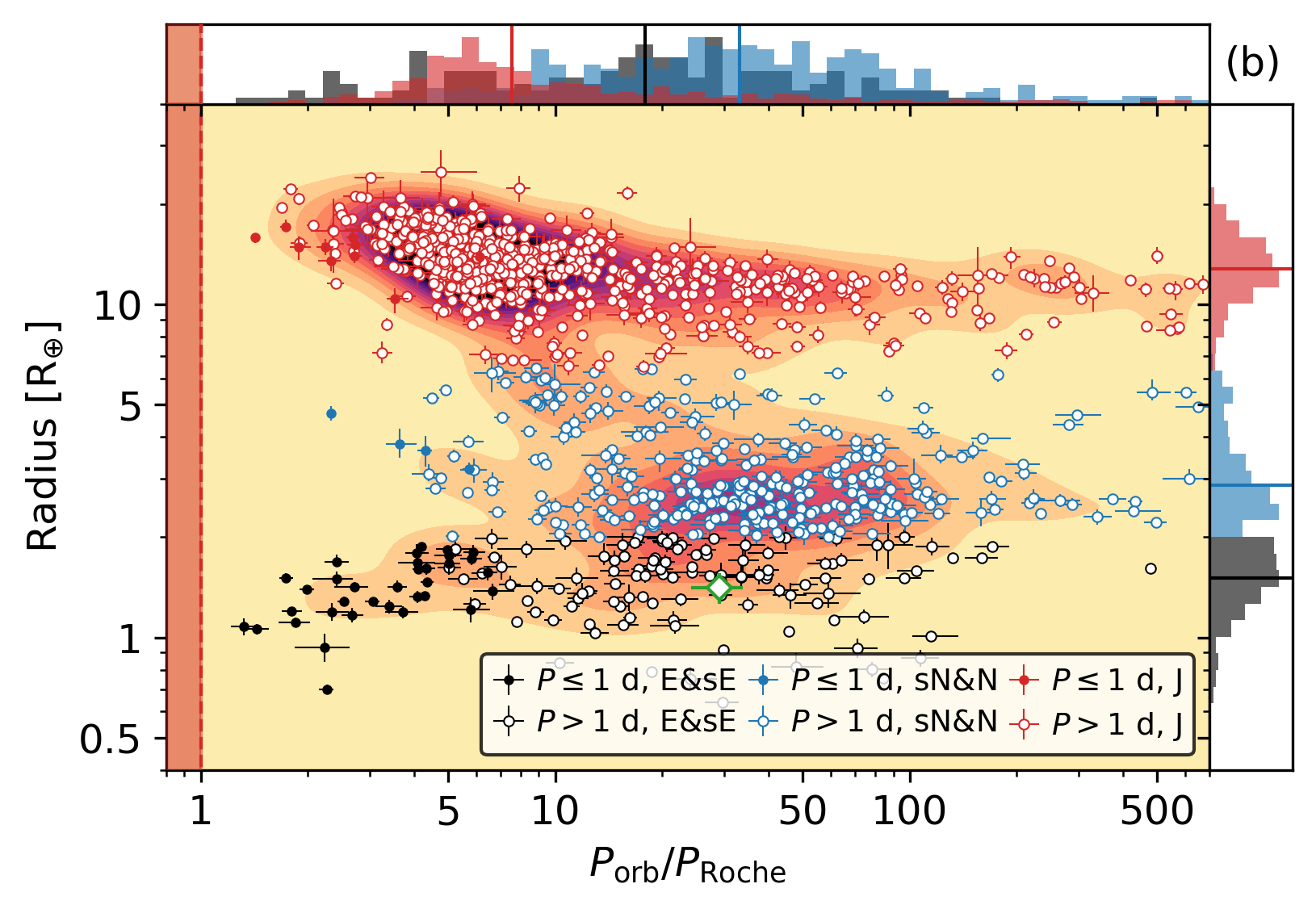}
    \includegraphics[width=0.49\linewidth]{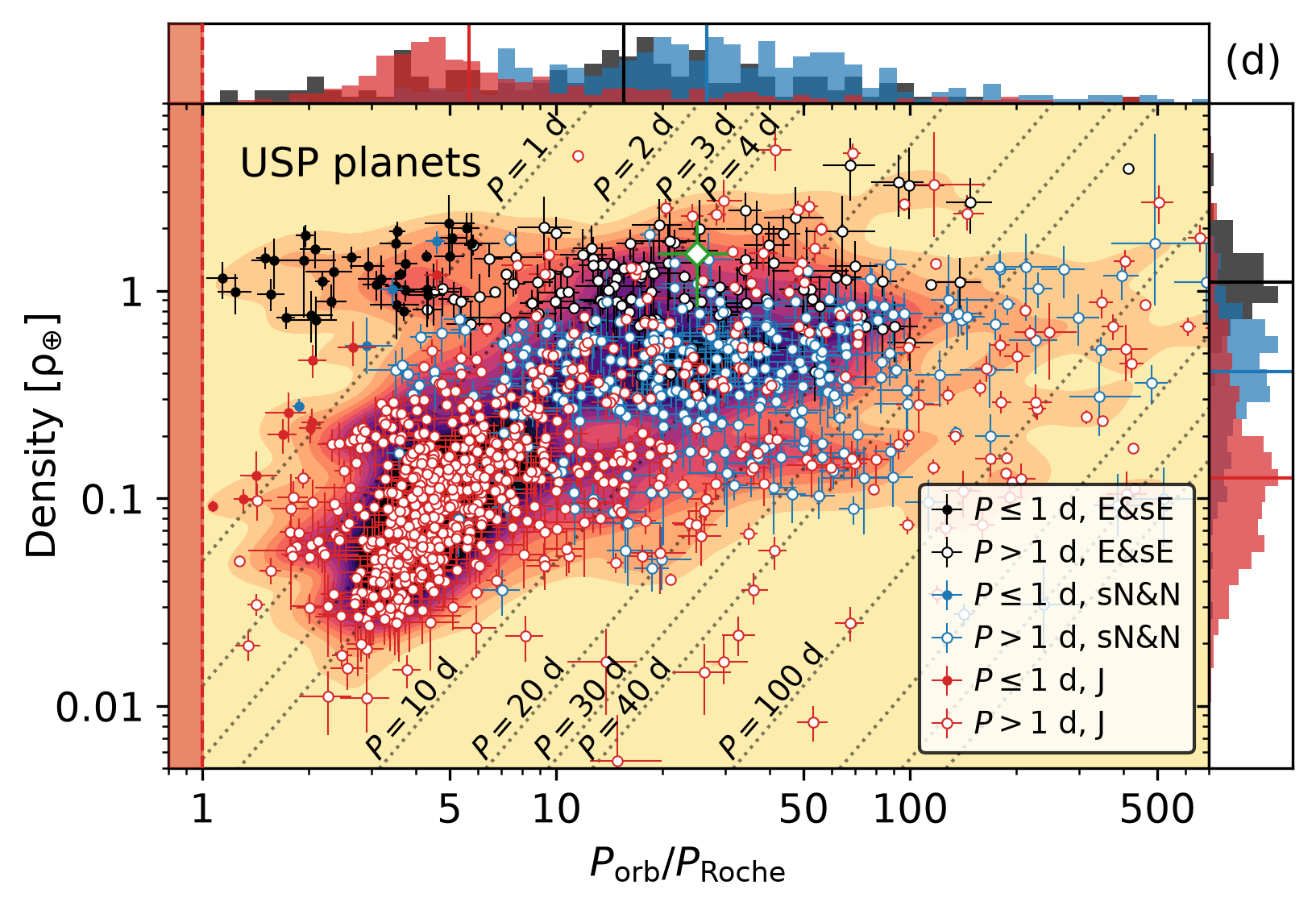}
    \caption{\planetc{} (green diamonds) compared to other planets with observationally measured masses and radii in different planes: radius vs. orbital period (a), radius vs. orbital period in units of the Roche period (b), and density vs. orbital period in Roche units (c/d). Along the two axes of~each panel, normalised histograms of the three classes are shown, colour-coded as the data points, with lines indicating the class medians. Panel (a) shows the boundaries of the Neptune desert according to \cite{Mazeh2016} and \cite{Castro-Gonzalez2024}, indicated by the dash-dotted and dashed lines. The dotted lines in panel (d) indicate the locations of selected orbital periods~($P$) for $\rho_{\rm 0p}/\rho_{\rm p}$\,=\,1 (black), while those in panel (c) correspond to $\rho_{\rm 0p}/\rho_{\rm p}$\,=\,2.5 (grey), 4 (blue), and 6 (red). The data were retrieved from the NASA Exoplanet Archive \citep{Christiansen2025} on~9~June 2026.}\label{figure-ev-USPs-pla}
\end{figure*}

To characterise the proximity of each planet to tidal disruption, we expressed the orbital period relative to the Roche period, where $P_{\rm orb}/P_{\rm Roche}$\,=\,1 corresponds to the Roche limit, below which a~planet disintegrates under tidal forces. \planetc{} and the comparison sample are shown in Fig.~\ref{figure-ev-USPs-pla} (panels a, b, and c) in various parameter spaces, including iso-density contours of the point distribution. Planets are colour-coded according to the radius-based classification (E\&sE, sN\&N, and J). With \pxporbR\,=\,\pcporbv{}, \pxr\,=\,\pcrv{}, \pxden\,=\,\pcdenEv{}, and $P_{\rm orb}/P_{\rm Roche}$\,=\,\pcporbprochev{}, \planetc{} falls within the parameter space typically occupied by E\&sE objects.

Beyond placing individual planets such as \planetc{}, these parameter spaces also reveal population-level features of the close-in planet distribution. A prominent example is the Neptune desert \citep{Szabo2011, Mazeh2016}, a deficit of Neptunian planets at short orbital periods, defined in the radius versus orbital period and mass versus orbital period planes, of which the former is shown in Fig.~\ref{figure-ev-USPs-pla}~(a). Even when based on largely the same dataset, the derived boundaries of the desert differ between studies depending on the adopted method \citep{Mazeh2016,Castro-Gonzalez2024}. This sensitivity motivated us to study the population through an alternative parametrisation based on the orbital period relative to the Roche period (Fig.~\ref{figure-ev-USPs-pla}~b). The comparison of planets by orbital period alone is complicated by the fact that at a fixed period, the physical distance from the star depends on the stellar mass. Expressing the orbital period in units of the Roche period instead places each planet on a common scale set by its own tidal disruption limit, providing a dimensionless quantity for comparison across systems. The effect of this rescaling is apparent in the median values of the three classes, marked in Fig.~\ref{figure-ev-USPs-pla}~(a/b). The median $P_{\rm orb}/P_{\rm Roche}$ of J planets ($\sim$7.5) is lower than that of the E\&sE population ($\sim$18.0), whereas their median orbital period is higher ($\sim$4.0 d versus $\sim$3.3 d). As a result, the ordering of the planet classes in this plane differs from that in orbital period: J planets, despite their longer periods, lie closer to their Roche limit, owing to their lower mean densities and correspondingly longer Roche periods. Notably, the sN\&N class retains the highest median $P_{\rm orb}/P_{\rm Roche}$ ($\sim$33.2), just as it has the highest median orbital period in the radius versus orbital period plane ($\sim$10.1 d), so that Neptune-sized planets stand out in both representations.

Motivated by the use of density as a population diagnostic \citep{Luque2022}, we replace in Fig.~\ref{figure-ev-USPs-pla}~(c) the planet radius on the vertical axis of panel (b) with the mean density, keeping $P_{\rm orb}/P_{\rm Roche}$ on the horizontal axis. Because $P_{\rm Roche}$ depends on~the mean density (Eq.~\ref{eq:Roche3}), at a fixed orbital period and $\rho_{\rm 0p}/\rho_{\rm p}$, the two coordinates $y$\,$\equiv$\,$\rho_{\rm p}$~[\ppdenu{}] and $x$\,$\equiv$\,$P_{\rm orb}/P_{\rm Roche}$ satisfy
\begin{equation}
y =\ x^2 \left( \frac{12.6\ {\rm h}}{P_{\rm orb}} \right)^2 \left( \frac{\rho_{\rm 0p}}{\rho_{\rm p}}  \right)^{-0.32}.
\label{eq:y1}
\end{equation}
In logarithmic variables $Y$\,$\equiv$\,$\log_{10} y$ and $X$\,$\equiv$\,$\log_{10} x$, this becomes a straight line,
\begin{equation}
Y = 2X + 2 \log_{10} \left( \frac{12.6\ {\rm h}}{P_{\rm orb}} \right) - 0.32 \log_{10} \left( \frac{\rho_{\rm 0p}}{\rho_{\rm p}}  \right).
\label{eq:y2}
\end{equation}
The slope was fixed at 2, while the orbital period and $\rho_{\rm 0p}/\rho_{\rm p}$ only shift the line vertically, through the intercept. In panel (c) we draw these lines for several orbital periods, indicated in the panel, allowing the absolute orbital period to be read off despite the normalised horizontal axis. Every period line has three versions, one per planet class, computed with its value of $\rho_{\rm 0p}/\rho_{\rm p}$ (2.5, 4, and 6). They lie close together, reflecting the weak dependence on the ratio.

Together, this representation (Fig.~\ref{figure-ev-USPs-pla}~c) displays the mean density, $P_{\rm orb}/P_{\rm Roche}$, and, through the constant-period lines, the orbital period itself. It shows that the low densities of the J~planets stretch their Roche limit toward longer orbital periods. At~the median density of each class ($\sim$1.11, $\sim$0.41, and $\sim$0.13 \Dea{} for E\&sE, sN\&N, and J), an orbital period of one day gives $P_{\rm orb}/P_{\rm Roche}$ of about 5, 3, and 2, respectively. The same orbital period can therefore correspond to different distances from tidal disruption. Ordering planets by $P_{\rm orb}/P_{\rm Roche}$ thus offers an alternative view of their distribution.

For $P_{\rm orb}/P_{\rm Roche}$\,$\lesssim$\,2.5, close to the Roche limit, the planets are almost all E\&sE or J, with only a single sN\&N case. As a result, the sample leaves a roughly triangular region empty: its base spans mean densities from $\sim$0.1 to $\sim$1\,\Dea{} at $P_{\rm orb}/P_{\rm Roche}$\,=\,1, narrowing to an apex at $P_{\rm orb}/P_{\rm Roche}$\,$\approx$\,2.5, near the median sN\&N density ($\sim$0.41 \Dea{}). As the constant-period lines show, this region lies at orbital periods below one day, so the absence concerns the USP population (marked with filled circles; the open circles show the remaining planets).

In a variant of panel (c), shown in panel (d) of Fig.~\ref{figure-ev-USPs-pla}, we set $\rho_{\rm 0p}/\rho_{\rm p}$\,=\,1 and recomputed $P_{\rm orb}/P_{\rm Roche}$ for all planets, treating them as homogeneous bodies with a density independent of radius. This uniform assumption simplified the analysis by removing the need for class-dependent values and reducing each orbital period to a single line. The planets shift slightly toward their Roche limit, but the picture remains qualitatively unchanged. Even in this most extreme case, all remain at $P_{\rm orb}/P_{\rm Roche}$\,>\,1, outside tidal disruption, with a few lying close to it. More generally, the $P_{\rm orb}/P_{\rm Roche}$ analysis provides a useful framework for~identifying planets near their Roche limit or bordering the observed empty regions, and it can help relate their locations to the tidal and evaporative processes shaping the population of~close-in planets.

\subsection{\host{}\,d in systems with confirmed outer companions}
\label{sec-demography-K2-223_d}
The discovery of a long-term signal in the radial velocities of~the star \host{}, tentatively attributed to planet d, prompted us to review the structures of other similar cases. Of the systems hosting a USP planet, 55 are multi-planet systems. \host{} is included in this sample due to its previously confirmed planets \host{} b and c. If the observed periodicity is of planetary origin, \host{} would be one of the systems showing the largest period difference between the inner and outer planet, with just two other systems having longer outer periods: Kepler-407 and 55 Cnc.

\section{Discussion and conclusions}
\label{sec-discussion}

\subsection{Internal composition of \planetc}
\label{sec-discussion-composition}
\begin{figure}[!h]
\centering
    \includegraphics[width=0.9\linewidth]{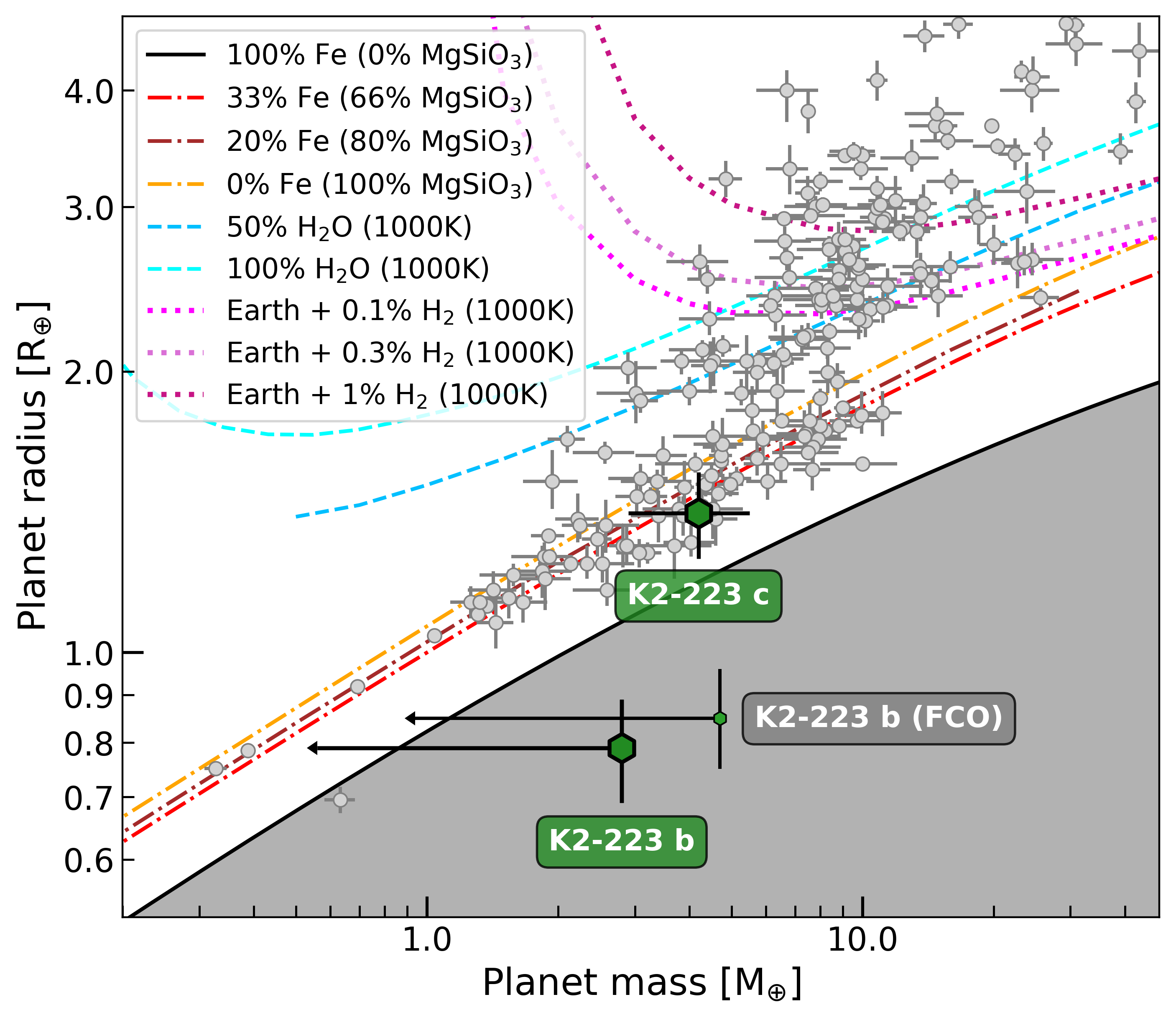}
    \caption{Radius as a function of mass for all known planets listed in the PlanetS catalogue \citep{Otegi2020, Parc2024}, with data retrieved on 9 June 2025. The curves marked with different colours correspond to theoretical planetary composition models from \cite{Zeng2019, Zeng2016} for rocky planets (with different iron and silicate compositions), Earth-like planets with hydrogen atmospheres at 1000 K, and water worlds at 1000 K. We mark the values of the radii and masses of~\host{} b, c determined in this work with green dots and indicate uncertainties. \label{figure-mvsr}}
\end{figure}

To constrain the internal composition of \planetc, we compared its position in a mass--radius diagram against the known exoplanet population from the PlanetS catalogue \citep{Otegi2020, Parc2024} using the {\tt mr-plotter}\footnote{\url{https://github.com/castro-gzlz/mr-plotter}} tool \citep[see our Fig.~\ref{figure-mvsr};][]{Castro-Gonzalez2023}. The resulting plot indicates that \planetc{} is predominantly rocky and highly compatible with Earth-like interior models. While alternative rocky configurations cannot be entirely excluded due to current observational uncertainties, nearly all non-rocky models (with the exception of a 100\% ${\mathrm{H_{2}O}}$ composition) evaluated at the planetary equilibrium temperature are firmly excluded beyond the 2$\sigma$ level. For~consistency, we also included \planetb{} in the mass--radius diagram, adopting its 3$\sigma$ upper mass limit.

These considerations, however, do not include the long-period candidate \host{}\,d, for which the global fit provided only a minimum mass of \pdmsiniv{}. No transits were identified in the available photometric measurements, preventing a determination of the radius and inclination for \host{}\,d.

\subsection{Habitable-zone estimates and prospects for an atmospheric characterisation}
\label{sec-discussion-prospects_for_atm_char}
We calculated the habitable-zone boundaries for this system using the formalism from \cite{Kopparapu2013, Kopparapu2014}. For a one-Earth-mass planet, the habitable zone extends from $\sim$0.93\,AU (runaway greenhouse) to $\sim$1.64\,AU (maximum greenhouse). For a five-Earth-mass planet, the inner boundary shifts to $\sim$0.90\,AU. \host{} b and c, located at \pbav{} and \pcav{}, receive extreme stellar irradiation due to their close orbits, with equilibrium temperatures of \pbteqv{} and \pcteqv{}, respectively (assuming zero albedo). Under these conditions, the planets are not expected to retain substantial volatile envelopes and are consistent with predominantly rocky compositions (see Sect.~\ref{sec-discussion-composition}). \host{}\,d, at \pdav{}, would lie beyond the outer habitable-zone boundary.

To assess the feasibility of characterising the possible atmosphere of K2-223 planets using the \textit{James Webb} Space Telescope, we calculated the transmission spectroscopy metric (TSM) and the emission spectroscopy metric (ESM), as defined by \citet{Kempton_et_al-2018PASP..130k4401K}. Unfortunately, the small sizes of the planets (\pbr{} = \pbrv{}, \pcr{} = \pcrv{}) imply that the TSM (equal to \pbtsmv[] and to \pctsmv[] for \planetb{} and \planetc{}, respectively) and ESM (equal to \pbesmv[] and to~\pcesmv[] for \planetb{} and \planetc{}, respectively) are below the thresholds proposed by \citet{Kempton_et_al-2018PASP..130k4401K} for terrestrial planets (TSM\,>\,10 and ESM\,>\,7.5 for \ppr{}\,<\,1.5\,\Rea{}).

\subsection{Conclusions}
We characterised the multi-planet system orbiting the solar-type star \host{} using photometric data from \ktwo{} and RVs from the ESPRESSO and HARPS-N spectrographs. Through precise measurements, we determined the fundamental parameters of the star and obtained constraints on the masses of the two transiting planets, \hostb{} (\pbm{}\,${<}$\,\pbmtsulv{}) and \hostc{} (\pcm{}\,=\,\pcmv{}). The mass of \planetc{} allowed us to estimate its bulk density (\pcden{}\,=\,\pcdenv{}) and make tentative inferences about its composition. Furthermore, our analysis revealed a third non-transiting tentative Jupiter-like companion on a significantly longer orbital period (\pdporb{}\,=\,\pdporbv{}), with a well-determined minimum mass (\pdmsini{}\,=\,\pdmsiniv{}). Our estimation of relativistic and tidal effects confirmed the validity of our adopted model.

\section*{Data availability}
Tables \ref{table-C10_1452-tng_harpn-0071-drs-complete_output}, \ref{table-C10_1452-tng_harpn-0071-srv-complete_output}, \ref{table-C10_1452-vlt_espresso-0025-drs-complete_output} and \ref{table-C10_1452-vlt_espresso-0025-srv-complete_output} are only available in electronic form at the CDS via anonymous ftp to cdsarc.u-strasbg.fr (130.79.128.5) or via \url{http://cdsweb.u-strasbg.fr/cgi-bin/qcat?J/A+A/}.

\begin{acknowledgements}
This paper includes data collected by the \tess{} mission. Funding for the \tess{} mission is provided by the NASA Explorer Program. We acknowledge the use of public TOI Release data from pipelines at the \tess{} Science Office and at the \tess{} Science Processing Operations Center. Resources supporting this work were provided by the NASA High-End Computing (HEC) Program through the NASA Advanced Supercomputing (NAS) Division at Ames Research Center for the production of the SPOC data products. This research has made use of the Exoplanet Follow-up Observation Program website, which is operated by the California Institute of Technology, under contract with the National Aeronautics and Space Administration under the Exoplanet Exploration Program.
This work has made use of data from the European Space Agency (ESA) mission {\it Gaia} (\url{https://www.cosmos.esa.int/gaia}), processed by the {\it Gaia} Data Processing and Analysis Consortium (DPAC, \url{https://www.cosmos.esa.int/web/gaia/dpac/consortium}). Funding for the DPAC has been provided by national institutions, in particular the institutions participating in the {\it Gaia} Multilateral Agreement.
Based on observations made with the Italian Telescopio Nazionale Galileo (TNG) operated on the island of La Palma by the Fundaci\'on Galileo Galilei of the INAF (Istituto Nazionale di Astrofisica) at the Spanish Observatorio del Roque de los Muchachos of the Instituto de Astrofisica de Canarias under programs OPT18B\_52, A38TAC\_26, CAT19A\_162, CAT21A\_119, CAT23A\_52, CAT23B\_74, CAT24B\_20, and CAT25A\_76.
The participation of D.J. in this research was supported by the Polish National Agency for Academic Exchange (NAWA) under the PROM Programme – short-term academic exchange, call 2024 (project FERS.01.05‑IP.08‑0218/23).
D.J. and G.N. gratefully acknowledges the Centre of Informatics Tricity Academic Supercomputer and networK (CI TASK, Gda\'nsk, Poland) for computing resources (grant no. PT01187). K.G. acknowledges the support and CPU resources provided by the Poznań Supercomputer Centre (PCSS, project PL0406-01).
We acknowledge financial support from the Agencia Estatal de Investigaci\'on of the Ministerio de Ciencia e Innovaci\'on MCIN/AEI/10.13039/501100011033 and the ERDF “A way of making Europe” through projects PID2021-125627OB-C32 and PID2024-158486OB-C32. This work is supported by the European Union (ERC AdvG SPEAR, GA 101200674). Views and opinions expressed are however those of the authors only and do not necessarily reflect those of the European Union or the European Research Council. Neither the European Union nor the granting authority can be held responsible for them.
T.M. acknowledges support from the Spanish Ministry of Science and Innovation with the proyecto plan nacional \textit{PLAtosonG} (grant no. PID2023-146453NB-100, PI: Beck).
Support for this work was provided by NASA through the NASA Hubble Fellowship grant a awarded by the Space Telescope Science Institute, which is operated by the Association of Universities for Research in Astronomy, Inc., for NASA, under contract NAS5-26555. 
R.L. acknowledges financial support from the Severo Ochoa grant CEX2021-001131-S funded by MCIN/AEI/10.13039/501100011033. 
R.L. is funded by the European Union (ERC, THIRSTEE, 101164189). Views and opinions expressed are however those of the author(s) only and do not necessarily reflect those of the European Union or the European Research Council. Neither the European Union nor the granting authority can be held responsible for them. 
F.M. acknowledges the financial support from the Agencia Estatal de Investigaci\'{o}n del Ministerio de Ciencia, Innovaci\'{o}n y Universidades (MCIU/AEI) through grant PID2023-152906NA-I00.
G.M. acknowledges financial support from the Severo Ochoa grant CEX2021-001131-S and from the Ramón y Cajal grant RYC2022-037854-I funded by MCIN/AEI/1144 10.13039/501100011033 and FSE+.
\end{acknowledgements}

\bibliographystyle{aa}
\bibliography{manuscript}

\begin{appendix}

\section{Relativistic and tidal interactions}
\label{sec-non-newton}
The extreme proximity of \hostb{} and \hostc{} to the parent star necessitates considering tidal and relativistic perturbations to Newtonian gravity when modeling their orbital motion.  The impact of these dynamical factors on modeling and interpreting observations of transiting inner planets -- particularly those secularly affected by a giant outer companion -- is a common subject of recent research \cite[e.g.][]{Wei2021, Maciejewski2024}. Here, we focus on short-term and intermediate orbital timescales driven by conservative interactions. As our model system, we adapted the best-fitting orbital elements listed in Table~\ref{tab:planet_bcd}.

\subsection{Timescales of tidal and GR perturbations}
Following \cite{Bernabo2024} and \cite{Csizmadia2019}, we consider a system of a star of mass $M_{\star}$ and a planet of mass $m_{\rm p}$ with radii $R_{\star}$ and $R_{\rm p}$, respectively. 
 We also assume that the star and the planet rotate normally to the orbital plane. For close-in planets, besides General Relativity (GR) perturbations, conservative and rotational tides arise from the static deformation (bulge) due to the differential gravitational force from another object. The GR and tidal forces cause the pericenter of the orbit to rotate, and this effect is quantified by the tidal Love numbers $k_\textrm{2,p}$ and  $k_{\rm 2,\star}$, attributed to the planet and the star, respectively.

The pericenter advance rate ($\dot{\varpi}_{\textrm{t}}$) combines tidal and rotational terms from the bulge raised on the planet by the star ($\dot{\varpi}_{\textrm{tp}}$) and on the star by the planet ($\dot{\varpi}_{\textrm{ts}}$). Explicit formulae in \citep{Bernabo2024,Csizmadia2019} read as follows, after rewriting and collecting terms for the star and planet contributions, respectively:
\begin{eqnarray}
\label{eq:td1}
\dot{\varpi}_\mathrm{tp} &=& 
\frac{n\,k_{2,\mathrm{p}}\,\beta_\mathrm{p}^5}{2 (1-e^2)^2}
\left[ \tau_\mathrm{p}^2 (1+\mu^{-1}) + 15 \mu^{-1} \frac{f(e)}{(1-e^2)^3} \right], \\
\label{eq:td2}
\dot{\varpi}_\mathrm{ts} &=& 
\frac{n\,k_{2,\mathrm{\star}}\,\beta_\mathrm{\star}^5}{2 (1-e^2)^2}
\left[ \tau_\mathrm{\star}^2 (1+\mu) + 15 \mu \frac{f(e)}{(1-e^2)^3} \right],
\end{eqnarray}
where index ``p'' is for the planet, $a$ and $e$ its semi-major axis and eccentricity, $G$ is gravitational constant; $P$, $P_\mathrm{rot,p}$ and $P_\mathrm{rot,\star}$ are the orbital period, and rotational periods of the planet and the star, respectively. The Keplerian mean motion $n = {2\pi} P^{-1}$ determines the III law of Kepler $n^2 a^3 = G (M_\mathrm{\star}+m_\mathrm{p})$. We also introduce non-dimensional length, mass and time coefficients
\[
\beta_i=\frac{R_i} {a},
\quad 
\mu=\frac{m_\mathrm{p}}{M_\star}, 
\quad
\tau_i = \frac{P}{P_\mathrm{rot,i}},
\quad
i=\mathrm{p,\star},
\]
and
$k_\mathrm{2,i} = 2 k_\mathrm{2,aps,i}$, which is
the second order Love fluid number expressed by the apsidal motion constant \citep{Csizmadia2019}. Finally, $f(e)$ is for the eccentricity factor
\[
f(e) = 1 + \frac{3}{2} e^2 + \frac{1}{8} e^4.
\]
Note that the first and second term in each of Eq.~\ref{eq:td1}-\ref{eq:td2} is for rotation and tidal components, respectively.  
Both terms are positive, independent on the spin inclination, and cause rotation of~the apsidal line in the same direction.
The period of tidal precession is then
$
P_\textrm{t,aps} = 
{2\pi}{(\dot{\varpi}_{\rm tp}+\dot{\varpi}_{\rm ts})^{-1}}.
$

The strong inverse dependence on semi-major axis $a$ can be important for USP objects, especially \host{}\,b. Moreover, given small planet mass $\mu \ll 1$, $e=0$, and $\tau_\textrm{p}=1$ for synchronously rotating planet $\dot{\varpi}_{\rm tp}$ dominates the stellar component, hence
\[
\frac{\dot{\varpi}_{\rm tp}}{\dot{\varpi}_{\rm ts}} \simeq  
\frac{16}{\tau_\mathrm{\star}} \frac{k_{2,\rm p}}{k_{2,\star}} \beta^5 \mu^{-3}.
\]
For a rocky planet with the mass of two Earth masses, $k_{2,p} \sim 0.3$, $R_{\rm p}\simeq 0.013 R_{\star}$ and $m_{\rm p} \simeq 2 M_{\odot}/333,030$, and $k_{2,\star} \simeq 0.03$. Then the precession frequencies ratio is roughly between $(5,50)$ for $P \in (0.5,5)$~days and \rm $P_\mathrm{\star} \simeq 22$~days (as found in Table~\ref{tab:planet_bcd}), so~the second term can be essentially omitted.

The GR pericenter precession rate $\dot{\varpi}_{{\rm GR}}$ for a small planet is \citep[e.g.][]{Bernabo2024}
\begin{equation}
\dot{\varpi}_\textrm{GR} = 
\frac{3 \, n \, G M_{\star} }{c^2 a (1-e^2)},
\end{equation}
where $c$ is the velocity of light. Now,
we can compare the dominant and tidal components of the apsidal advance.
\begin{equation}
\frac{\dot{\varpi}_{\rm tp}}{\dot{\varpi}_{\rm GR}} \simeq 
\frac{8}{3}  c^2 \, k_{2,\rm p}
\, \beta_\textrm{\rm p}^5 \, \mu
\frac{G M_{\star}}{a}.
\end{equation}

Table~\ref{tab:tGR} shows the pericenter advance frequencies for the tidal, rotational, and GR components, the total frequency and period for different types of planets and the Love number $k_\textrm{2,p}$. Clearly, the partition of the tidal and GR terms of the pericenter advance is not intuitive, given the complex interplay of model parameters.

Considering the \hostb{} and \hostc{} rocky planets with $k_{2,\textrm{p}} \simeq 0.3$, the GR-induced pericenter advance is approximately 50 times faster than the total tidally induced precession for \hostb{}, despite its extremely short orbital period. For \hostc{}, this ratio is approximately 200. It should be noted that for USP Jupiter planets, the ratio can be reverse. For instance, for WASP-19b, it is approximately 1/500, hence tidal interactions strongly dominate the GR-induced pericenter advance.

\subsection{Tidal, rotational, and GR corrections to the RV model}
Given significant tidal and GR perturbations, one could be concerned about the validity of the Keplerian RV model in Sect.~\ref{sec-modelling_and_results}. Therefore we aim to estimate how these perturbations affect the RV measurements model expressed by three terms:
\begin{equation}
V_\textrm{r} = V_\textrm{K} +
V_\textrm{GR+t} + V_{\textrm{tidal},\star},
\end{equation}
where $V_\textrm{K}$ is the classic, Keplerian component, $V_\textrm{GR+t}$ is the term stemming from non-Keplerian pericenter advance, and $V_{\textrm{tidal},\star}$ is the stellar term arising due to the tidal distortion of the star by a~close-in planet \cite[e.g.][]{Arras2012}. 

Following \cite{Csizmadia2019}, the two first orbital terms $V_\textrm{orb}=V_\textrm{K} + V_\textrm{GR+t}$, where the Keplerian, kinematic term, 
\begin{equation}
\label{eq:Vk}
V_\textrm{K} = 
K [ e \cos \varpi' + \cos( \nu+\varpi' ) ]
,\end{equation}
is modified by the GR and tidal interactions:
\begin{equation}
\label{eq:deltaV}
 V_\textrm{GR+t} = K \frac{\dot\varpi}{n} 
\frac{(1-e^2)^{3/2} \cos ( \nu + \varpi' )}{1+e \cos \nu}.
\end{equation}
Here $K$ is Keplerian semi-amplitude of the orbital RV signal, $\nu=\nu(t)$ is the true anomaly, and we introduced a new variable $\varpi'$, i.e. the pericenter longitude modified by the pericenter advance, hence $\varpi'(t) = \varpi_0 + \dot{\varpi} \Delta\,t$, where $\varpi'(t_0) = \varpi_0$ and $\Delta\,t=(t - t_0)$ for fixed epoch $t_0$. For circular orbits, $V_\textrm{orb}$ greatly simplifies:
\begin{equation}
\label{eq:dK}
V_\textrm{orb}(\Delta\,t,e=0) = 
K \left( 1+\frac{\dot{\varpi}}{n} \right) 
\cos(\nu + \varpi_0 + \dot{\varpi} \Delta\,t).
\end{equation}
This expression means a change of the Keplerian semi-amplitude $K$ of the RV signal and its orbital phase. Furthermore, Eq.~\ref{eq:dK} can be interpreted such that the mean motion $n$ expressing the phase $\nu(t) \equiv {\cal M}(t) = n \Delta\,t$ is corrected as $n' = n + \dot{\varpi}$. This means a change of the orbital period $P$, also introducing second-order correction to $K$ through $\dot{\varpi} n^{-1}$ (Eq.~\ref{eq:deltaV}). 

The combined effect of the apsidal rotation is therefore a change of $K (1+\Delta\,K)$ of the RV semi-amplitude and the ephemeris of the planet. When inspecting estimates for the  \host{}\,b, c planets in Table~\ref{tab:tGR}, we can conclude that $\Delta K \simeq 10^{-7}$ is negligible, deep below any present detection limit. 

The uncertainty of the photometric orbital period of \hostb{}  $\simeq 10^{-4}$\,d is much larger than $\simeq 4 \times 10^{-6}$\,d caused by the pericenter advance, making our Keplerian parameter estimates (the orbital period of \hostb{}) in Table~\ref{tab:planet_bcd} valid. However, the secular change of $\varpi(t)$ is more ``dangerous'' as it can affect, in~the long term, moments of transits and, for instance, TTV measurements. The TTV change over the present $\simeq 9$~yr observations window is substantial. The GR and tidally induced orbital phase shift for \hostb{} during this interval is approximately $6^{\circ}$ which translates to significant change of the mid-transit moments by $\simeq 12$~min., compared to transit width of $\simeq 70$~min.

The third, generally significant and measurable correction to~the RV is the planet-induced tidal bulge of the star that results in $V_{\textrm{tidal},\star}$ term of the total RV. While a derivation of this term complementing $V_\textrm{orb}$ can be found in \citep{Csizmadia2019}, an~excellent explanation and discussion of this effect give also \cite{Arras2012}. As they note, in the first approximation, for~circular orbits
$
V_{\textrm{tidal},\star} \simeq \mu \rho_{\star}^3 n R_\star, 
$
and it reaches $1-10$\,m\,s$^{-1}$ for~massive, close-in planets. With a more rigorous treatment, the tidally induced RV signal can be expressed through the semi-amplitude
\begin{equation}
\label{eq:tKs}
K_{\textrm{t},\star}
= \frac{3}{2} n \mu \rho^3_\mathrm{\star} R_\mathrm{\star} f_2 \sin^2 I,
\end{equation}
where $f_2 \simeq 1.2$ is the Eddington coefficient related to the limb darkening, and $I$ is the inclination. 

Again, inspection of planet parameters and data in Table~\ref{tab:tGR} assures us that this component of RV is essentially irrelevant for the \host{} model due to a small amplitude of $K_{\textrm{t},\star} \simeq 0.025$\,m\,s$^{-1}$ induced by \hostb{}. However, for hot Jupiters WASP-19b and WASP-18b listed in Table~\ref{tab:tGR} the tidal semi-amplitude $K_{\textrm{tidal},\star} \simeq 3$\,m\,s$^{-1}$.

\def\arcsec{[$''$]}
\begin{table*}
\centering
\caption{Apsidal advance frequencies and total rotation period for \host{}\,b,c planets and other exoplanets.}
\begin{tabular}{lcccccccccc}
\hline
\hline
Planet & $a$ & $P_{\rm orb}$ & $e$ & $k_{2,p}$ & $\dot{\varpi}_{\rm GR}$ & 
$\dot{\varpi}_{\rm tidal}$ & $\dot{\varpi}_{\rm rot}$ & $\dot{\varpi}_{\rm total
}$ & $P_{\rm aps, total}$ \\
 Type & (AU) & (d) & & & (\arcsec\,d$^{-1}$) & (\arcsec\,d$^{-1}$) & (\arcsec\,d$^{-1}$) & (\arcsec\,d$^{-1}$) & (years) \\
\hline
K2--223b     & 0.0135 & 0.554 & 0     & 0.10 & 5.49 & 0.0403 & 0.1434 & 5.675 &    625 \\
K2--223b     & 0.0135 & 0.554 & 0     & 0.30 & 5.49 & 0.1003 & 0.1474 & 5.740 &    618 \\
K2--223b     & 0.0135 & 0.554 & 0     & 1.50 & 5.49 & 0.4602 & 0.1714 & 6.123 &    579 \\
\hline
K2--223c     & 0.0551 & 4.610 & 0     & 0.10 & 0.16 & 0.0000 & 0.0010 & 0.160 &  22229 \\
K2--223c     & 0.0551 & 4.610 & 0     & 0.30 & 0.16 & 0.0001 & 0.0010 & 0.160 &  22221 \\
K2--223c     & 0.0551 & 4.610 & 0     & 1.50 & 0.16 & 0.0004 & 0.0010 & 0.160 &  22176 \\
\hline
Super--Earth & 0.0500 & 4.084 & 0.020 & 0.30 & 0.19 & 0.0004 & 0.0012 & 0.190 &  18713 \\
Neptune--like& 0.0800 & 8.265 & 0.010 & 0.39 & 0.06 & 0.0003 & 0.0002 & 0.059 &  60507 \\
Warm Jupiter & 0.1000 & 11.55 & 0.050 & 0.57 & 0.03 & 0.0013 & 0.0002 & 0.035 & 102017 \\
\hline
WASP--19b    & 0.0165 & 0.790 & 0.002 & 0.01 & 2.85 & 13.005 & 0.8743 & 16.73 &    212 \\
WASP--19b    & 0.0165 & 0.790 & 0.002 & 0.22 & 2.85 & 272.63 & 18.202 & 293.7 &   12.1 \\
WASP--19b    & 0.0165 & 0.790 & 0.002 & 1.50 & 2.85 & 1855.1 & 123.82 & 1982.&    1.8 \\
WASP--103b    & 0.0196 & 0.907 & 0     & 0.59 & 2.63 & 394.69 & 26.362 & 423.7 &    8.4 \\
WASP--18b    & 0.0203 & 0.945 & 0.009 & 0.62 & 2.50 & 16.680 & 0.9280 & 20.11 &    176 \\
\hline
\end{tabular}
\label{tab:tGR}
\tablefoot{Apsidal advance rates given by Eqs.~(\ref{eq:td1}) and (\ref{eq:td2}) and their original versions in \citep{Bernabo2024}. 
Stellar parameters assume main sequence relations with $k_{2,\star} = 0.03$ and $P_{\rm rot,\star} = 22$~days. Planet rotation is synchronous with its orbital period. We denote
$\dot{\varpi}_{\rm GR}$, 
$\dot{\varpi}_{\rm tidal}$, 
$\dot{\varpi}_{\rm rot}$ for GR-, tidally and rotationally-induced pericenter advance frequencies, respectively. Then
$T_{\rm aps,total} = 2\pi/\dot{\varpi}_{\rm total}$ denotes the total apsidal advance period in years. Parameters for \host{} planets are from this paper; for USP Jupiters: WASP-18b from \citep{Csizmadia2019}, for WASP-19b from \citep{Bernabo2024}, and for WASP-103b from  \citep{Barros2022}; other systems taken as generic.}
\end{table*}

\section{Dynamical stability of the system}
\label{sec-dynamic_analysis}

\subsection{Near 9:1 MMR of the inner pair }
To explore whether the close-in sub-system Earth-like planets might be near mean-motion resonances (MMRs) and to assess how dynamically stable this subsystem is, we employed direct $N$-body integrations with Newtonian forces. 
We conducted the integrations to compute dynamical fast indicator called the reversibility error method \citep[REM,][]{Panichi2017}, directly related to the maximum Lyapunov exponent (MLE). The REM algorithm relies on a time-reversible (symplectic) integration scheme to solve the equations of motion forward and backward in time for the same number of a constant time step. By normalising the difference between the initial and final state by the orbit size in the phase-space, we can differentiate between regular (stable) and chaotic (unstable) configurations. The REM is scaled  such that when $\log\mathrm{\widehat{REM}}=0$ (strong chaos), it indicates that the difference between starting and final states compares to~the orbit size.

\begin{figure}[!h]
\centering
    \includegraphics[width=\linewidth]{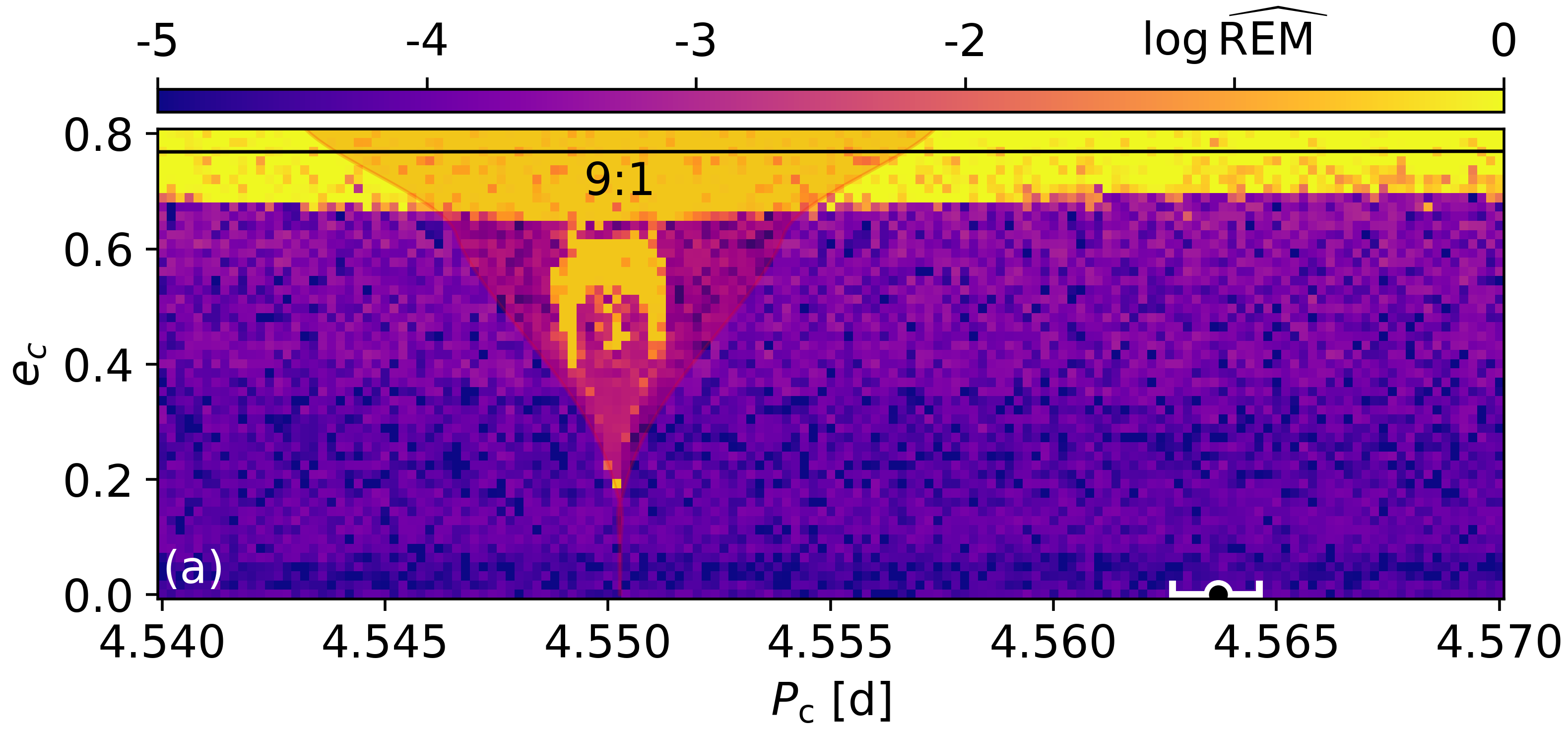}
    \includegraphics[width=\linewidth]{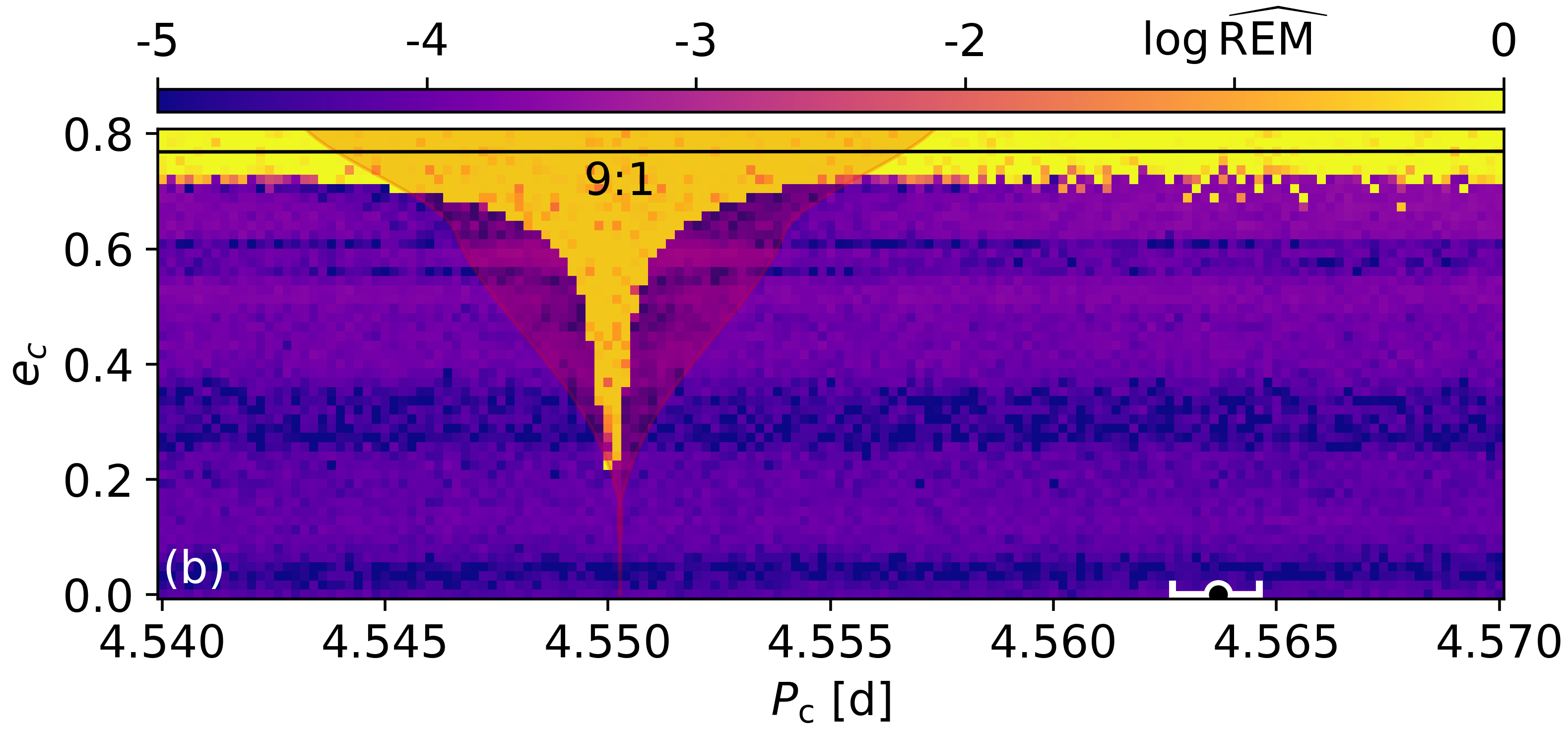}
    \caption{REM dynamical maps in the (\pcporb, \pce)-plane comparing (a) purely gravitational dynamics with (b) dynamics including relativistic and tidal effects. Small values of $\log \mathrm{\widehat{REM}}$ indicate stable, long-term regular solutions, shown in dark blue or purple regions. Chaotic dynamical solutions are represented by progressively lighter colours, up to yellow. The red area corresponds to resonance widths calculated using the semi-analytic algorithm in \cite{Gallardo2021}. The black curve represents the geometric collision of orbits, defined by the condition: $a_{\rm b}(1 + e_{\rm b}) = a_{\rm c}(1 - e_{\rm c})$. The white circle with a horizontal line represents the best-fit solution with $3\sigma$ uncertainties for the parameter \pcporb{} ($e_{\rm c}=0$ fixed). The map resolution is 151~$\times$~51 pixels.\label{figure-remAE}}
\end{figure}

We computed the REM with the \texttt{WHFAST} integrator and the 17th order corrector.
The \texttt{WHFAST} is the leap-frog scheme implemented in the \texttt{REBOUND}\footnote{\url{https://github.com/hannorein/rebound}} package \citep{Rein2012,Rein2015}. We conducted integrations with a fixed time step of 0.02\,d and the time interval $\simeq 1\times 10^6$ orbital periods of \hostc{}. The resulting dynamical map shown in Fig.~\ref{figure-remAE} (a) demonstrates rigorously stable and non-resonant orbit of~the close-in planet system. Their orbital period ratio of $\simeq 9.03$ could imply increased interactions due to proximity of the high-order 9:1 MMR. However, the nominal best-fitting model is at~a~completely safe distance of several $\sigma$ from this MMR structure shown in the $(a_{\rm c},e_{\rm c})$-plane. The 9:1 MMR can be identified both by the chaotic V-shape region in terms of the REM indicator and also determined with a semi-analytic algorithm by \cite{Gallardo2021}, and their \texttt{plares} code\footnote{\url{http://www.fisica.edu.uy/~gallardo/atlas/plares.html}}.

\subsection{Could Lidov-Kozai oscillations be present?}
A large gap between the inner planets and the tentative outer companion raises some questions about the dynamical architecture and stability of the \host{} system. In this analysis, we treat the outer signal \host{}\,d as planetary in nature. While the configuration of two small planets and the outer component is hierarchical with period ratios as large as a few $10^3$ and could be anticipated as stable, the unconstrained inclination of \host{}\,d and its true mass can still significantly affect the orbital evolution of the system over long timescales due to the Lidov-Kozai (L-K) resonance \citep{Naoz2016}. Briefly,
the L-K effect emerges when
the outermost, relatively distant planet forming a hierarchical configuration with the close-in planets attains particular mutual inclination ($i_{\rm mut})$ of the orbits:
\[
\cos i_{\rm mut} = \cos i_{\rm c} \cos i_{\rm d} + 
\sin i_{\rm c} \cos i_{\rm d} \cos\Delta\Omega, 
\]
where $\Delta\Omega=\Omega_{\rm d} - \Omega_{\rm c,b}$.  If the mutual inclination reaches the $\sim$$(39^{\circ}, 141^{\circ})$ range, the inner eccentricity can oscillate with large amplitude reaching even $e_{\rm in} \simeq 1$, in anti-phase with $i_\textrm{mut}$. The period of the L-K oscillations can be approximated to the quadrupole order of the ($a_{\rm in}/a_{\rm out})$ ratio as \cite[e.g.][]{Kiseleva1998}
\begin{equation}
\label{eq:LKperiod}
\tau_{\rm in} = \frac{2 P_{\rm out}^2}{3 \pi P_{\rm in}} \frac{M_{\star}+M_{\rm in} + M_{\rm out}}{M_{\rm out}} (1-e_{\rm out}^2)^{3/2}.
\end{equation}
where $P_{\rm out}$, $P_{\rm in}$, $M_{\rm out}$, $M_{\rm in}$, and $e_{\rm out}$, $e_{\rm in}$ are the orbital periods, masses and eccentricity of the outer perturber planet and the close-in companion, respectively. Using this formulae and precession advance frequencies (Table~\ref{tab:tGR}), we can estimate the period $\tau_{\idm c} \simeq 55,000\ P_{\rm d}$ of the L-K cycles for \hostc{} and $\tau_{\rm b}$ which is $\simeq 20$ times longer for the innermost planet. We consider such intervals as the intermediate timescale of the orbital evolution.  

In order to quantify such long-term L-K effect, the averaging techniques are usually employed \citep[e.g.][]{Naoz2016}. However, in this paper, similarly to previous studies \citep{Maciejewski2024,Subjak2025}, we used non-averaged dynamical model through direct numerical integrations covering several approximate periods of the L-K oscillations, i.e. $\simeq 4\times 10^5$ orbital periods of the outermost planet \hostd{}. We used the \texttt{WHFAST} integrator with the time step of $0.1$ days. To reduce CPU overhead, in the first attempt to quantify the orbital evolution in the presence of the outer perturber \hostd{}, we excluded the close-in \hostb{} from the model and focused only on the orbital evolution of planet \hostc{}. As noted above, the inner pair is only weakly gravitationally coupled, given significant dynamical separation from nearby 9:1~MMR. 

Due to the proximity of \host{}\,c and \host{}\,b to the star, in this experiment spanning long-term evolution, we also included post-Newtonian and tidal corrections in the equations of motion. To account for the GR, we added simplest, conservative radial potential mimicking the relativity (GR) correction as proposed by \cite{Nobili1986}. This potential closely reproduces the pericenter advance with a minimal CPU overhead of a few per cent. To perform computations, we used the \texttt{REBOUNDx}\footnote{\url{https://github.com/dtamayo/reboundx}} package \citep{Tamayo2020} and its \texttt{gr\_potential} function implementing the GR correction in \citep{Nobili1986}. 

\begin{figure*}[!h]
\centering
    \includegraphics[width=0.3175\linewidth]{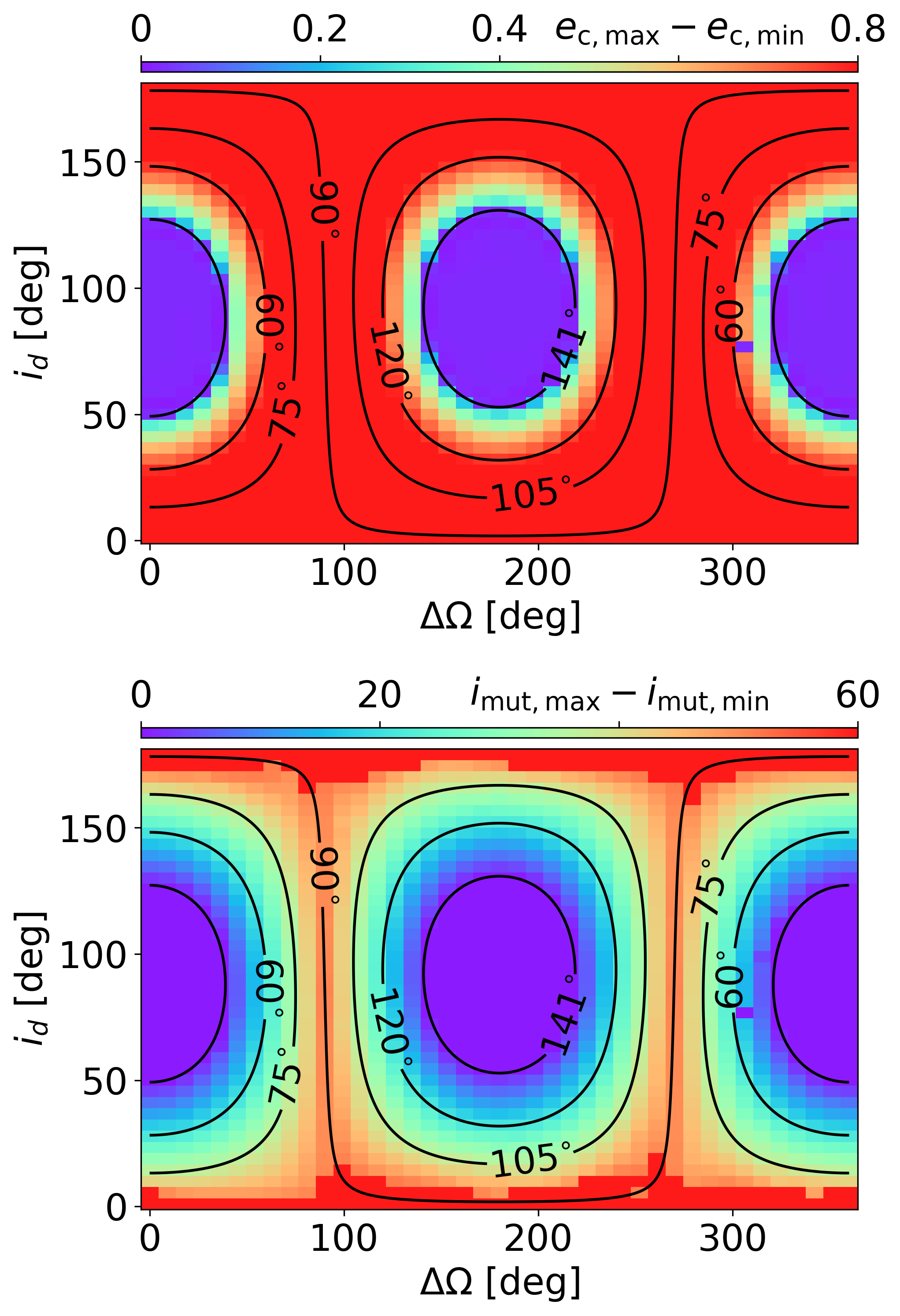}
    \includegraphics[width=0.33\linewidth]{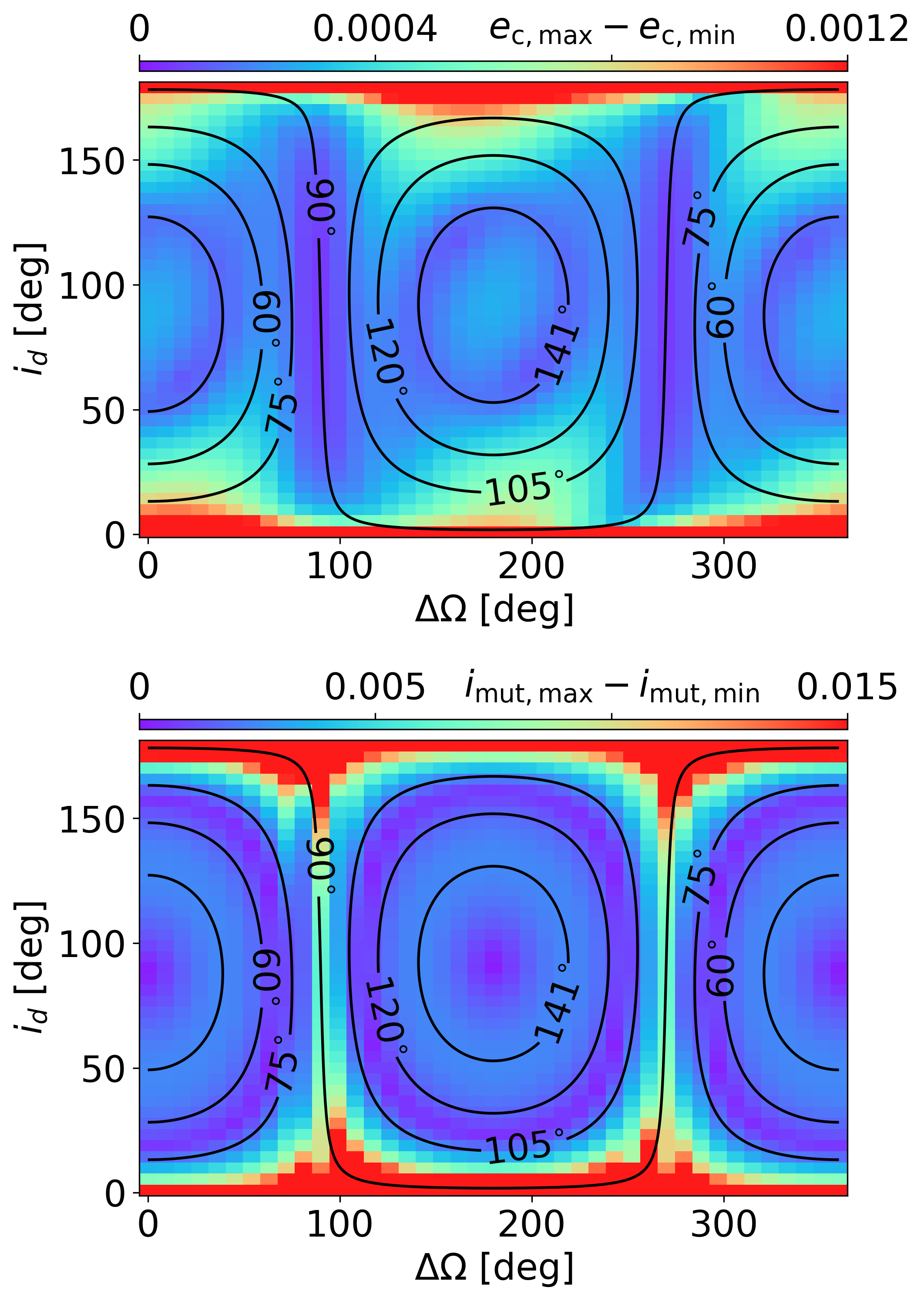}
    \includegraphics[width=0.33\linewidth]{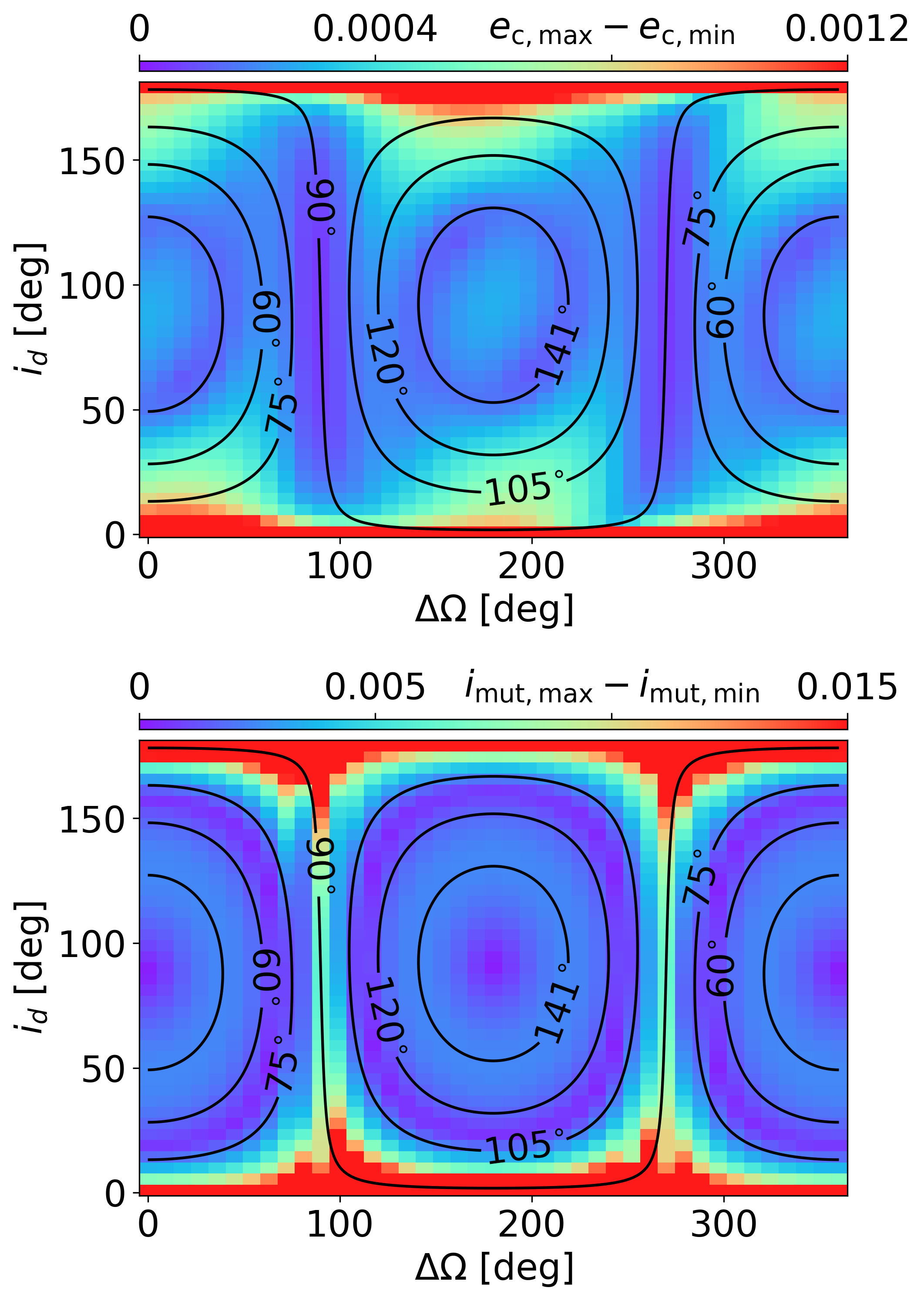}
    \caption{Eccentricity variations of \hostc{} and mutual orbital inclination over several oscillation cycles in the ($\Delta\Omega$, \pdi)-plane, where $\Delta\Omega \in [0^{\circ}, 360^{\circ}]$ and $i_{\rm d} \in [1^{\circ}, 179^{\circ}]$. \textit{Left column}: Newtonian gravitational effects. \textit{Middle}: With GR perturbations taken into account. \textit{Right}: With both GR perturbations and tidal effects in the equations of motion taken into account. The black curves represent specific initial levels of mutual orbital inclination. The resolution of the maps is 41~$\times$~41 points. \label{figure-mutual}}
\end{figure*}

The results shown in the middle column of Fig.~\ref{figure-mutual} differ significantly from those obtained when considering Newtonian gravitational forces alone (left column). The changes in eccentricity and  inclination of \hostc{} in the model with GR perturbation are orders of magnitude smaller for most tested solutions. This holds even for inclinations of \hostd{} around $0^\circ$ and $180^{\circ}$, respectively, which would imply large masses of the perturber planet. The suppression of the L-K oscillations by relatively fast GR pericenter advance is well known \citep{Naoz2016}. Our numerical experiment fully confirms analytical predictions based on averaging theory.

 To model the tidal perturbations, we again used the \texttt{REBOUNDx} package, incorporating the \texttt{tides\_constant\_time\_lag} conservative potential \citep{Baronett2022, Bolmont2015, Hut1981}. We focused on the tides generated by \host{} and \host{}\,c by assigning them the physical radii from Tables~\ref{tab:stellar_params} and \ref{tab:planet_bcd}, respectively, and the second-order tidal Love number $k_{2,\rm p}$. For main sequence solar-like stars, such as \host{}, $k_{2,\star} \approx 0.03$ \citep{Claret1995}. For \host{} c, we assume a rocky, Earth-like interior and adopt $k_{2,\rm p}$\,$\approx$\,0.3 \citep{Amorim2024} as a representative value for the tidal Love number. Given the dominance of the GR-induced rotation of the pericenter, estimated in Table~\ref{tab:tGR}, the results illustrated in the right column of Fig.~\ref{figure-mutual} confirm the lack of differences between the model with tides and GR alone. 

We recall that the orbits of small-mass \hostb{} and \hostc{} were considered as decoupled in the simulations of mutual inclination of the outermost \host{}\,d, as demonstrated with the dynamical REM map. For planet \host{}\,b, which was not included in simulations for the pair \hostc{}--\hostd{}, the approximate timescale of L-K oscillations is nearly 10 times longer than for \hostc{}, i.e. $\sim 500,000$ \pdporb{}. To cover several complete L-K cycles in this case, we would have to integrate the equations of motion for tens of Myrs. This is not justified, given a huge CPU overhead and non-adequate direct numerical approach to the system with orbital period ratio $\simeq 3600$. Based on the L-K analysis and theoretical developments in the literature, we conclude that due to even faster combined GR and tidal pericenter advance of the innermost planet than for \hostc{}, the L-K oscillations for \hostb{} will also remain suppressed.  A detailed study of the long-term dynamics of the whole three-planet system is beyond the scope of this paper, as it would require even more realistic force field, such as dissipative tides.

A general conclusion flowing from the direct $N$-body integration experiments is the absence of dynamical stability constraints that could help to estimate limits on the true mass of~\hostd{}.

For consistency with our treatment of the system throughout this work, we created a sister map to Fig.~\ref{figure-remAE}a that incorporates both GR corrections and tidal interactions, the results are shown in Fig.~\ref{figure-remAE}b. We adopted $k_{2,\rm p} \approx 1.5$ for planet b. While the overall stability structure remains similar to the purely gravitational case, these additional physical effects cause minor changes in the resonance architecture visible in the map.

\section{Figures and tables}
\begin{figure}[!h]
\centering
    \includegraphics[width=0.9\linewidth]{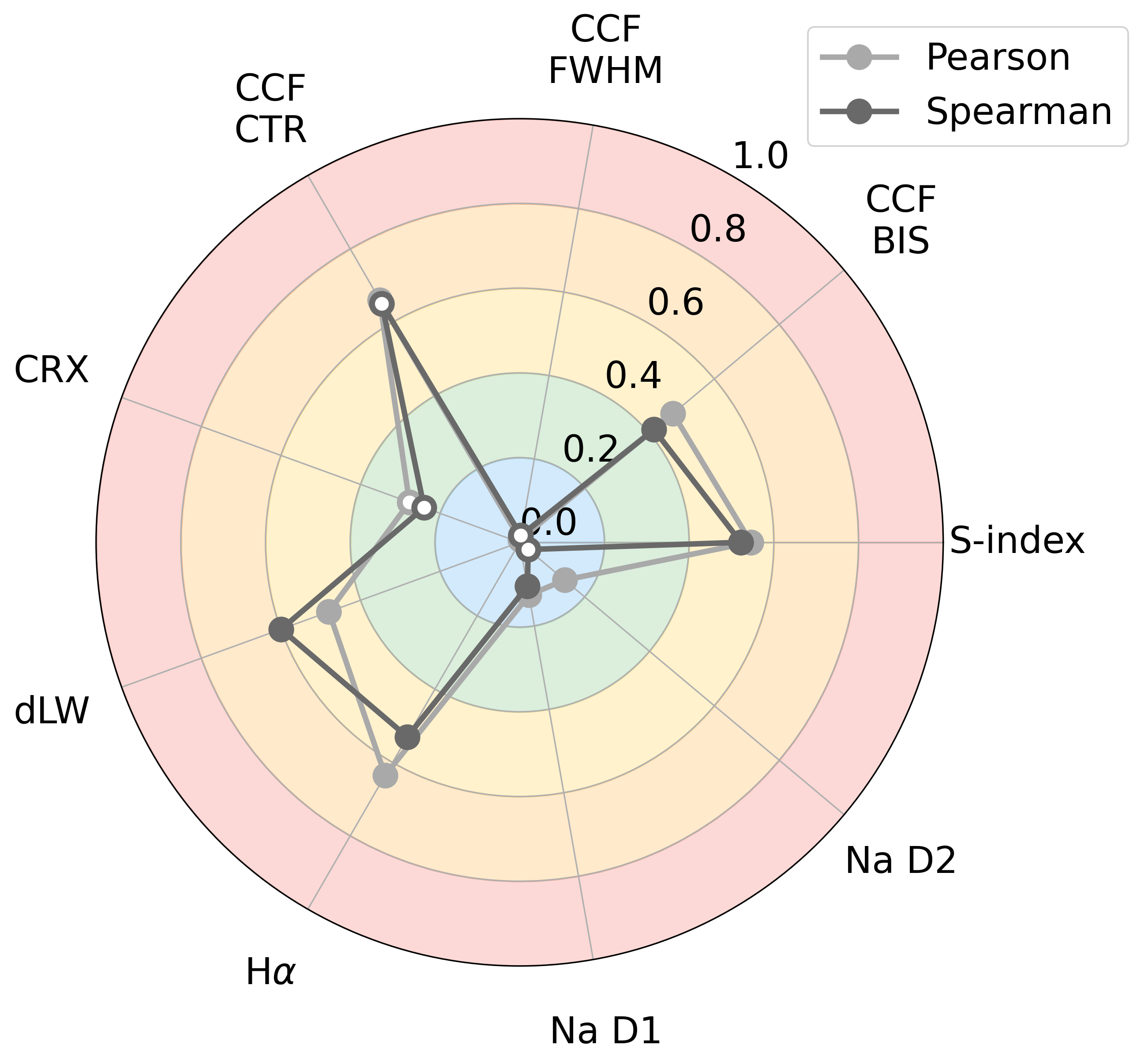}
    \caption{Radar chart of the correlation between RVs and activity indicators for the HARPS-N data. The plotted values correspond to the absolute Pearson and Spearman correlation coefficients, with white-filled markers indicating negative correlations. \label{figure-ind-coeff}}
\end{figure}

\begin{figure}[!h]
\centering
    \includegraphics[width=1\linewidth]{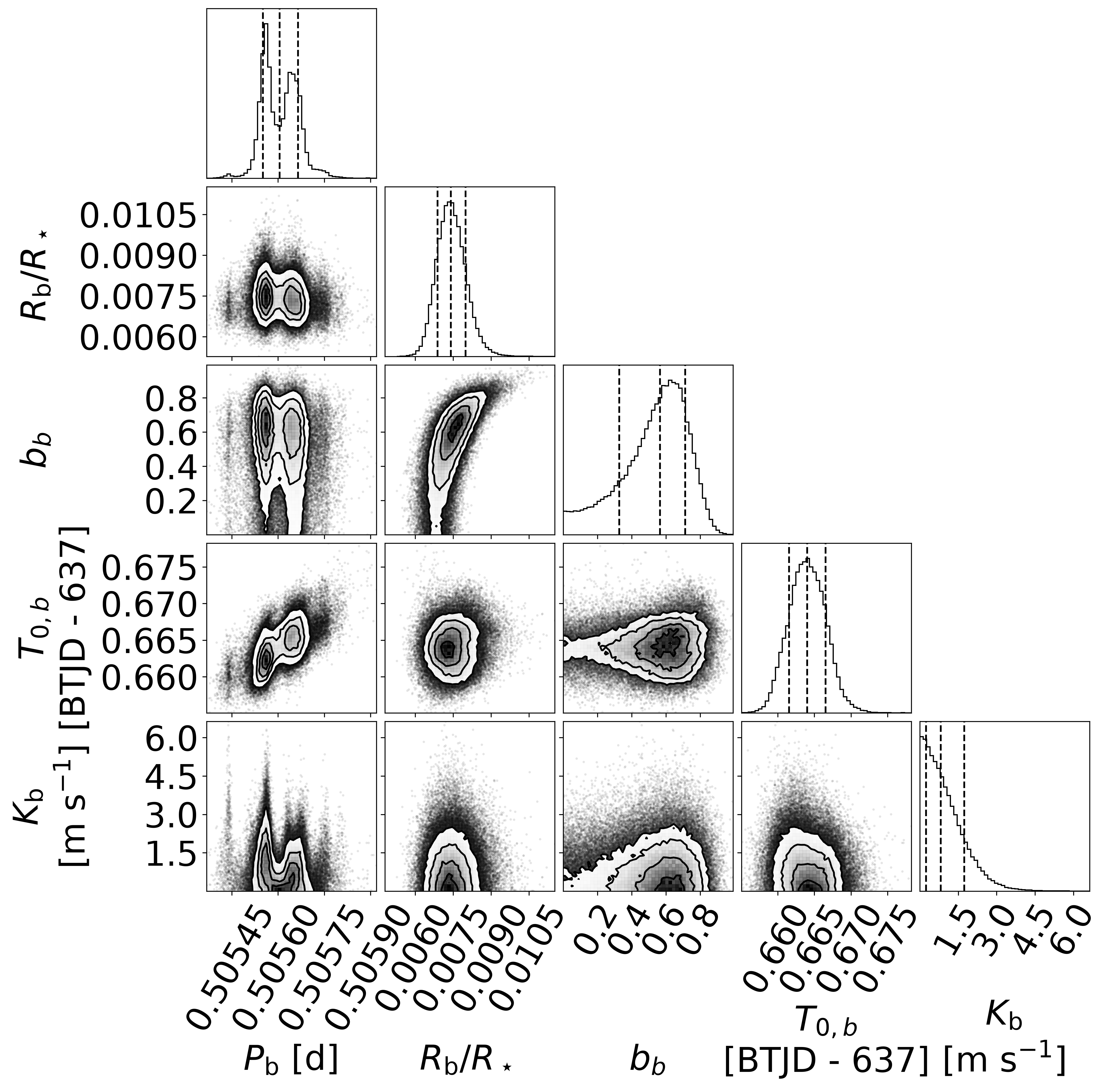}
    \caption{MCMC histograms for planet \host{}\,b using the FCO method. The dashed lines mark the 16th, 50th, and 84th percentiles of~each marginalised posterior distribution. \label{figure-planet_b-mcmc}}
\end{figure}

\begin{figure}[!h]
\centering
    \includegraphics[width=0.9\linewidth]{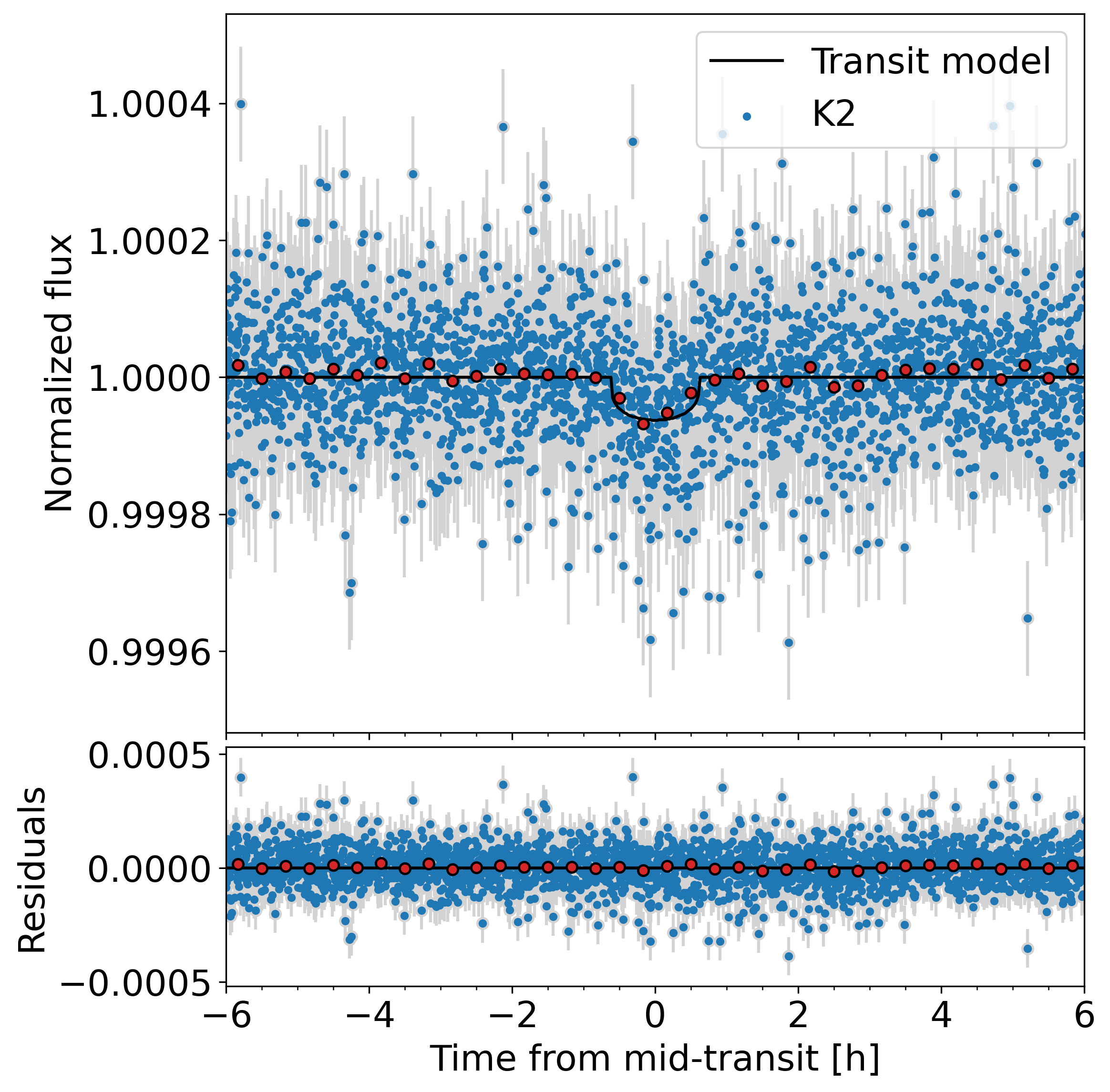}
    \includegraphics[width=0.8\linewidth]{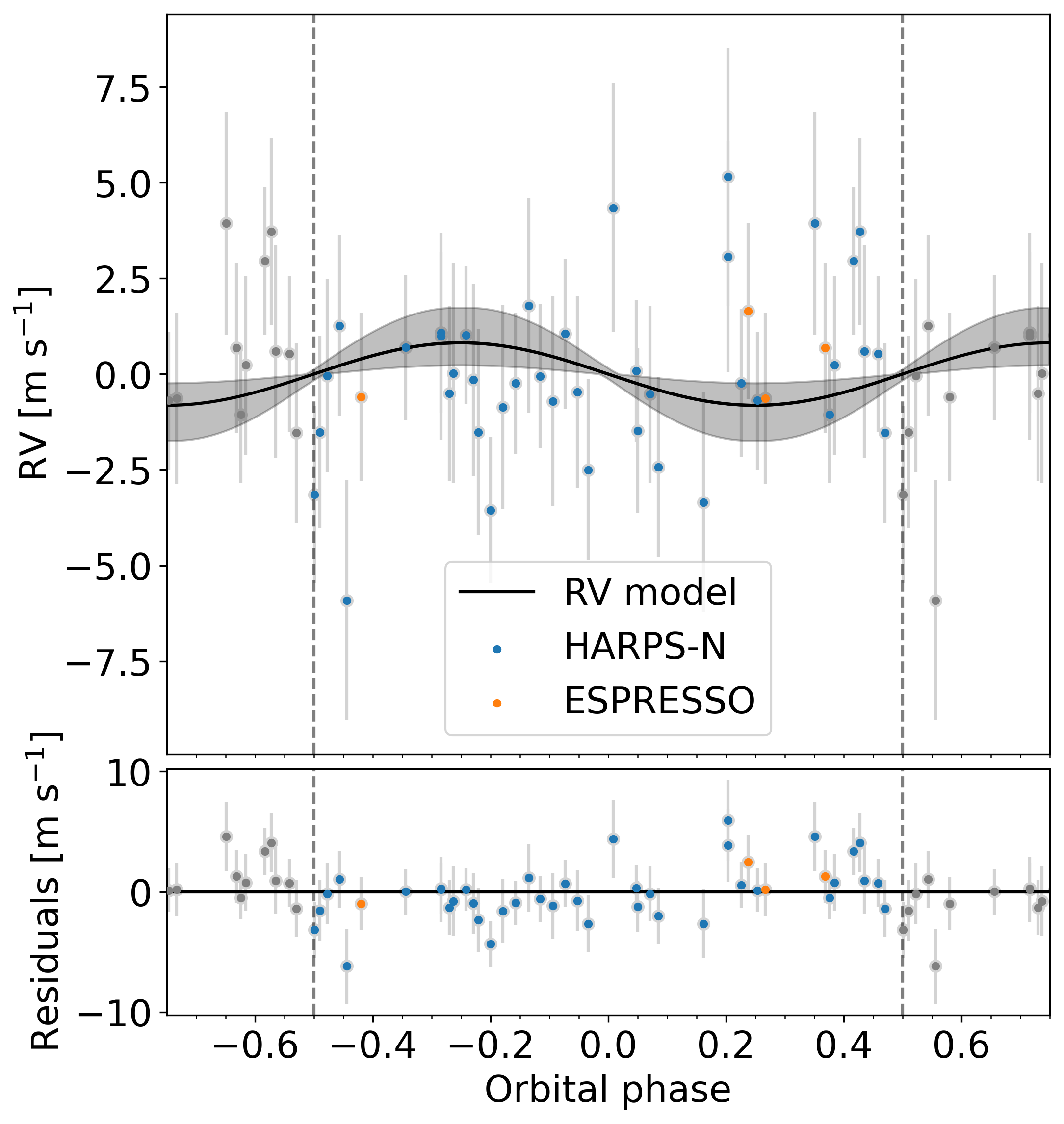}
    \caption{Phased model marked with a black line for the photometric curve (\textit{top}) and radial velocities (\textit{bottom}) of planet \host{}\,b using the FCO method. The red dots represent the median value calculated every 20 minutes from the centre of the transit. The grey area for radial velocities represents the 1$\sigma$ uncertainties of all fitted orbital parameters. \label{figure-planet_b-phased}}
\end{figure}

Supplementary material for the main text. Figure~\ref{figure-ind-coeff} is discussed in Sect.~\ref{sec-modelling_and_results-planets_in_periodograms}, Table~\ref{tab:planet_bc_phot} in Sect.~\ref{sec-modelling_and_results-transits}, Figs.~\ref{figure-planet_b-mcmc} and~\ref{figure-planet_b-phased} in Sect.~\ref{sec-modelling_and_results-planet_b}, and Table~\ref{tab:planet_bcd} in Sects. ~\ref{sec-modelling_and_results-planet_b} and \ref{sec-modelling_and_results-planet_bcd}.

\begin{table}[ht!]
    \centering
    \caption{\host{} system parameters obtained in this work using photometric data from the \ktwo{} mission.}
    \begin{tabular}{lcc}
    \hline
    \hline
    \noalign{\smallskip}
        Parameter & \host{}\,b & \host{}\,c \\
    \noalign{\smallskip}
    \hline
    \noalign{\smallskip}
        \multicolumn{3}{c}{Model parameters} \\
    \noalign{\smallskip}
    \hline
    \noalign{\smallskip}
        \pxporb{} [d] & $0.505556_{-0.000046}^{+0.000048}$ & $4.56365_{-0.00058}^{+0.00048}$ \\
        \pxe{} & 0 (fixed) & 0 (fixed) \\
        \pxw{} [$^{\circ}$] & 90 (fixed) & 90 (fixed) \\
        \pxpsrr{} & $0.00708_{-0.00061}^{+0.00068}$ & $0.01235_{-0.00056}^{+0.00062}$ \\
        \pxb{} & $0.65_{-0.17}^{+0.12}$ & $0.38_{-0.23}^{+0.17}$ \\
        \pxtzero{} [BTJD] & $637.6618_{-0.0023}^{+0.0029}$ & $628.6156_{-0.0028}^{+0.0034}$ \\
    \noalign{\smallskip}
    \hline
    \noalign{\smallskip}
        \multicolumn{3}{c}{Derived parameters} \\
    \noalign{\smallskip}
    \hline
    \noalign{\smallskip}
        \pxar{} & $2.78_{-0.23}^{+0.17}$ & $12.04_{-1.00}^{+0.72}$ \\
        \pxi{} [$^{\circ}$] & $76.3_{-3.6}^{+3.9}$ & $88.2_{-1.0}^{+1.1}$ \\
        \pxr{} [\Rea{}] & $0.81_{-0.10}^{+0.11}$ & $1.41$\,$\pm$\,$0.15$ \\
        \pcttof{} [d] & $0.0470_{-0.0058}^{+0.0048}$ & $0.1130_{-0.0059}^{+0.0051}$ \\
        \pctttt{} [d] & $0.0458_{-0.0061}^{+0.0050}$ & $0.1096_{-0.0062}^{+0.0052}$ \\
        \pxa{} (\sden{}, \sr{}) [AU] & $0.0134$\,$\pm$\,$0.0016$ & $0.0581$\,$\pm$\,$0.0071$ \\
        \pxteq{} [K] & $2484_{-77}^{+104}$ & $1194_{-37}^{+50}$ \\
    \noalign{\smallskip}
    \hline
    \noalign{\smallskip}
        \multicolumn{3}{c}{Common parameters} \\
    \noalign{\smallskip}
    \hline
    \noalign{\smallskip}
        \sden{} [$\rho_\odot$] & \multicolumn{2}{c}{$1.12_{-0.26}^{+0.21}$} \\
        $q_{1, \ktwo{}}$ & \multicolumn{2}{c}{$0.53$\,$\pm$\,$0.10$} \\
        $q_{2, \ktwo{}}$ & \multicolumn{2}{c}{$0.19$\,$\pm$\,$0.16$} \\
        $\sigma_{\rm \ktwo{}}$ & \multicolumn{2}{c}{$0.0000634$\,$\pm$\,$0.0000015$} \\
    \noalign{\smallskip}
    \hline
    \hline
    \end{tabular}
    \label{tab:planet_bc_phot}
\end{table}

\begin{table*}[ht!]
    \centering
    \caption{\host{} system parameters obtained in this work using photometric data from the \ktwo{} mission and radial velocities data from VLT/ESPRESSO + TNG/HARPS-N.}
    \resizebox{\textwidth}{!}{%
    \begin{tabular}{lc|c|ccc}
    \hline
    \hline
    \noalign{\smallskip}
        Parameter & Prior\tablefootmark{a} (\host{} b/c/d) & \planetb{} (FCO) & \planetb{} & \planetc{} & \host{} d candidate \\
    \noalign{\smallskip}
    \hline
    \noalign{\smallskip}
        \multicolumn{5}{c}{Model parameters} \\
    \noalign{\smallskip}
    \hline
    \noalign{\smallskip}
        \pxporb{} [d] & \ldots\,/\,\ldots\,/\,\ldots\tablefootmark{b} & $0.505605_{-0.000054}^{+0.000061}$ & $0.505608_{-0.000057}^{+0.000073}$ & $4.56395_{-0.00033}^{+0.00034}$ & $1653_{-124}^{+118}$\\
        \pxe{} & -\,/\,-\,/\,$\mathcal{U}[0, 0.5]$ & 0 (fixed) & 0 (fixed) & 0 (fixed) & $0.355_{-0.056}^{+0.057}$ \\
        \pxw{} [$^{\circ}$] & - & 90 (fixed) & 90 (fixed) & 90 (fixed) & $-127$\,$\pm$\,$12$ \\
        \pxpsrr{} & - & $0.00741_{-0.00053}^{+0.00058}$ & $0.00687_{-0.00053}^{+0.00059}$ & $0.01233_{-0.00054}^{+0.00059}$ & - \\
        \pxb{} & $\mathcal{U}[0, 1.1]$\,/\,$\mathcal{U}[0, 1.1]$\,/\,- & $0.57_{-0.24}^{+0.15}$ & $0.57_{-0.21}^{+0.15}$ & $0.38_{-0.22}^{+0.17}$ & - \\
        \pxtzero{} [BTJD] & \ldots\,/\,\ldots\,/\,\ldots\tablefootmark{b} & $637.6640$\,$\pm$\,$0.0025$ & $637.6646_{-0.0030}^{+0.0028}$ & $628.6173_{-0.0033}^{+0.0025}$ & $3787_{-50}^{+52}$ \\
        \pxk{} [\mps{}] & $\mathcal{U}[0, 10]$\,/\,$\mathcal{U}[0, 10]$\,/\,$\mathcal{U}[0, 50]$ & $<3.6$ & $<2.1$ & $1.54$\,$\pm$\,$0.49$ & $22.7_{-2.2}^{+3.0}$ \\
    \noalign{\smallskip}
    \hline
    \noalign{\smallskip}
        \multicolumn{5}{c}{Derived parameters} \\
    \noalign{\smallskip}
    \hline
    \noalign{\smallskip}
        \pxar{} & - & $2.70_{-0.26}^{+0.21}$ & $2.79_{-0.23}^{+0.16}$ & $12.10_{-1.00}^{+0.68}$ & $611_{-55}^{+49}$ \\
        \pxi{} [$^{\circ}$] & - & $78.0_{-4.7}^{+5.5}$ & $78.1_{-4.1}^{+4.7}$ & $88.2_{-1.0}^{+1.1}$ & - \\
        \pxm{} [\Mea{}] & - & $<4.7$ & $<2.8$ & $4.2$\,$\pm$\,$1.3$ & - \\
        \pxmsini{} [\Mjup{}] & - & - & - & - & $1.29_{-0.17}^{+0.23}$ \\
        \pxr{} [\Rea{}] & - & $0.85_{-0.10}^{+0.11}$ & $0.79$\,$\pm$\,$0.10$ & $1.41$\,$\pm$\,$0.15$ & - \\
        \pcttof{} [d] & - & $0.0519_{-0.0044}^{+0.0043}$ & $0.0502_{-0.0053}^{+0.0045}$ & $0.1131_{-0.0060}^{+0.0045}$ & - \\
        \pctttt{} [d] & - & $0.0507_{-0.0047}^{+0.0045}$ & $0.0491_{-0.0055}^{+0.0046}$ & $0.1098_{-0.0063}^{+0.0046}$ & - \\
        \pxa{} (\sm{}) [AU] & - & $0.01270_{-0.00028}^{+0.00027}$ & $0.01271_{-0.00028}^{+0.00027}$ & $0.0551$\,$\pm$\,$0.0012$ & $2.80$\,$\pm$\,$0.15$ \\
        \pxa{} (\sden{}, \sr{}) [AU] & - & $0.0131$\,$\pm$\,$0.0017$ & $0.0135$\,$\pm$\,$0.0016$ & $0.0584_{-0.0071}^{+0.0070}$ & $2.96_{-0.38}^{+0.39}$ \\
        \pxteq{} [K] & - & $2521_{-99}^{+122}$ & $2480_{-72}^{+103}$ & $1191_{-34}^{+50}$ & $168_{-7}^{+8}$ \\
        \pxden{} [\Dea{}] & - & $<7.6$ & $<5.7$ & $1.50$\,$\pm$\,$0.67$ & - \\
        \pxden{} [\ppdenu{}] & - & $<42.2$ & $<31.6$ & $8.3$\,$\pm$\,$3.7$ & - \\
    \noalign{\smallskip}
    \hline
    \noalign{\smallskip}
        \multicolumn{5}{c}{Common parameters} \\
    \noalign{\smallskip}
    \hline
    \noalign{\smallskip}
        \sden{} [$\rho_\odot$] & $\mathcal{N}[0.92, 0.28]$ & $1.04_{-0.27}^{+0.26}$ & & $1.14_{-0.26}^{+0.20}$ \\
        $q_{1, \ktwo{}}$ & $\mathcal{U}[0.51, 0.10]$ & $0.54$\,$\pm$\,$0.10$ & & $0.525$\,$\pm$\,$0.096$ \\
        $q_{2, \ktwo{}}$ & $\mathcal{U}[0.15, 0.17]$ & $0.19$\,$\pm$\,$0.16$ & & $0.19$\,$\pm$\,$0.16$\\
        $\sigma_{\rm \ktwo{}}$ & - & $0.0000651$\,$\pm$\,$0.0000015$ & & $0.0000598$\,$\pm$\,$0.0000015$ \\ 
        $\sigma_{\rm ESPRESSO}$ [\mps{}] & - & $2.1_{-1.4}^{+5.2}$ & & $2.46_{-0.51}^{+0.59}$ \\
        $\sigma_{\rm HARPS-N}$ [\mps{}] & - & $0.76_{-0.51}^{+0.68}$ & & $1.63_{-0.66}^{+0.60}$ \\
    \noalign{\smallskip}
    \hline
    \noalign{\smallskip}
        \multicolumn{5}{c}{GP parameters} \\
    \noalign{\smallskip}
    \hline
    \noalign{\smallskip}
        \sprot{} [d] & $\mathcal{U}[16, 36]$ & - & & $26.90_{-1.10}^{+0.16}$ & \\
        \spdec{} [d] & $\mathcal{U}[32, 5000]$ & - & & $517_{-233}^{+292}$ & \\
        \soamp{} & $\mathcal{U}[0.01, 0.6]$\,\&\,$\mathcal{N}[0.31, 0.08]$ & - & & $0.392$\,$\pm$\,$0.068$ & \\
    \noalign{\smallskip}
    \hline
    \hline
    \end{tabular}
    }
    \label{tab:planet_bcd}
    \tablefoot{
    The determination of \host{} b parameters using the floating chunk offset (FCO) method is described in Sect.~\ref{sec-modelling_and_results-planet_b}.\\
    \tablefoottext{a}{$\mathcal{U}[a, b]$ and $\mathcal{N}(\mu, \sigma)$ denote a uniform prior between $a$ and $b$, and a Gaussian prior with mean $\mu$ and standard deviation $\sigma$, respectively.}\\
    \tablefoottext{b}{\pxporb{} [d]: $\mathcal{U}[0.5050, 0.5065]$\,/\,$\mathcal{U}[4.561, 4.566]$\,/\,$\mathcal{U}[1400, 2100]$, \pxtzero{} [BTJD]: $\mathcal{U}[637.655, 637.705]$\,/\,$\mathcal{U}[628.59, 628.64]$\,/\,$\mathcal{U}[3600.0, 4000.0]$.}
    }
\end{table*}

\section{Radial velocity data}
Radial velocity and activity indices measurements from the TNG/HARPS-N (Tables~\ref{table-C10_1452-tng_harpn-0071-drs-complete_output} and \ref{table-C10_1452-tng_harpn-0071-srv-complete_output}) and VLT/ESPRESSO (Tables~\ref{table-C10_1452-vlt_espresso-0025-drs-complete_output} and \ref{table-C10_1452-vlt_espresso-0025-srv-complete_output}).

\newpage

\begin{sidewaystable}
\begin{tiny}
\begin{center}
\tablefirsthead{
\hline
\hline
\noalign{\smallskip}
\multicolumn{1}{r}{BJD$_\mathrm{TDB}$} &
\multicolumn{1}{r}{RV} &
\multicolumn{1}{r}{$\sigma_\mathrm{RV}$} &
\multicolumn{1}{r}{CCF-BIS} &
\multicolumn{1}{r}{$\sigma_\mathrm{CCF-BIS}$} &
\multicolumn{1}{r}{CCF-FWHM} &
\multicolumn{1}{r}{CCF-CTR} &
\multicolumn{1}{r}{$\mathrm{\log{R^{`}_{HK}}}$} &
\multicolumn{1}{r}{$\mathrm{\sigma_{\log{R^{`}_{HK}}}}$} &
\multicolumn{1}{r}{$\mathrm{S_{MW}}$} &
\multicolumn{1}{r}{$\mathrm{\sigma_{S_{MW}}}$} &
\multicolumn{1}{r}{S/N} &
\multicolumn{1}{r}{$\mathrm{T_{exp}}$}\\ 
\multicolumn{1}{r}{-2457000} &
\multicolumn{1}{r}{($\mathrm{m\,s^{-1}}$)} &
\multicolumn{1}{r}{($\mathrm{m\,s^{-1}}$)} &
\multicolumn{1}{r}{($\mathrm{m\,s^{-1}}$)} &
\multicolumn{1}{r}{($\mathrm{m\,s^{-1}}$)} &
\multicolumn{1}{r}{($\mathrm{km\,s^{-1}}$)} &
\multicolumn{1}{r}{(\%)} &
\multicolumn{1}{r}{---} &
\multicolumn{1}{r}{---} &
\multicolumn{1}{r}{---} &
\multicolumn{1}{r}{---} &
\multicolumn{1}{r}{@550nm} &
\multicolumn{1}{r}{(s)}\\ 
\noalign{\smallskip}
\hline
}
\tablehead{
\multicolumn{11}{c}
{{\bfseries \tablename\ \thetable{}} -- continued from previous page.}\\
\hline
\hline
\noalign{\smallskip}
\multicolumn{1}{r}{BJD$_\mathrm{TDB}$} &
\multicolumn{1}{r}{RV} &
\multicolumn{1}{r}{$\sigma_\mathrm{RV}$} &
\multicolumn{1}{r}{CCF-BIS} &
\multicolumn{1}{r}{$\sigma_\mathrm{CCF-BIS}$} &
\multicolumn{1}{r}{CCF-FWHM} &
\multicolumn{1}{r}{CCF-CTR} &
\multicolumn{1}{r}{$\mathrm{\log{R^{`}_{HK}}}$} &
\multicolumn{1}{r}{$\mathrm{\sigma_{\log{R^{`}_{HK}}}}$} &
\multicolumn{1}{r}{$\mathrm{S_{MW}}$} &
\multicolumn{1}{r}{$\mathrm{\sigma_{S_{MW}}}$} &
\multicolumn{1}{r}{S/N} &
\multicolumn{1}{r}{$\mathrm{T_{exp}}$}\\ 
\multicolumn{1}{r}{-2457000} &
\multicolumn{1}{r}{($\mathrm{m\,s^{-1}}$)} &
\multicolumn{1}{r}{($\mathrm{m\,s^{-1}}$)} &
\multicolumn{1}{r}{($\mathrm{m\,s^{-1}}$)} &
\multicolumn{1}{r}{($\mathrm{m\,s^{-1}}$)} &
\multicolumn{1}{r}{($\mathrm{km\,s^{-1}}$)} &
\multicolumn{1}{r}{(\%)} &
\multicolumn{1}{r}{---} &
\multicolumn{1}{r}{---} &
\multicolumn{1}{r}{---} &
\multicolumn{1}{r}{---} &
\multicolumn{1}{r}{@550nm} &
\multicolumn{1}{r}{(s)}\\ 
\noalign{\smallskip}
\hline
}
\tabletail{
\hline
\multicolumn{11}{c}{{Continued on next page}}\\
\hline
}
\tablelasttail{
\hline
}
\tablecaption{Radial velocities and spectral activity indicators measured from TNG/HARPS-N spectra measured with DRS.\label{table-C10_1452-tng_harpn-0071-drs-complete_output}}
\begin{supertabular}{rrrrrrrrrrrrr}
     \noalign{\smallskip}
     1493.68912 &        9371.649 &           2.184 &          -5.742 &           3.089 &           7.033 &          49.378 &         -4.7867 &          0.0201 &         0.2105 &          0.0058 &            44.3 &          2400.0\\
     
     1493.76708 &        9371.935 &           1.785 &          -7.580 &           2.524 &           7.033 &          49.440 &         -4.7941 &          0.0145 &         0.2084 &          0.0041 &            52.3 &          2400.0\\
     1494.68854 &        9378.519 &           1.697 &          -4.550 &           2.400 &           7.025 &          49.464 &         -4.7841 &          0.0136 &         0.2113 &          0.0039 &            54.5 &          2500.0\\
     \ldots & \ldots & \ldots & \ldots & \ldots & \ldots & \ldots & \ldots & \ldots & \ldots & \ldots & \ldots & \ldots \\
     \noalign{\smallskip}
\end{supertabular}
\end{center}
\end{tiny}
\medskip
\begin{tiny}
\begin{center}
\tablefirsthead{
\hline
\hline
\noalign{\smallskip}
\multicolumn{1}{r}{BJD$_\mathrm{TDB}$} &
\multicolumn{1}{r}{RV} &
\multicolumn{1}{r}{$\sigma_\mathrm{RV}$} &
\multicolumn{1}{r}{CRX} &
\multicolumn{1}{r}{$\sigma_\mathrm{CRX}$} &
\multicolumn{1}{r}{dlW} &
\multicolumn{1}{r}{$\sigma_\mathrm{dlW}$} &
\multicolumn{1}{r}{$\mathrm{H_{\alpha}}$} &
\multicolumn{1}{r}{$\mathrm{\sigma_{H_{\alpha}}}$} &
\multicolumn{1}{r}{$\mathrm{NaD_{1}}$} &
\multicolumn{1}{r}{$\mathrm{\sigma_{NaD_{1}}}$} &
\multicolumn{1}{r}{$\mathrm{NaD_{2}}$} &
\multicolumn{1}{r}{$\mathrm{\sigma_{NaD_{2}}}$}\\ 
\multicolumn{1}{r}{-2457000} &
\multicolumn{1}{r}{($\mathrm{m\,s^{-1}}$)} &
\multicolumn{1}{r}{($\mathrm{m\,s^{-1}}$)} &
\multicolumn{1}{r}{($\mathrm{m\,s^{-1}\,Np^{-1}}$)} &
\multicolumn{1}{r}{($\mathrm{m\,s^{-1}\,Np^{-1}}$)} &
\multicolumn{1}{r}{($\mathrm{m^2\,s^{-2}}$)} &
\multicolumn{1}{r}{($\mathrm{m^2\,s^{-2}}$)} &
\multicolumn{1}{r}{---} &
\multicolumn{1}{r}{---} &
\multicolumn{1}{r}{---} &
\multicolumn{1}{r}{---} &
\multicolumn{1}{r}{---} &
\multicolumn{1}{r}{---}\\
\noalign{\smallskip}
\hline
}
\tablehead{
\multicolumn{13}{c}
{{\bfseries \tablename\ \thetable{}} -- continued from previous page.}\\
\hline
\hline
\noalign{\smallskip}
\multicolumn{1}{r}{BJD$_\mathrm{TDB}$} &
\multicolumn{1}{r}{RV} &
\multicolumn{1}{r}{$\sigma_\mathrm{RV}$} &
\multicolumn{1}{r}{CRX} &
\multicolumn{1}{r}{$\sigma_\mathrm{CRX}$} &
\multicolumn{1}{r}{dlW} &
\multicolumn{1}{r}{$\sigma_\mathrm{dlW}$} &
\multicolumn{1}{r}{$\mathrm{H_{\alpha}}$} &
\multicolumn{1}{r}{$\mathrm{\sigma_{H_{\alpha}}}$} &
\multicolumn{1}{r}{$\mathrm{NaD_{1}}$} &
\multicolumn{1}{r}{$\mathrm{\sigma_{NaD_{1}}}$} &
\multicolumn{1}{r}{$\mathrm{NaD_{2}}$} &
\multicolumn{1}{r}{$\mathrm{\sigma_{NaD_{2}}}$}\\ 
\multicolumn{1}{r}{-2457000} &
\multicolumn{1}{r}{($\mathrm{m\,s^{-1}}$)} &
\multicolumn{1}{r}{($\mathrm{m\,s^{-1}}$)} &
\multicolumn{1}{r}{($\mathrm{m\,s^{-1}\,Np^{-1}}$)} &
\multicolumn{1}{r}{($\mathrm{m\,s^{-1}\,Np^{-1}}$)} &
\multicolumn{1}{r}{($\mathrm{m^2\,s^{-2}}$)} &
\multicolumn{1}{r}{($\mathrm{m^2\,s^{-2}}$)} &
\multicolumn{1}{r}{---} &
\multicolumn{1}{r}{---} &
\multicolumn{1}{r}{---} &
\multicolumn{1}{r}{---} &
\multicolumn{1}{r}{---} &
\multicolumn{1}{r}{---}\\ 
\noalign{\smallskip}
\hline
}
\tabletail{
\hline
\multicolumn{13}{c}{{Continued on next page}}\\
\hline
}
\tablelasttail{
\hline
}
\tablecaption{Radial velocities and spectral activity indicators measured from TNG/HARPS-N spectra measured with SERVAL.\label{table-C10_1452-tng_harpn-0071-srv-complete_output}}
\begin{supertabular}{rrrrrrrrrrrrr}
     \noalign{\smallskip}
     1493.68912 &          16.423 &           1.421 &          -0.679 &          11.345 &          -6.468 &           3.087 &          0.4305 &          0.0019 &          0.2559 &          0.0020 &          0.3729 &          0.0026\\
     1493.76708 &          21.453 &           1.589 &           8.883 &          12.174 &         -14.003 &           2.538 &          0.4261 &          0.0015 &          0.2570 &          0.0016 &          0.3790 &          0.0021\\
     1494.68854 &          26.208 &           1.439 &         -21.116 &          11.037 &         -11.974 &           2.037 &          0.4302 &          0.0016 &          0.2461 &          0.0016 &          0.3718 &          0.0021\\
     \ldots & \ldots & \ldots & \ldots & \ldots & \ldots & \ldots & \ldots & \ldots & \ldots & \ldots & \ldots & \ldots \\
     \noalign{\smallskip}
\end{supertabular}
\end{center}
\end{tiny}
\end{sidewaystable}
\begin{sidewaystable}
\begin{tiny}
\begin{center}
\tablefirsthead{
\hline
\hline
\noalign{\smallskip}
\multicolumn{1}{r}{BJD$_\mathrm{TDB}$} &
\multicolumn{1}{r}{RV} &
\multicolumn{1}{r}{$\sigma_\mathrm{RV}$} &
\multicolumn{1}{r}{CCF-BIS} &
\multicolumn{1}{r}{$\sigma_\mathrm{CCF-BIS}$} &
\multicolumn{1}{r}{CCF-FWHM} &
\multicolumn{1}{r}{CCF-CTR} &
\multicolumn{1}{r}{CCF-ASYM} &
\multicolumn{1}{r}{$\sigma_\mathrm{CCF-ASYM}$} &
\multicolumn{1}{r}{$\mathrm{S_{MW}}$} &
\multicolumn{1}{r}{$\mathrm{\sigma_{S_{MW}}}$}\\
\multicolumn{1}{r}{-2457000} &
\multicolumn{1}{r}{($\mathrm{m\,s^{-1}}$)} &
\multicolumn{1}{r}{($\mathrm{m\,s^{-1}}$)} &
\multicolumn{1}{r}{($\mathrm{m\,s^{-1}}$)} &
\multicolumn{1}{r}{($\mathrm{m\,s^{-1}}$)} &
\multicolumn{1}{r}{($\mathrm{km\,s^{-1}}$)} &
\multicolumn{1}{r}{(\%)} &
\multicolumn{1}{r}{($\mathrm{km\,s^{-1}}$)} &
\multicolumn{1}{r}{($\mathrm{km\,s^{-1}}$)} &
\multicolumn{1}{r}{---} &
\multicolumn{1}{r}{---}\\
\noalign{\smallskip}
\hline
}
\tablehead{
\multicolumn{11}{c}
{{\bfseries \tablename\ \thetable{}} -- continued from previous page.}\\
\hline
\hline
\noalign{\smallskip}
\multicolumn{1}{r}{BJD$_\mathrm{TDB}$} &
\multicolumn{1}{r}{RV} &
\multicolumn{1}{r}{$\sigma_\mathrm{RV}$} &
\multicolumn{1}{r}{CCF-BIS} &
\multicolumn{1}{r}{$\sigma_\mathrm{CCF-BIS}$} &
\multicolumn{1}{r}{CCF-FWHM} &
\multicolumn{1}{r}{CCF-CTR} &
\multicolumn{1}{r}{CCF-ASYM} &
\multicolumn{1}{r}{$\sigma_\mathrm{CCF-ASYM}$} &
\multicolumn{1}{r}{$\mathrm{S_{MW}}$} &
\multicolumn{1}{r}{$\mathrm{\sigma_{S_{MW}}}$}\\
\multicolumn{1}{r}{-2457000} &
\multicolumn{1}{r}{($\mathrm{m\,s^{-1}}$)} &
\multicolumn{1}{r}{($\mathrm{m\,s^{-1}}$)} &
\multicolumn{1}{r}{($\mathrm{m\,s^{-1}}$)} &
\multicolumn{1}{r}{($\mathrm{m\,s^{-1}}$)} &
\multicolumn{1}{r}{($\mathrm{km\,s^{-1}}$)} &
\multicolumn{1}{r}{(\%)} &
\multicolumn{1}{r}{($\mathrm{km\,s^{-1}}$)} &
\multicolumn{1}{r}{($\mathrm{km\,s^{-1}}$)} &
\multicolumn{1}{r}{---} &
\multicolumn{1}{r}{---}\\
\noalign{\smallskip}
\hline
}
\tabletail{
\hline
\multicolumn{11}{c}{{Continued on next page}}\\
\hline
}
\tablelasttail{
\hline
}
\tablecaption{Radial velocities and spectral activity indicators measured from VLT/ESPRESSO spectra measured with DRS.\label{table-C10_1452-vlt_espresso-0025-drs-complete_output}}
\begin{supertabular}{rrrrrrrrrrrrr}
     \noalign{\smallskip}
     3295.82419 &        9248.034 &           0.998 &          -85.050 &           1.996 &           7.719 &          62.556 &         0.0371 &          0.0017 &         0.1852 &          0.0020 \\
     3297.82759 &        9247.546 &           0.951 &          -87.867 &           1.902 &           7.720 &          62.630 &         0.0373 &          0.0017 &         0.1820 &          0.0018 \\
     3302.82478 &        9247.186 &           0.848 &          -85.695 &           1.697 &           7.721 &          62.646 &         0.0370 &          0.0015 &         0.1870 &          0.0015 \\
     \ldots & \ldots & \ldots & \ldots & \ldots & \ldots & \ldots & \ldots & \ldots & \ldots & \ldots \\
     \noalign{\smallskip}
\end{supertabular}
\end{center}
\end{tiny}
\medskip
\begin{tiny}
\begin{center}
\tablefirsthead{
\hline
\hline
\noalign{\smallskip}
\multicolumn{1}{r}{BJD$_\mathrm{TDB}$} &
\multicolumn{1}{r}{RV} &
\multicolumn{1}{r}{$\sigma_\mathrm{RV}$} &
\multicolumn{1}{r}{CRX} &
\multicolumn{1}{r}{$\sigma_\mathrm{CRX}$} &
\multicolumn{1}{r}{dlW} &
\multicolumn{1}{r}{$\sigma_\mathrm{dlW}$} &
\multicolumn{1}{r}{$\mathrm{H_{\alpha}}$} &
\multicolumn{1}{r}{$\mathrm{\sigma_{H_{\alpha}}}$} &
\multicolumn{1}{r}{$\mathrm{NaD_{1}}$} &
\multicolumn{1}{r}{$\mathrm{\sigma_{NaD_{1}}}$} &
\multicolumn{1}{r}{$\mathrm{NaD_{2}}$} &
\multicolumn{1}{r}{$\mathrm{\sigma_{NaD_{2}}}$}\\ 
\multicolumn{1}{r}{-2457000} &
\multicolumn{1}{r}{($\mathrm{m\,s^{-1}}$)} &
\multicolumn{1}{r}{($\mathrm{m\,s^{-1}}$)} &
\multicolumn{1}{r}{($\mathrm{m\,s^{-1}\,Np^{-1}}$)} &
\multicolumn{1}{r}{($\mathrm{m\,s^{-1}\,Np^{-1}}$)} &
\multicolumn{1}{r}{($\mathrm{m^2\,s^{-2}}$)} &
\multicolumn{1}{r}{($\mathrm{m^2\,s^{-2}}$)} &
\multicolumn{1}{r}{---} &
\multicolumn{1}{r}{---} &
\multicolumn{1}{r}{---} &
\multicolumn{1}{r}{---} &
\multicolumn{1}{r}{---} &
\multicolumn{1}{r}{---}\\
\noalign{\smallskip}
\hline
}
\tablehead{
\multicolumn{13}{c}
{{\bfseries \tablename\ \thetable{}} -- continued from previous page.}\\
\hline
\hline
\noalign{\smallskip}
\multicolumn{1}{r}{BJD$_\mathrm{TDB}$} &
\multicolumn{1}{r}{RV} &
\multicolumn{1}{r}{$\sigma_\mathrm{RV}$} &
\multicolumn{1}{r}{CRX} &
\multicolumn{1}{r}{$\sigma_\mathrm{CRX}$} &
\multicolumn{1}{r}{dlW} &
\multicolumn{1}{r}{$\sigma_\mathrm{dlW}$} &
\multicolumn{1}{r}{$\mathrm{H_{\alpha}}$} &
\multicolumn{1}{r}{$\mathrm{\sigma_{H_{\alpha}}}$} &
\multicolumn{1}{r}{$\mathrm{NaD_{1}}$} &
\multicolumn{1}{r}{$\mathrm{\sigma_{NaD_{1}}}$} &
\multicolumn{1}{r}{$\mathrm{NaD_{2}}$} &
\multicolumn{1}{r}{$\mathrm{\sigma_{NaD_{2}}}$}\\ 
\multicolumn{1}{r}{-2457000} &
\multicolumn{1}{r}{($\mathrm{m\,s^{-1}}$)} &
\multicolumn{1}{r}{($\mathrm{m\,s^{-1}}$)} &
\multicolumn{1}{r}{($\mathrm{m\,s^{-1}\,Np^{-1}}$)} &
\multicolumn{1}{r}{($\mathrm{m\,s^{-1}\,Np^{-1}}$)} &
\multicolumn{1}{r}{($\mathrm{m^2\,s^{-2}}$)} &
\multicolumn{1}{r}{($\mathrm{m^2\,s^{-2}}$)} &
\multicolumn{1}{r}{---} &
\multicolumn{1}{r}{---} &
\multicolumn{1}{r}{---} &
\multicolumn{1}{r}{---} &
\multicolumn{1}{r}{---} &
\multicolumn{1}{r}{---}\\ 
\noalign{\smallskip}
\hline
}
\tabletail{
\hline
\multicolumn{13}{c}{{Continued on next page}}\\
\hline
}
\tablelasttail{
\hline
}
\tablecaption{Radial velocities and spectral activity indicators measured from VLT/ESPRESSO spectra measured with SERVAL.\label{table-C10_1452-vlt_espresso-0025-srv-complete_output}}
\begin{supertabular}{rrrrrrrrrrrrr}
     \noalign{\smallskip}
     3295.82419 &          0.455 &           0.736 &           6.102 &          4.493 &          -0.586 &           0.228 &          0.4153 &          0.0011 &          0.2535 &          0.0013 &          0.3358 &          0.0015\\
     3297.82759 &          0.466 &           0.658 &           6.038 &          4.019 &          -0.412 &           0.241 &          0.4119 &          0.0011 &          0.2491 &          0.0012 &          0.3379 &          0.0015\\
     3302.82478 &         -0.368 &           0.611 &           8.176 &          3.657 &          -0.563 &           0.215 &          0.4129 &          0.0010 &          0.2509 &          0.0011 &          0.3372 &          0.0013\\
     \ldots & \ldots & \ldots & \ldots & \ldots & \ldots & \ldots & \ldots & \ldots & \ldots & \ldots & \ldots & \ldots \\
     \noalign{\smallskip}
\end{supertabular}
\end{center}
\end{tiny}
\end{sidewaystable}
\end{appendix}

\end{document}